\documentclass[10pt]{article}

\usepackage{lineno}
\usepackage{todonotes}
\usepackage{xargs} 
\usepackage[toc]{appendix}
\usepackage{lscape}
\usepackage{float}
\usepackage[flushleft]{threeparttable}
\usepackage{placeins}
\usepackage{array,booktabs,ragged2e}
\usepackage{caption,subcaption}
\usepackage{footnote, longtable}
\usepackage{setspace}
\usepackage{color,soul}
\usepackage{graphicx}
\usepackage{amssymb}
\usepackage{amsthm}
\usepackage{amsmath}
\usepackage{url}
\usepackage{moresize}
\usepackage{epstopdf}
\usepackage{bbm}
\usepackage{booktabs}
\usepackage{multirow}
\usepackage{footmisc}
\usepackage{setspace}
\usepackage{rotating}
\usepackage{textgreek}
\usepackage{longtable}
\usepackage{enumitem}
\usepackage{verbatim}
\usepackage{xr-hyper}
\usepackage{hyperref}
\usepackage{svg}
\usepackage{authblk}
\usepackage{caption}
\usepackage{float} 
\usepackage[export]{adjustbox}
\usepackage{nomencl}
\usepackage{algorithm2e}
\usepackage{textcomp}
\usepackage{pifont} 
\usepackage{csquotes} 
\usepackage{siunitx} 

\usepackage[backend=biber,
            style=numeric-comp,
            sorting=none,
            maxcitenames=1]{biblatex}

\hypersetup{
    colorlinks=true,       
    linkcolor=red,          
    citecolor=black,        
    filecolor=blue,      
    urlcolor=blue           
}

\usepackage{xr}
\makeatletter
\newcommand*{\addFileDependency}[1]{
    \typeout{(#1)}
    \@addtofilelist{#1}
    \IfFileExists{#1}{}{\typeout{No file #1.}}
}
\makeatother

\title{A multi-model framework for identifying affordable low-carbon electricity pathways for India under uncertainty} 



\date{\today}

\author[1,2,*]{Guillermo Terr\'en-Serrano}
\author[1,2,3,*,$\dagger$]{Ranjit Deshmukh}
\author[4]{Srihari Dukkipati}
\author[2]{Measrainsey Meng}
\author[2,3,5]{Shradhey Prasad}
\author[6]{Ana Mileva}
\author[4]{Ashwin Gambhir}

\affil[1]{\small{Environmental Studies Program, University of California Santa Barbara, Santa Barbara, CA, USA.}}
\affil[2]{\small{Environmental Markets Lab (emLab), University of California Santa Barbara, Santa Barbara, CA, USA.}}
\affil[3]{\small{Bren School of Environment Science \& Management, University of California Santa Barbara, Santa Barbara, CA, USA.}}
\affil[4]{\small{Prayas (Energy Group), Pune, MH, India.}}
\affil[5]{\small{Marine Science Institute, University of California Santa Barbara, Santa Barbara, CA, USA}}
\affil[6]{\small{Sylvan Energy Analytics, Portland, OR, USA.}}
\affil[*]{\small{These authors contributed equally.}}
\affil[$\dagger$]{\small{Corresponding author: \texttt{rdeshmukh@ucsb.edu}}}

\usepackage[font=small]{caption}

\usepackage[margin=.65in]{geometry}

\usepackage{pdfpages}

\begin{document}


\maketitle

\begin{abstract}

    To meet climate goals, countries must decarbonize their electricity systems. Yet, uncertainty in future technology costs, electricity demand, and renewable energy generation complicates power system planning and makes the affordability of clean electricity uncertain. Here, we develop a multi-model framework to examine cost-optimal pathways for generation, storage, and transmission expansion in India under alternative technology costs, electricity demand projections, and clean energy and carbon-emission targets. Across all scenarios, real average system costs remain below 2020 levels throughout 2030-2050, even when carbon emissions decline linearly to 90\% below current levels by 2050. Solar PV comprises over half to three-quarters of total installed capacity by 2050 supported by large-scale deployment of short-duration battery storage. Expanding green hydrogen, pumped hydro storage, and nuclear reduces costs by less than 2\%, whereas demand response lowers costs by up to 10\%. Declining renewable and battery costs, together with demand response, make deep electricity decarbonization affordable across a range of realistic scenarios.

\end{abstract}




\section{Introduction}


Decarbonizing the electricity sector is a major strategy for meeting climate goals because of the declining costs of wind, solar, and battery storage technologies, and the potential for electrifying the transportation, industry, and other sectors \cite{williams_technology_2012,larson_net-zero_nodate,das_pathways_2023}.  However, uncertainties in technology cost trajectories and electricity demand forecasts, in addition to the weather-dependency of wind and solar resources, introduce new challenges to planning and operating the electricity system. More importantly, climate targets need to be balanced against energy affordability, access, and economic growth, especially in emerging economies. This challenge is particularly salient for India, the world's most populous country, fastest growing major economy, and the third largest greenhouse gas emitter (7.6\% of annual global greenhouse gas (GHG) emissions \cite{european_commission_joint_research_centre_ghg_2025}), but with an average gross domestic product (GDP) and GHG emissions per capita substantially lower than the world average \cite{world_bank_world_2025}. With an average per capita energy consumption also below world average, India's total emissions could substantially grow in the future depending on how the country provides basic energy services, powers its growing economy, and adapts to climate change through space cooling and other energy services. While India has set a goal of economy-wide carbon neutrality by 2070, the country will need relatively ambitious medium and long-term clean energy and climate policies to decarbonize its electricity sector earlier by 2050, which will in turn help decarbonize the rest of the economy. In this study, developing a multi-model framework, we examine how existing and alternative clean energy and climate policies in India will balance electricity costs against GHG emissions and new infrastructure investments under technology cost and electricity demand uncertainty across 2030 to 2050.

Previous studies of India's energy transition through 2040 and beyond have examined technology costs and performance, clean energy and climate policy, or both. Rose et al., Rudnick et al., and Barbar et al. modeled the effects of technology cost projections, a key driver of decarbonization \cite{rose_least-cost_2020, rudnick_decarbonization_2022, barbar_impact_2023}, and Barbar et al. further used bottom-up demand projections to assess how storage costs and air-conditioner efficiency affect variable renewable energy (VRE) and storage deployment and emissions through 2050 \cite{barbar_impact_2023}. Beyond solar and wind, other studies have examined green hydrogen \cite{song_deep_2022} and nuclear energy \cite{bhattacharya_bending_2024}. Policy-driven scenarios explored include India's 500~GW non-fossil capacity target and 2070 net-zero pledge \cite{abhyankar_indias_2023}, varying renewable energy shares over time \cite{lu_indias_2020, rodrigues_indias_2023}, and a carbon-tax-driven 100\% clean energy system by 2050 \cite{gulagi_role_2022}. These studies diverge in projected infrastructure investment, system costs, and emissions, reflecting differences in (1) input assumptions (technology costs and availability, demand projections, and clean energy and climate targets) and (2) model structure, including temporal and spatial resolution and technical constraints. We improve upon these studies by developing technology cost projection models using multiple forecasts and empirical data; bottom-up electricity demand projections representing impacts of economic growth, climate change, and emerging patterns of agricultural load; and hourly generation profiles of hundreds of solar and wind projects to capture the weather dependency of these resources. Our open-source models, which we describe below, have the finest spatial  (state-level representation) and temporal resolution (hourly) across all previous studies, and realistic representation of reserve margins and outages, which is important for capturing transmission congestion and reliability in low-carbon systems. We compare these studies along both dimensions in Supplementary Tables~\ref{stab:literature_review} and \ref{stab:literature_review_continue}.

In this study, we develop a multi-model framework to evaluate low-cost and low-carbon investment pathways for India under uncertainty in electricity demand and technology cost projections. This framework includes an open-source electricity system planning and operations model---\emph{GridPath-India}---built on the open-source GridPath power systems modeling platform \cite{blue_marble_analytics_gridpath_2025}. This model co-optimizes generation, storage, and transmission investments along with hourly operations of the system across four periods (2020, 2030, 2040, and 2050), each representing 10~years, under several economic, technical, and policy constraints. To capture transmission constraints in India's power system, we model 35 load zones with an hourly demand profiles and existing, planned, and candidate generation and storage projects. Load zones are connected by transmission corridors. For technology costs, we develop low, mid, and high projections across 2020-2050 for all candidate clean energy technologies, including residential and utility-scale solar PV, onshore and offshore wind, battery storage, and hydrogen, using India-specific and global data, and combine them with two coal power plant capital cost levels (see Supplementary Note~\ref{si:cost_projections}). To represent the uncertainty in future electricity demand growth and patterns, we deploy a bottom-up electricity demand model, \textit{PIER}, which includes additional demand from economic growth, agricultural demand shift to solar hours, increased cooling loads, and climate impacts \cite{sreenivas_pier_2021}. To capture the uncertainty in weather-dependent solar and wind resources, we use the multi-criteria for planning renewable energy (MapRE) and the REZoning models to spatially identify over 1,300 candidate onshore and offshore wind and solar sites and develop hourly generation profiles using weather data \cite{deshmukh_geospatial_2019}. We design three sets of scenarios to examine the tradeoffs between electricity costs against GHG emission mitigation under uncertainty---(1) technology cost projections, (2) electricity demand projections, and (3) clean energy and climate policy targets, which includes India's near-term (2030) clean energy capacity target of 500 GW and a set of potential linearly-decreasing carbon emission targets until 2050 (Table~\ref{tab:scenarios}).

\begin{table*}[htb!]
    \centering
    \footnotesize
    \caption{\textbf{Technology cost, policy, and demand scenarios}. Technology costs for Variable Renewable Energy (VRE) and Energy Storage Systems (ESS) are given in Supplementary Note~\ref{si:cost_projections}. Bottom-up electricity demand projections are from \emph{PIER v2.0} \cite{prayas_energy_group_pier_2025}. Linearly-scaled electricity demand projections match the annual bottom-up energy projections but maintain the 2019 hourly demand profiles. Policy scenarios include the 500~GW clean-energy or non-fossil fuel generation capacity target by 2030 and a target of 80\%, 90\%, or 100\% CO$_2$ emission reduction from 2020 levels by 2050, and a 50\% renewable purchase obligation (RPO) for the 22 states with the highest electricity demand in 2040. Combination in bold is the reference scenario (\textit{REF}).}
    \setlength{\tabcolsep}{4pt} 
    \renewcommand{\arraystretch}{1} 
    \begin{tabular}{c|c|c|c|c|c|c}
    \toprule
    \multicolumn{2}{c|}{\textbf{Technology Cost}} & \multicolumn{2}{c|}{\textbf{Demand profile and growth}}  & \multicolumn{3}{c}{\textbf{Policy}}\\
    \midrule
    \textbf{VRE/ESS} & \textbf{Coal} & \textbf{Bottom-Up} & \textbf{Linearly-Scaled}& \textbf{2030 Clean-Energy Target} & \textbf{2050 Carbon Target} & \textbf{2040/2050 RPO} \\
    \midrule
    low & \textbf{low}& low growth & low growth  & No & No &  \textbf{50\%} \\
    \textbf{mid} & high & \textbf{mid} & mid  & \textbf{500~GW} & 80\% \\
    high & & high & high & & \textbf{90\%} &  \\
    & & & & & 100\% &  \\
    \bottomrule
\end{tabular}
    \label{tab:scenarios}
\end{table*}

\section{Methods}\label{sec:methods}
\begin{figure}[htb!]
    \centering
    \includegraphics[scale = .725, trim = {0cm 0cm 0cm 0cm}, clip]{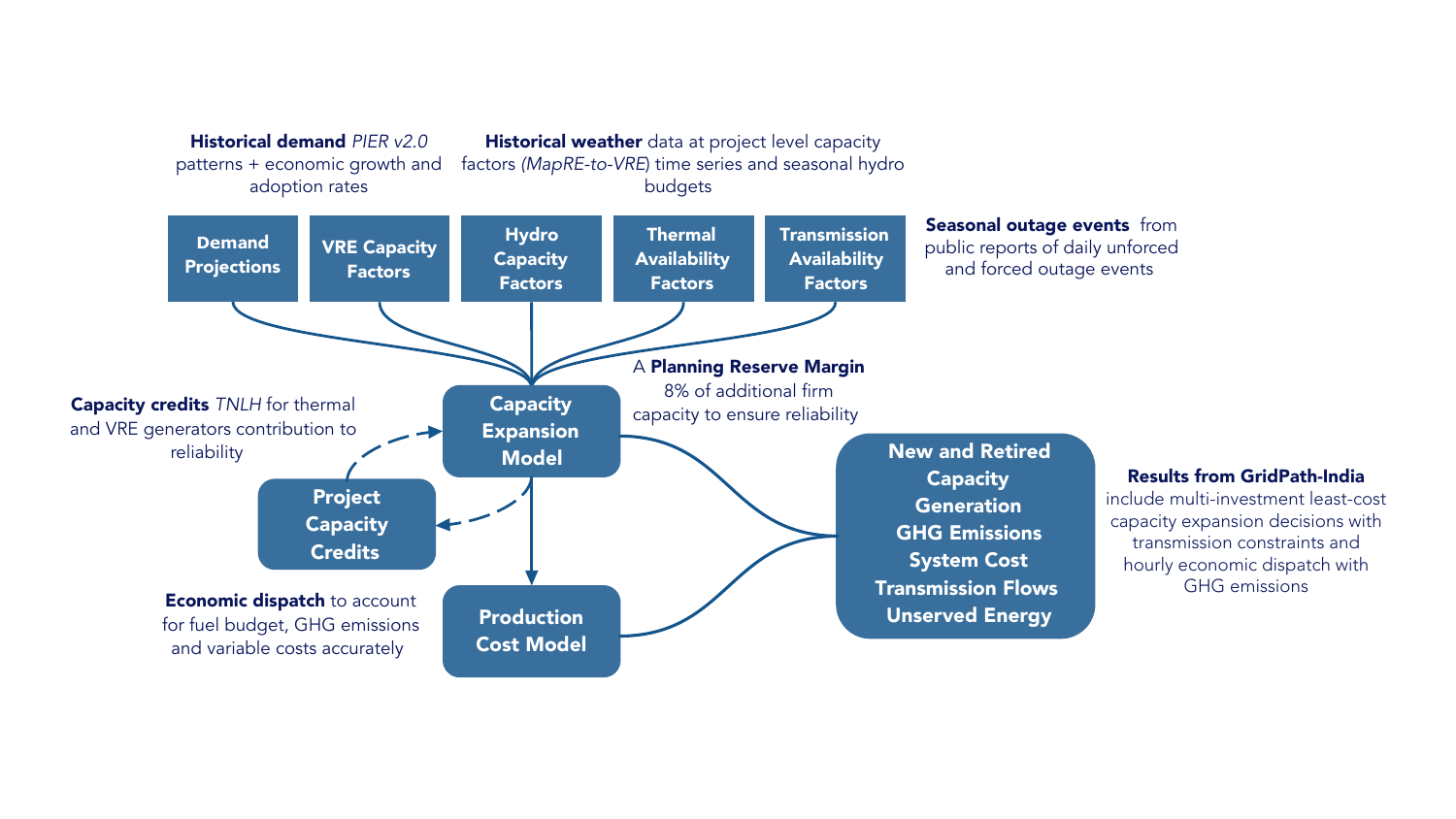}
    \caption{\textbf{Multi-model Framework}. \emph{GridPath-India} is an open-source model of India's national power system built on the \emph{GridPath} power system planning and operation modeling platform. The Perspective on Indian Energy based on Rumi (\emph{PIER}) is an energy systems model to develop sectoral energy demand projections for India. The Multi-criteria Analysis for Planning Renewable Energy (MapRE) model develops spatially-explicit variable renewable energy (VRE) zones and generation data. Capacity credits are contributions of generators and storage during top net load hours (TNLH) that meet the planning reserve margin constraint in the \emph{GridPath-India} capacity expansion model.}
    \label{fig:workflow}
\end{figure}

To identify cost-optimal electricity infrastructure investments, costs, and emissions of India's future electricity system under technical, economic, and policy constraints, we developed a multi-model framework (Figure~\ref{fig:workflow}), that includes an open-source electricity system (capacity expansion and production cost) model, a spatial renewable resource assessment model, and an electricity demand projection model.

\begin{figure}[htb!]
    \includegraphics[scale = .2, trim = {-1.8cm 1cm 0cm 0cm}, clip]{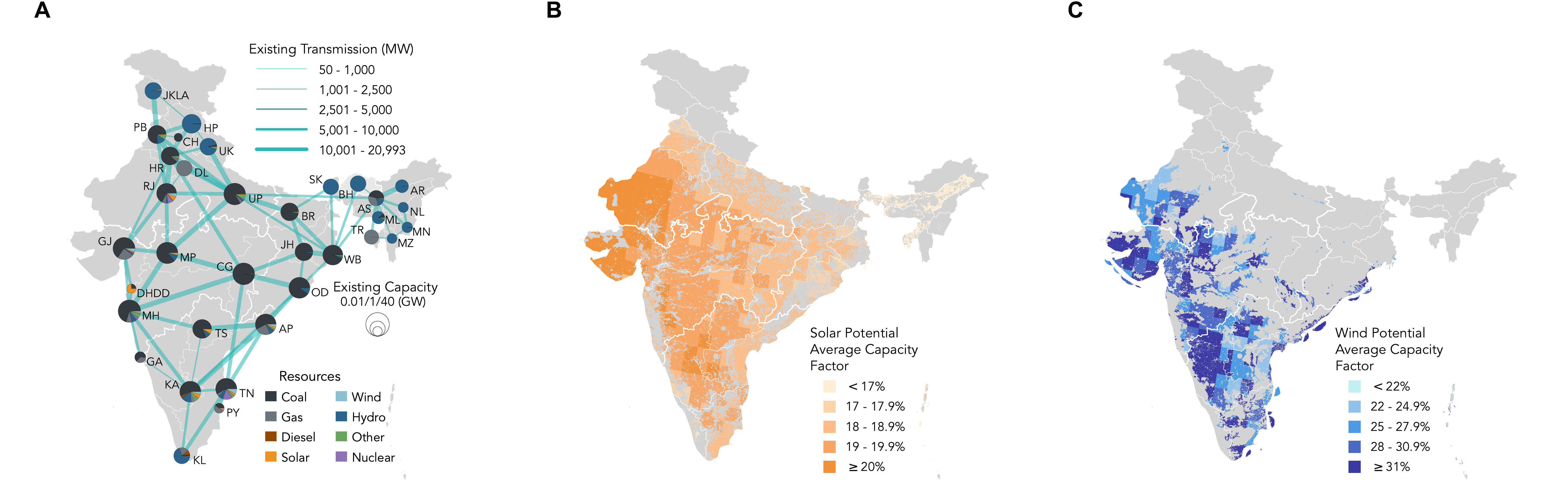}
    \includegraphics[scale = .1975, trim = {-.25cm 0cm 0cm 0cm}, clip]{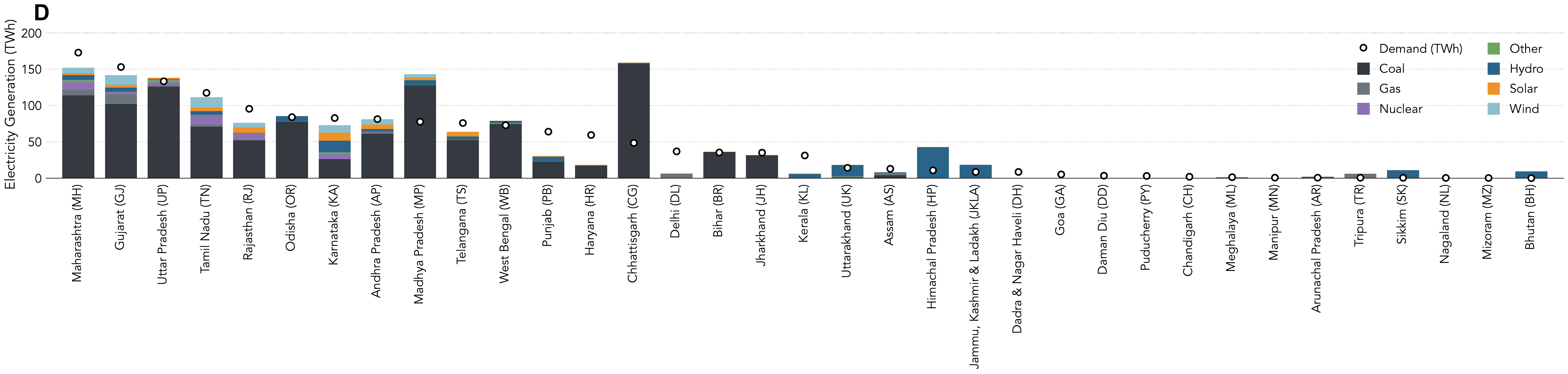}
    \includegraphics[scale = .2, trim = {0cm 0cm 0cm 0cm}, clip]{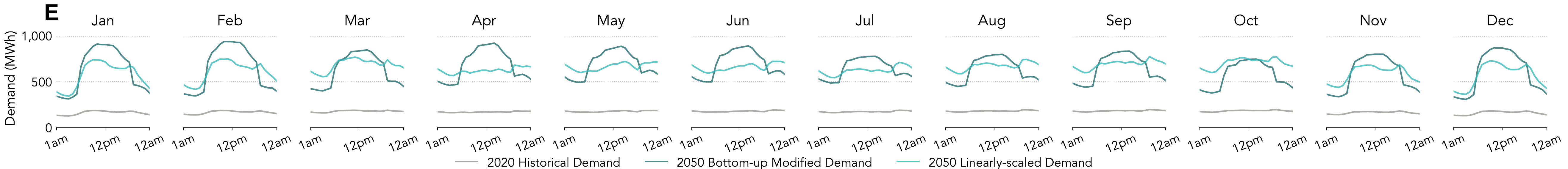}
    \includegraphics[scale = .2, trim = {0cm 0cm 0cm -.75cm}, clip]{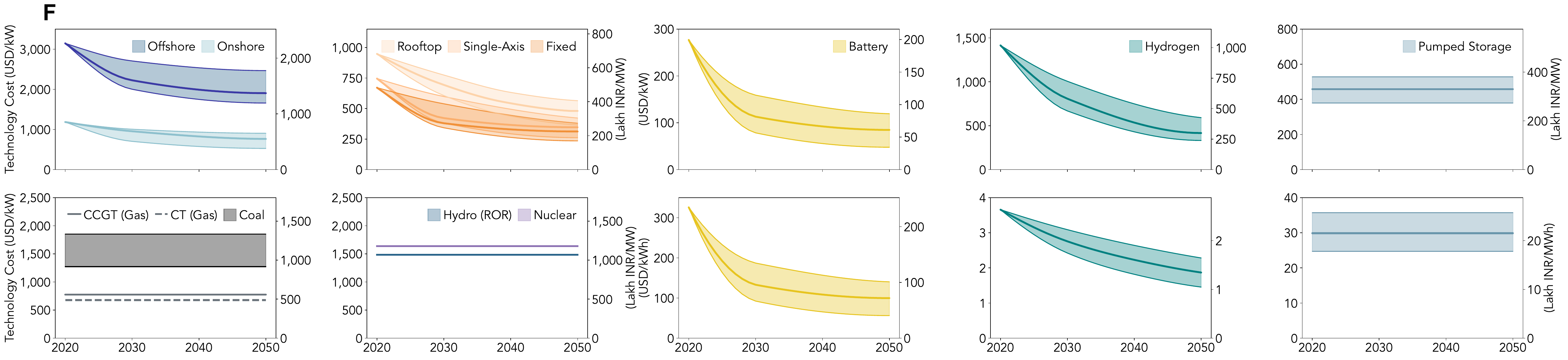}
    \caption{(A) Existing generation capacity mix (2020) and interstate transmission corridors, (B) Solar PV resource average capacity factors, (C) Onshore and offshore wind resource average capacity factors, (D) State-wise electricity generation by resource (stacked bars) and electricity demand (circles) in 2020, (E) National-level average hourly electricity demand by month in 2020 and in 2050 for two projection scenarios (bottom-up modeled and linearly-scaled demand, and (F) Capital cost projections from 2020-2050 for onshore and offshore wind, rooftop, single-axis, and fixed-tilt solar PV, coal, combined cycle gas turbine, combustion turbine, hydropower (run-of-river), nuclear, battery storage, hydrogen storage, and pumped hydro storage.}
    \label{fig:electricity_system_overview}
\end{figure}

\subsection{Electricity system model}\label{sec:electricity_system_model}

To develop a model of India's national power system, we use \emph{GridPath}, an open-source production cost and capacity expansion modeling platform \cite{blue_marble_analytics_gridpath_2025}, which defines a Mixed-Integer Linear programming (MILP) problem solved with a commercial solver \emph{Gurobi} \cite{gurobi_optimization_gurobi_2024}; see formulation in Supplementary Note~\ref{si:formulation}. The India model (\emph{GridPath-India}) has a spatial resolution of 35 load zones, each roughly representing a state, union territory, or the neighboring country of Bhutan (Supplementary Note~\ref{si:representation} and Table~\ref{stab:load_zones}). Each load zone is associated with an hourly electricity demand profile and existing and candidate generator and storage projects (Figure~\ref{fig:electricity_system_overview}A). 

Load zones are connected by interstate transmission corridors with limits on transmission transfer capacities (Figure~\ref{fig:electricity_system_overview}A, Supplementary Table~\ref{stab:spec_tx}, \cite{spencer_t_rodrigues_n_pachouri_r_thakre_s__renjith_g_renewable_2020}). Transmission corridors connect the most populous cities in load zones. Transmission losses are linearly correlated with distance and per unit distance loss factors. Transmission loss factors, line and substation costs, and line lifetimes are given in Supplementary Note~\ref{si:representation}. We ignore transmission and distribution constraints within states for computational tractability.

Over four investment periods (2020, 2030, 2040, and 2050) representing four decades, the model co-optimizes hourly operating costs and investments in new system infrastructure including generation, storage, and transmission, while meeting hourly electricity demand and meeting reliability and policy goals. Each investment period has an hourly temporal resolution and two representative days (representing peak and median demand) for each month to balance representation of seasonal and diurnal variability patterns with model tractability (Supplementary Note~\ref{si:demand_projections}). To examine system reliability across all hours of a year, we also run an hourly operations or production cost \emph{GridPath-India} model  

The inputs to \emph{GridPath} include projected hourly electricity demand for each investment period, existing and candidate generator, storage, and transmission capacities, hourly capacity factors for weather-dependent generators, and techno-economic, policy, and system reliability parameters (Supplementary Notes~\ref{si:formulation} and~\ref{si:representation}).

\subsection{Generators}

\paragraph{Existing generators.}\label{sec:existing_generators}

Existing generator capacity is from the fiscal year 2019-2020 (April 2019 - March 2020), see Figure~\ref{fig:electricity_system_overview}A,D. We aggregate this capacity by vintage (commission year), state, and technology type for model tractability. Technology types include supercritical coal (unit capacity $\ge$600~MW), subcritical coal large (unit capacity $\ge$250~MW $\ge$600~MW), subcritical coal small (unit capacity $<$250~MW), combined cycle gas turbine (CCGT), diesel, biomass (bagasse or sugarcane co-generation units), waste heat recovery (WHR), hydro storage, hydro run-of-the-river (ROR), pumped storage hydropower (PSH), onshore wind,, and solar photovoltaic (PV), (see Supplementary Table~\ref{stab:specified_capacity} for state-wise capacity). Each coal type is associated with a different heat rate (Supplementary Table~\ref{stab:heat_emissions_rates}). Local and imported liquefied natural gas have different costs. Load zones that host natural gas generators use one or the other based on historical usage (Supplementary Table~\ref{stab:spec_projects_char}). Hydro storage projects balance energy across each month with a fixed monthly energy budget based on historical generation. Hydro ROR is treated as must run, generating at historical average daily generation levels. (Supplementary Note~\ref{si:capacity_factors}). Hydro-electric plants in Bhutan are assumed to be ROR. We assume all existing solar capacity is fixed tilt even though small percentages of single axis and rooftop PV were deployed by 2020 \cite{rahul_kamat_track_2017, pc_division_state-wise_2020}.

Existing coal generators are retired exogenously based on age (45 years \cite{minister_of_power_draft_2023} Supplementary Table~\ref{stab:spec_projects_char}). Because of their relatively low installed capacity, existing projects of all other technologies are assumed to remain online through the entire study period. All new projects are retired endogenously after their operational lifetime (Supplementary Table~\ref{stab:new_projects_char}).

\paragraph{New generators.}\label{sec:planned_and_proposed_capacity}

We include hydro ROR (13.3~GW), supercritical coal (36.5~GW) \cite{kashish_shah_overestimated_2021}, and nuclear (4.8~GW) projects \cite{wna_nuclear_2025} that are in the pipeline, i.e., commissioned after 2020 or currently under construction (Supplementary Table~\ref{stab:pipeline_projects}). We also consider new candidate projects for different technologies to be cost-optimally selected by the model (Supplementary Table~\ref{stab:new_projects}). We allow candidate supercritical coal projects in 17 states, which have high electricity demand or existing coal capacity. We allow new gas projects, CCGT and Combustion Turbines (CT), only in 12 states with existing natural gas generation capacity (except in Assam) and allow the expansion of only imported liquefied natural gas because of constraints on domestic supplies of natural gas \cite{international_energy_agency_india_2025}. Beyond projects in the pipeline, nuclear capacity is allowed to expand cost-optimally in certain scenarios in states with existing capacity. Candidate PSH projects have a total power capacity potential of 138~GW and an energy capacity of 839~GWh \cite{cea_roadmap_2026}. New battery storage projects are considered in 21 states with existing coal or natural gas projects. For long duration storage, we consider green hydrogen \cite{thema_power--gas_2019}, limiting candidate projects to states where suitable underground nonporous cavern formations are available. We do not place limits on the energy capacity of those sites \cite{king_overview_2021} (Supplementary Table~\ref{stab:new_projects}). 


\paragraph{Operational characteristics.}\label{sec:existing_generators_op_char}

Subcritical and supercritical coal, natural gas CCGT and CT, and diesel generators are dispatchable with constraints on minimum up and down time hours, and minimum stable level fraction considering unit sizes. Nuclear generators are dispatchable but relatively inflexible with a 90\% minimum generation level. Hydro storage projects are must-run generators constrained by average monthly capacity factors across monthly balancing horizons and by their rated capacity and minimum historical generation level for maximum and minimum instantaneous generation. PSH and battery storage have daily balancing horizons with the same state-of-charge at the end of the daily balancing cycle. Biomass, hydro ROR, solar PV, and wind generators are non-dispatchable and hourly dispatch is based on historical generation. Biomass and hydro ROR are considered not curtailable, while solar PV and wind are treated as curtailable generation sources. Monthly capacity factors for hydro and biomass generators are derived from historical generation reports published by the CEA \cite{operation_performance_monitoring_division_monthly_nodate, renewable_energy_project_monitoring_division_monthly_nodate}, as detailed in Supplementary Note~\ref{si:capacity_factors}. Operational characteristics, heat rates, and CO\textsubscript{2} emission rates by technology are given in Supplementary Tables~\ref{stab:new_projects_char} and~\ref{stab:heat_emissions_rates}.

\subsection{Wind and solar resources}\label{sec:wind_and_solar_resources}

For identifying candidate onshore wind and solar sites (Figure~\ref{fig:electricity_system_overview}B,C), we adopted the Multi-criteria Analysis for Planning Renewable Energy (\emph{MapRE}) modeling framework \cite{wu_strategic_2017} and the \emph{REZoning} platform for candidate offshore wind sites \cite{esmap_and_ucsb_renewable_2023}. \emph{REZoning} is the global scale version of \emph{MapRE}, including offshore wind siting capabilities (Supplementary Note~\ref{si:mapre}).

We identify multiple suitable candidate sites at the state level by considering social (e.g.,~population, UNESCO sites), infrastructural (e.g.,~roads, ports, and water access), and environmental (e.g.,~protected area, terrain, slope) constraints to determine the available capacity at each site for renewable energy deployment. We select 556 solar PV sites, 300 onshore wind sites, and 46 offshore sites with suitable solar (global tilt irradiance $\ge4$~kWh/m$^2$ per day) or wind resources (wind speed at 100~m $\ge5$~m/s). To represent both utility-scale solar PV technologies, we assign 80\% of potential capacity at each candidate site to single-axis tracking systems and 20\% to fixed-tilt systems, totaling 1,112 solar PV sites. We add a rooftop solar PV candidate site in each of the 35 load zones, with generation profiles from the largest city in each state.

We simulate the capacity factor time series for each solar and wind candidate site, using the ERA5 reanalysis for wind and the National Solar Radiation Database (NSRDB) for solar \cite{sengupta_national_2018, hersbach_era5_2020}. Because the NSRDB PSM~V3 model extends only until 2019, we use ERA5 and PSM~V3 for the Fiscal Year 2018-19 (FY19). To overcome the limitation of low resolution of wind resource (onshore and offshore), we adjust the wind velocity time series data from ERA5 for each location such that the annual averages match the average wind speeds from the nearest locations in the Global Wind Atlas (GWA) \cite{davis_global_2023}. We further adjust the existing annual average capacity factors to match FY19 wind generation aggregated at the state level. Module and turbine specifications assumed for each solar PV and wind technology, and the conversion of solar radiation to capacity factors using \emph{PVWatts} \cite{national_renewable_energy_laboratory_nrel_ssc_2024}, are given in Supplementary Note~\ref{si:capacity_factors}.

\subsection{Technology costs}\label{sec:technology_costs}

Annualized capital investment plus fixed operation and maintenance (O\&M) costs along with the fuel and variable O\&M costs for existing thermal generators (coal, gas, and diesel) are from tariff orders of 12 states for the FY18 or FY19 \cite{telangana_state_electricity_regulatory_commission_order_2018, andhra_pradesh_electricity_regulatory_commission_retail_2018, andhra_pradesh_power_generation_corporation_limited_retail_2018, maharashtra_electricity_regulatory_commission_orders_2017, north_bihar_power_distribution_company_limited_nbpdcl_tariff_2018, south_bihar_power_distribution_company_limited_sbpdcl_tariff_2018,madhya_pradesh_electricity_regulatory_commission_mperc_tariff_2017,madhya_pradesh_electricity_regulatory_commission_mperc_tariff_2018,punjab_state_power_corporation_limited_pspcl_tariff_2017,tamil_nadu_electricity_regulatory_commission_tnerc_generation_2017,assam_electricity_regulatory_commission_aerc_tariff_2019,himachal_pradesh_electricity_regulatory_commission_hperc_retail_2018,uttar_pradesh_electricity_regulatory_commission_uperc_petitions_2017,rajasthan_electricity_regulatory_commission_rerc_review_2018,rajasthan_electricity_regulatory_commission_rerc_tariff_2019}. Project-level data not available in tariff orders were collected from the ``Merit Order of Despatch for the Rejuvenation of Income and Transparency (MERIT)'' program in 2020 \cite{ministry_of_power_merit_2020}.

For new generation and storage technologies, we construct India-specific low, medium, and high cost projections by adjusting international projections from the 2023 Annual Technology Baseline (ATB) \cite{mirletz_2023_2023, vimmerstedt_2022_2022} to India-specific sources, namely the 2022 Indian Technology Catalogue (ITC) \cite{cea_and_dea_indian_2022}, the 2024 ITC Power Storage \cite{cea_itc_2024-1} and Green Fuels \cite{cea_itc_2024} catalogues, and IRENA auction prices in India \cite{international_renewable_energy_agency_irena_renewable_2024} (Figure~\ref{fig:electricity_system_overview}F and Supplementary Note~\ref{si:cost_projections}). New coal (supercritical), Combustion Turbine (CT), CCGT, and nuclear projects are mature technologies and thus, we assume constant real cost projections from 2020 to 2050. Across all scenarios, we assume mid-cost estimates for variable O\&M costs from the same sources. We assume international liquefied gas prices for new CT and CCGT generators.

We use a 7\% social discount rate to compute electricity system's net present value (NPV) from a central planning perspective, consistent with macroeconomic benchmarks such as the Reserve Bank of India’s FY20 policy rate. To annualize overnight capital costs, we apply a 9\% financing rate reflecting new projects weighted average cost of capital (WACC) which is within the range (8.8\%-10\% post-tax) seen in India’s power sector \cite{international_energy_agency_iea_cost_2021}. The financing rate is higher because it reflects the private cost of capital required to attract investment, including risk premium and equity return requirements, whereas the social discount rate represents a planner’s intertemporal valuation of system costs and does not embed risk specific to a project. All costs are reported in 2020 USD (Supplementary Note~\ref{si:cost_projections}).

\subsection{Demand projections}\label{sec:demand_data}

To explore the uncertainty in future demand projections, we develop two sets of demand profiles, a bottom-up set and a linearly-scaled set. Within each set, we use three different demand growth projections (mid-growth scenarios shown in Figure~\ref{fig:electricity_system_overview}E).

We use Perspectives on Indian Energy based on Rumi (\emph{PIER v2.0}), an energy systems model that develops bottom-up, state-wise electricity demand projections for India \cite{prayas_energy_group_pier_2025}, with hourly granularity for a representative (average) day for each month. This model captures sectoral energy use by integrating bottom-up energy service-based methods, top-down GDP elasticity projections, and trend analysis. The remaining demand is projected using historical trends and calibrated with reported consumption data \cite{central_electricity_authority_all_2023}.

To develop hourly demand profiles for an entire year, we use electricity demand from Fiscal Year 2019 (FY19), reported by Grid-India, the electricity system operator \cite{grid-india_demand_2023}, and NITI Aayog, the Indian government think tank \cite{niti_aayog_india_2023}. We select FY19 profiles to preserve the weather synchronization of demand with weather-dependent energy generation. While industrial and commercial demand profiles are projected to remain relatively stable, the agricultural sector is assumed to shift toward daytime consumption driven by the ``solarization'' of agricultural power supply. For demand growth projections, we adapt three contrasting demand scenarios developed by \emph{PIER v2.0}: ``Reference'' (mid), ``Vikasit Bharat'' (high), and ``Vichalit Bharat'' (low), described in Supplementary Note~\ref{si:pier_scenarios}.

We construct linearly-scaled demand profiles using state-level hourly demand data from FY19 \cite{grid-india_demand_2023}. For each state and month, we first linearly scale the FY19 hourly profile so that total energy demand matches the projections for a future period in \emph{PIER v2.0} (following the low, mid, and high demand scenarios). We then adjust the hourly profile to match the projected hourly peak demand for each month in in \emph{PIER v2.0} while preserving the total monthly energy projection. This procedure is described in detail in Supplementary Note~\ref{si:demand_projections}.


\subsection{Resource adequacy and operational reliability}\label{sec:reliability}

To account for potential bias in the investment portfolio introduced by using representative days in the capacity expansion model and to address interannual variability in generation resources, we include a Planning Reserve Margin (PRM) constraint, which is the additional capacity above peak demand required to maintain a reliable electricity supply \cite{nrel_explained_2024}. We assume a PRM level of 8\%.

Each VRE asset (i.e.,~solar, wind, and run-of-river hydro) contributes to the PRM a capacity equivalent to its Effective Load-Carrying Capability (ELCC), or capacity credit, estimated as its average capacity factor during the top net load hours (TNLH) \cite{shaju_draft_2022}. We generate national hourly demand for 18 weather years (2000--2014, 2017--2019) by weather-synchronizing two historical years (2018--2019) of state-level hourly demand data. We combine this hourly demand with hourly capacity factors for all selected VRE assets to calculate weather-synchronized national net demand over the 18 years (see Supplementary Note~\ref{si:nlcc}). For all other generation and storage assets, we use the average historical availability factor as the capacity credit.


For availability factors for each technology and state, we use average monthly reports of historical capacity factors and outage rates from FY19 to FY25 \cite{niti_aayog_india_2023}. The proposed availability factors account for limitations in fuel availability to differentiate between states that consume local gas or coal and those that import them (see Supplementary Note~\ref{si:availability_factors}).

To ensure operational reliability, we apply the operational reserve constraint at the state and regional levels as specified by the national system operator, Grid-India. We model secondary reserves as Spinning Reserves (SR) and tertiary reserves as Frequency Response (FR), and we aggregate the reserve requirements at the regional level to limit the number of constraints. We do not model primary reserves, whose inertia-based response is faster than the temporal resolution of our model. To determine reserve levels, we follow the method outlined in the ``Detailed Procedure on Interim Methodology for Estimation of Reserves Under CERC (Ancillary Services) Regulations 2022'' \cite{grid-india_detailed_2022}, and we project the reserve requirements to future investment periods (see Supplementary Note~\ref{si:reserves_projections}).

We run a production cost operational model for 8,760 hours of the entire year after fixing the new generator, storage, and transmission investments in each investment period, to ensure reliable operation of the electricity system.

\subsection{Policy goals}

We include three different clean energy and climate policy goals. We impose India’s official goal of 500 GW of clean-energy capacity by 2030 (i.e.,~clean target). Clean-energy includes VRE, large hydropower, and nuclear projects. We enforce the clean target as system-level upper bound constraint because of the favorable economics of solar and wind technologies, which lead to higher cost-optimal deployment than the target. We also enforce India's official renewable portfolio obligation (RPO) target that sets a minimum share of VRE generation to geographically balance VRE deployment across investment periods. Official annual RPO targets are specified from 2025 to 2030. We extend these targets exogenously to 2040 and 2050, reaching approximately 50\% of total generation by 2040 and holding this level constant until 2050. To reduce the computational load, we enforce the RPO as lower bound constraint only in the 21 states with high demand (Supplementary Table~\ref{stab:new_projects}). Finally, we impose a national-level linearly decreasing annual carbon emissions (i.e.,~carbon) target that enforces a 80\%, 90\%, and 100\% reduction by 2050 compared to carbon emissions in 2020. 

\section{Results}

\subsection{Costs, emissions, and clean energy shares}

We evaluate our scenarios based on system costs and GHG emissions because of their relevance to consumer costs and climate change mitigation. Average system costs normalize total system costs as electricity demand grows over subsequent investment periods. Costs include only generation, storage, and interstate transmission investment and operations costs, and do not include intrastate transmission and distribution costs. Costs are expressed in real 2020 U.S. Dollar (USD) or Indian Rupee (INR) i.e., without accounting for inflation between periods. Total GHG emissions allow comparison with 2020 emissions whereas the GHG emissions intensity provides a measure of the climate impact per unit of energy generation. Lastly, clean energy share estimates the share of non-fossil fuel sources, including solar, wind, biomass, hydropower, and nuclear, in total electricity generation. 

\begin{figure*}[htb!]
    \centering
    \includegraphics[scale = .24, trim = {0cm 0cm 0cm 0cm}, clip]{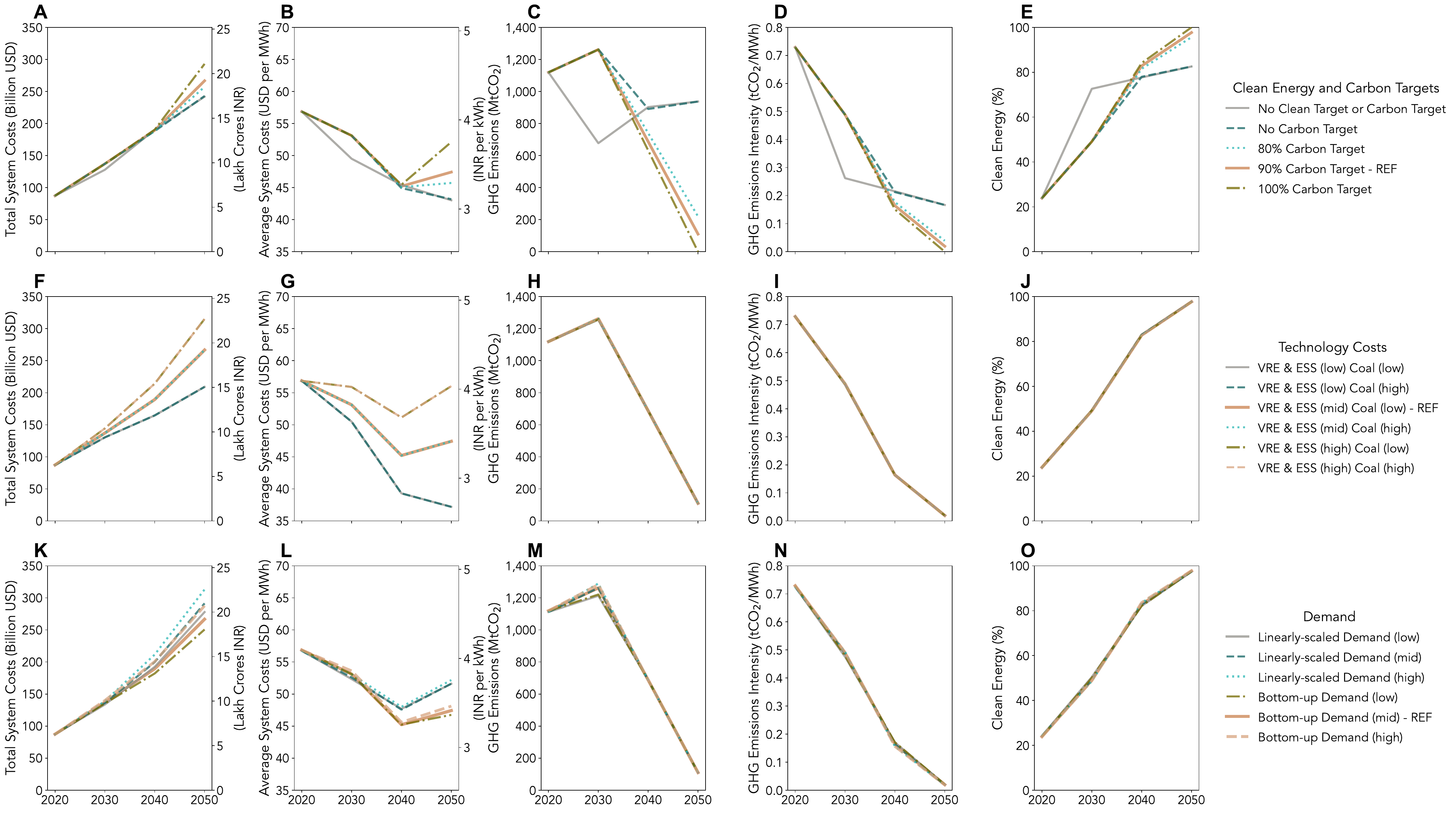}
    \caption{\textbf{System costs, greenhouse gas (GHG) emissions, and clean energy generation shares for climate policy scenarios with bottom-up demand (A-E), technology cost scenarios with a 90\% carbon emissions reduction target by 2050 (F-J), and demand projection scenarios with bottom-up and linearly-scaled demand profiles (K-O).} Results include total system costs (A, F, K), average system costs (B, G, L), total GHG emissions (C, H, M), GHG emissions intensity (D, I, N), and clean energy generation shares (E, J, O) for the 2020 to 2050 periods. Costs expressed in 2020 real U.S. dollars (USD) and Indian Rupee (INR) with a conversion rate of 72 INR to USD. Costs include only generation, storage, and interstate transmission investment and operations costs and exclude intrastate transmission and distribution costs. Clean generation share includes generation from renewable energy, large hydropower, and nuclear energy technologies. See Supplementary Tables~\ref{stab:pier_summary}-\ref{stab:demand_summary} for the underlying data.}
    \label{fig:summary}
\end{figure*}

We first examine the effects of clean energy and carbon targets on India's future electricity system (Figure~\ref{fig:summary}A-E). In the near-term (2030), meeting India's official target of 500~GW of clean energy capacity results in a clean energy share of 45\% in line with India's renewable purchase obligation (RPO) trajectory (43\% by 2029-2030), but a 13\% increase in carbon emissions above 2020 levels (see all clean energy and carbon targets scenarios except the \textit{No Clean Target or Carbon Target} scenario in Figure~\ref{fig:summary}A-E). Without a clean energy capacity target (\textit{No Clean Target or Carbon Target} scenario), a cost-optimal investment substantially exceeds the official clean energy capacity target (790~GW vs. 500~GW), while achieving lower carbon emissions (40\% below 2020 levels) and incurring a 6.8\% lower cost compared to the official target (Figure~\ref{fig:summary}C,D,E). However, because of on-the-ground constraints on land availability, relative slower pace of equipment manufacturing, and transmission interconnection bottlenecks, which could all restrict the rate of clean energy deployment, the near-term cost optimal result of the \textit{No Clean Target or Carbon Target} scenario is unlikely to unfold. Thus, across all other scenarios, we fix the clean energy capacity target at a maximum deployment (upper limit) of 500~GW in 2030. Importantly, in spite of fixing the clean energy capacity target under these scenarios, real average system costs in 2030 decline to USD 53.1 per MWh (INR 4.1 per kWh) compared to USD 56.9 per MWh (INR 3.8 per kWh) in 2020.


In the medium-term (2040), under the \textit{No Carbon Target} scenario where no carbon cap is applied in 2040, total GHG emissions fall by only 20\% relative to 2020 because of tradeoffs between clean technology cost declines and an increase in electricity demand. When a carbon target of 39\% below 2020 levels is imposed, consistent with a linear carbon emission reduction trajectory from 2030 emissions to the 90\% target by 2050, the average system costs are remarkably similar (less than 1\% higher) to the \textit{No Carbon Target} scenario (Figure~\ref{fig:summary}B). Meeting this carbon emission reduction target and assuming mid-trajectories for technology costs and electricity demand, the average system cost and GHG emissions intensity decline from 2030 levels to USD 45 per MWh (INR 3.3 per kWh) and 0.16 tCO2 per MWh, respectively, 21\% and 78\% lower than in 2020. Clean energy capacity rises to 1,640~GW meeting 77\% of electricity demand in 2040.


In the long-term (2050), cost and emissions differ substantially across scenarios because of varying carbon targets applied in 2050. With \textit{No Carbon Target} in 2050, GHG emissions remain relatively unchanged compared to 2040 and decline by only 16\% compared to 2020 (Figure~\ref{fig:summary}C). Applying a carbon target in 2050, however, imposes additional costs on the average system costs in 2050 increasing costs by 7.2\%, 11\%, and 22\% for the \textit{80\%}, \textit{90\%}, and \textit{100\% Carbon Target} scenarios, respectively, compared to the \textit{No Carbon Target} scenario (Figure~\ref{fig:summary}B). Yet, across all carbon target scenarios, the 2050 average system costs are lower than in 2020. For example, the \textit{90\% Carbon Target} in 2050 results in costs of USD 47 per MWh (INR 3.4 per kWh), which is 17\% lower than in 2020. 

Next, we evaluate the effects of VRE and storage technology costs and demand projections on India's electricity system. Because of the lower unit costs in 2050 compared to 2020 across all technology cost projections and demand scenarios and its alignment with India's economy-wide carbon neutrality goal by 2070, we set the \textit{90\% Carbon Target} by 2050 as the reference scenario and the default climate policy for the technology costs and electricity demand scenarios, thus fixing the upper limit for GHG emissions in 2040 and 2050. 

VRE and storage technology costs impact system costs across all investment periods (Figure~\ref{fig:summary}F,G). In contrast, coal capital and variable costs have little effect on system costs because of coal's diminishing share of capacity and energy generation across all investment periods. Total system costs vary substantially across low to high VRE and storage cost trajectories, increasing by 1.5-1.7 times (USD 130-140 billion or INR 9-10 lakh crore) in 2030, 1.9-2.5 times (USD 170-210 billion or INR 12-15  lakh crore) in 2040, and 2.4-3.6 times (USD 210-320 billion or INR 15-23 lakh crore) in 2050 compared to 2020. At the same time, average system costs are lower in all future investment periods relative to 2020 across all technology cost projections including the \textit{VRE \& ESS (high)} cost scenarios. 


While demand growth projections (assumed in this study) have a limited effect on system costs, demand patterns have a substantial effect (Figure~\ref{fig:summary}K-L). Within each set of demand projections---\textit{bottom-up} and \textit{linearly-scaled} demand patterns---total costs vary with annual demand growth projections but average system costs are similar. However, demand projections based on \textit{bottom-up} sectoral modeling result in lower system costs (about 5.1-5.3\% in 2040 and 8.1-9.8\% in 2050) compared to the \textit{linearly-scaled} demand projections that assume no change to 2020 hourly demand patterns (Figure~\ref{fig:summary}K,L). This decrease in costs is mainly driven by a shift in agricultural demand to solar hours, a demand response program that enables a lower total generation capacity to meet total electricity demand (see Supplementary Figure~\ref{sfig:demand-capacity_generation} and Table~\ref{stab:demand_capacity}). Results for climate policy scenarios with linearly-scaled demand profiles are provided in Supplementary Figure~\ref{sfig:iced-summary} and Table~\ref{stab:iced_summary}.

\subsection{New capacity investments and energy generation}



\begin{figure*}[htb!]
    \centering
    \includegraphics[scale = .29, trim = {0cm 0cm 0cm 0cm}, clip]{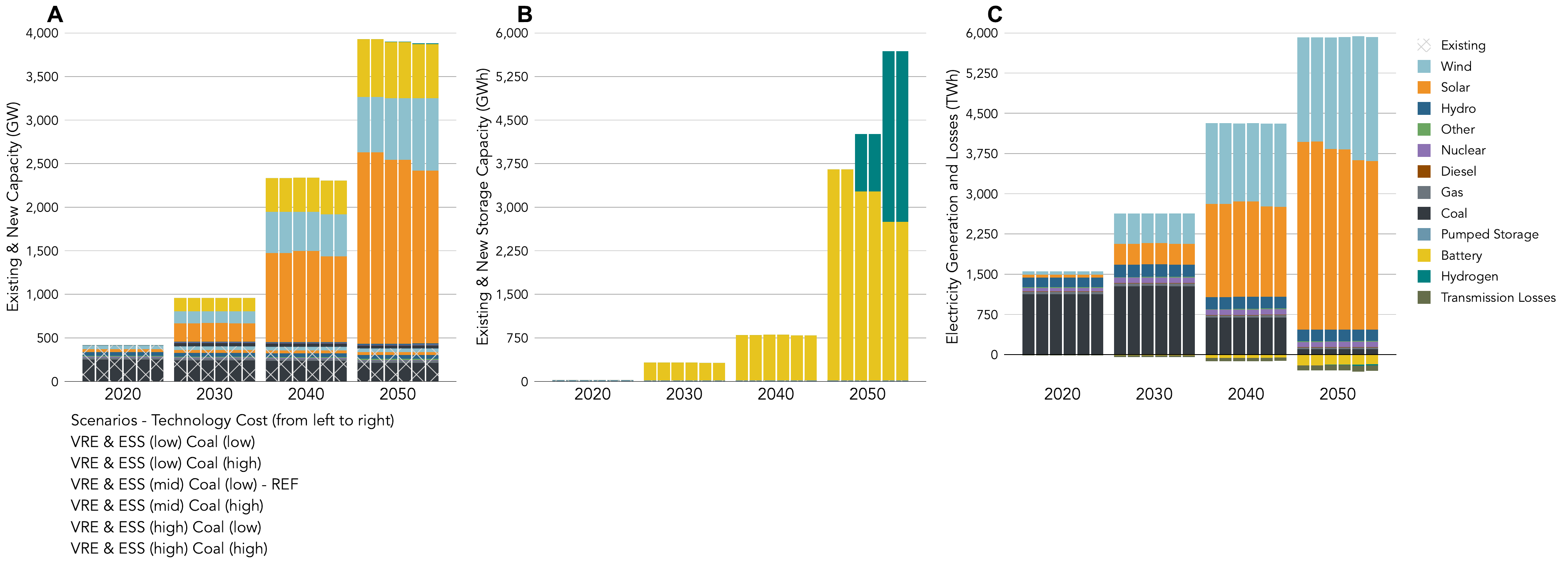}
    \caption{\textbf{Technology capacity investments and energy generation and losses for technology cost projection scenarios with a 90\% carbon emissions reduction target by 2050.} The new and existing generating capacity (MW) per technology (A), new and existing energy capacity (MWh) for storage technologies (B), and electricity generation (positive), and transmission and storage losses (negative) by technology (C) is shown for each period from 2020 to 2050. The hatch pattern differentiates existing from new capacity (A, B). See Supplementary Table~\ref{stab:costs_capacity} for the underlying data.}
    \label{fig:cost-capacity_generation}
\end{figure*}

Coal dominates India's electricity system, accounting for 235~GW (or 69\%)  of total installed capacity and 990~TWh (or 71\%) of total energy generation in 2020 \cite{cea_all_2021} (Figure~\ref{fig:electricity_system_overview}A,D). An additional 15.5~GW of coal capacity was built from 2020 until 2025 and 21~GW is expected to be constructed by 2030 \cite{niti_aayog_india_2023}, all of which we include in the 2030 coal capacity in our model. The model then builds additional coal capacity endogenously based on least-cost principles in addition to other technologies including solar, wind, natural gas combined cycle gas turbines, batteries, and depending on the scenario, hydrogen storage, nuclear, and pumped storage hydro. Coal power plants are assumed to retire at 45 years, which results in coal capacity retirements of 33.4~GW by 2040 and 50~GW by 2050. We first focus on the new capacity investments and energy generation across the technology costs scenarios, which result in the most divergence in system costs (Figure~\ref{fig:cost-capacity_generation}). This set of scenarios assumes a 90\% carbon target in 2050 and electricity demand with the bottom-up profile and mid-growth projection. Results for the other climate targets and demand projection scenarios are shown in Supplementary Figures~\ref{sfig:pier-capacity_generation} and~\ref{sfig:demand-capacity_generation} and Tables~\ref{stab:pier_capacity} and~\ref{stab:demand_capacity}, respectively.  

Across the technology cost scenarios, limited fossil fuel generation capacity is built. Specifically, only 36.5~GW of new coal capacity and 4~GW of new gas capacity is built by 2050 (Figure~\ref{fig:cost-capacity_generation}A). In fact, all the new coal capacity is assumed in the pipeline (either planned or under construction) and no additional coal capacity is cost-optimally built in any period. Capital costs of new coal power plants has no effect on optimal new coal deployment. While coal capacity is increasingly used to meet electricity demand during non-solar and non-windy hours, total coal energy generation falls as the share of low variable cost renewable energy generation increases over time. As a result, capacity factors of coal increase to 53\% in 2030 before falling to 29\%, and 4.7\% in 2040 and 2050, respectively (Supplementary Figure~\ref{sfig:cost-coal_capacity_factor} and Table~\ref{stab:costs_capacity}). Commensurately, the share of coal generation decreases to 50\%, 17\%, and 1.8\% from 2030 to 2050 even as electricity demand quadruples by 2050. 

If hourly demand profiles remain unchanged, i.e., linearly-scaled, which results in high demand peaks in the evenings, then only a relatively small (up to 3~GW) additional new coal capacity is cost-optimally deployed whereas a substantially large new gas capacity (59-74~GW or 3 times the gas capacity in 2020) is installed (Supplementary Figure~\ref{sfig:demand-capacity_generation} and Table~\ref{stab:demand_capacity}). Natural gas power plants with their lower capital cost relative to coal power plants do not generate much energy (less than 10\% capacity factor) and are mainly used for balancing and reliability. Because India has not witnessed much expansion of new gas capacity in the past few decades, we also examined a case where no new gas capacity is allowed. In this scenario (\emph{No New Gas \& ALT}), the lack of new natural gas capacity is compensated by additional solar (125~GW more solar and 81~GW less wind) and battery storage (47~GW/390~GWh) but with minimal impact on system costs (less than a 0.3\% increase) (Supplementary Figure~\ref{sfig:additional_6-capacity_summary} and Table~\ref{stab:additional_6_capacity}). 



In contrast to coal capacity, solar PV, wind, and battery storage technologies dominate new infrastructure investments across all scenarios under the carbon target, with differences in their levels and mixes driven mainly by their cost projections. VRE (wind and solar) capacity increases from an existing 72~GW in 2020 to 421~GW in 2030, which meets a large share of India's official non-fossil or clean energy target of 500~GW. From 2030, VRE capacity more than triples (3.7 times) to 1,540-1,570~GW by 2040, and then doubles (1.9 times) to 2,890-2,900~GW to achieve a 90\% carbon emissions reduction by 2050. This VRE capacity meets 75\% and 92\% of energy generation in 2040 and 2050, respectively. Because of the projected substantial decline in cost of not only solar PV but also battery storage, which enables the daily balancing of solar energy, solar PV forms an increasing share of VRE installed capacity, from 66-69\% in 2040 to 70-77\% in 2050 with higher shares in the low VRE and battery storage cost (\textit{VRE \& ESS (low) Coal (low)}) scenario. Compared to its shares in installed capacity, shares of solar in total VRE energy generation are lower, ranging from 52-55\% in 2040 and 58-64\% in 2050, because of solar PV's lower capacity factors relative to wind. Within wind technologies, no offshore wind capacity is built (except in the high electricity demand scenarios, which build only 3.5~GW of new offshore wind capacity) because of its higher capital costs relative to onshore wind. The system absorbs all VRE generation in 2030, while curtailment increases to 186 GWh~in 2040 and 249~GWh in 2050 (Supplementary Table~\ref{stab:unserved_energy}). As the VRE share of generation increases, the curtailment rate is 4.5\% in 2040 and 2050, which is likely an underestimation because of a lack of representation of transmission line-level congestion in this study.



Energy storage system (ESS) or battery storage, which is limited to short-term (daily) balancing energy in this study, plays a major role in balancing the diurnal variability of VRE generation, especially solar. In the reference scenario, battery storage capacity (expressed in both power and energy capacity) rises from near-zero in 2020 to 150~GW/310~GWh in 2030, 390~GW/790~GWh in 2040, and 650~GW/3,250~GWh in 2050, equivalent to storage purchase obligation (SPO) levels of 3.3\%, 7\%, and 26.7\% by 2030, 2040, and 2050, respectively (Supplementary Table~\ref{stab:unserved_energy}). The high power capacity relative to energy capacity in 2030 and 2040 is driven by battery storage's contribution to the planning reserve margin (PRM), which ensures adequate total system installed capacity to reliably meet demand. This contribution is shown in Supplementary Figure~\ref{sfig:additional_4-capacity_summary} and Table~\ref{stab:additional_4_capacity}, which compares the battery capacity deployment in the reference scenario with and without PRM. In 2050, in scenarios with a 90\% carbon emissions target, the battery energy capacity increases substantially---6.6-22 times the energy capacity in 2030---highlighting the greater need for balancing services as the share of weather-dependent energy sources like wind and solar increases above 90\%. The variation in battery energy capacities in 2050 is driven by technology costs (Figure~\ref{fig:cost-capacity_generation}) and electricity demand projections (Supplementary Figure~\ref{sfig:demand-capacity_generation}).  

\subsection{Effect of alternate clean energy technologies and demand patterns}

\begin{figure*}[htb!]
    \centering
    \includegraphics[scale = .3125, trim = {0cm 0cm 0cm 0cm}, clip]{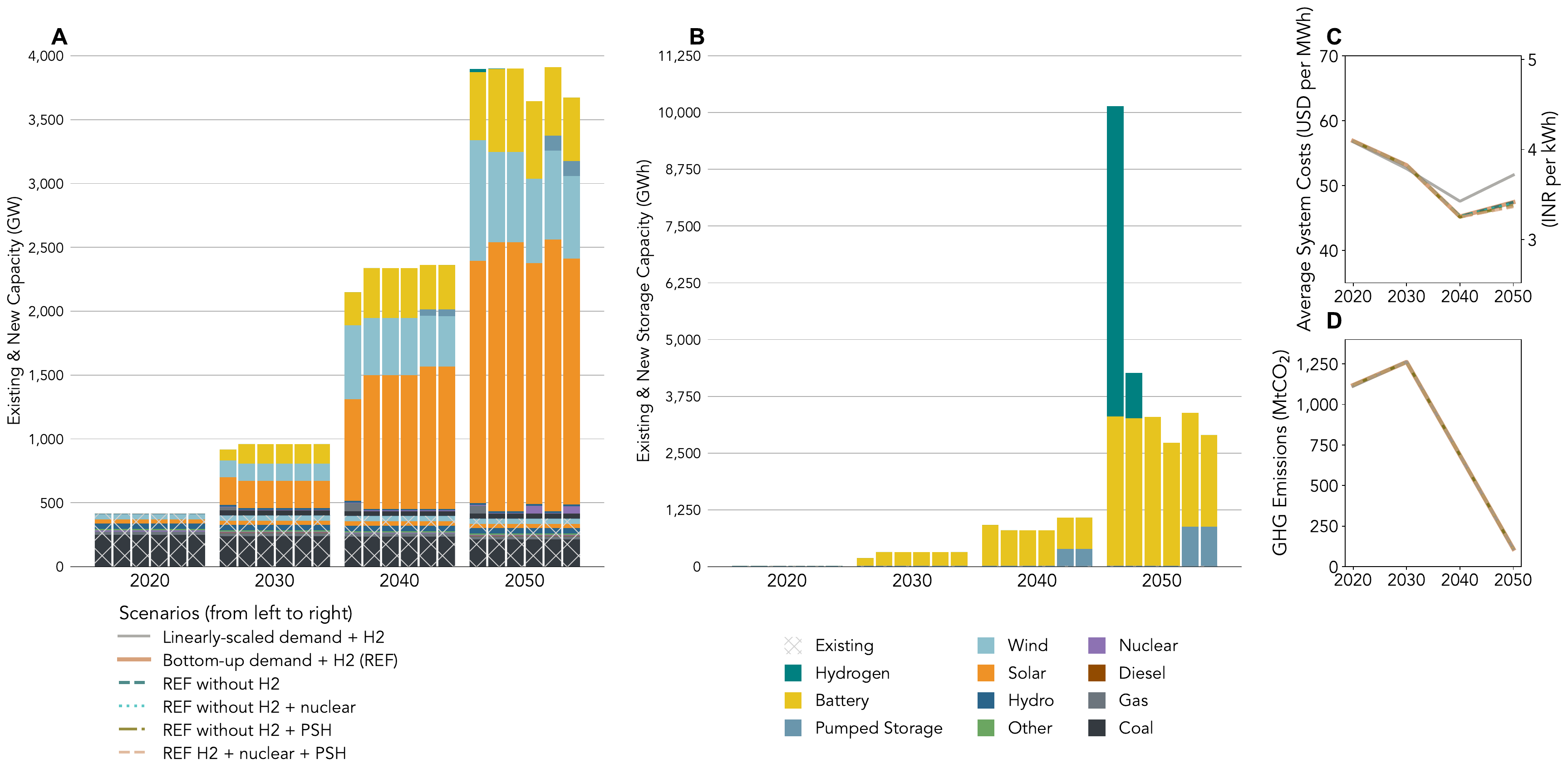}
    \caption{\textbf{Sensitivity of capacity investments, costs, and greenhouse gas (GHG) emissions to green hydrogen storage, nuclear, Pumped Storage Hydro (PSH), and demand profiles.} The reference scenario (REF), Bottom-up demand + long duration hydrogen storage, includes cost-optimal expansion of green hydrogen storage but no expansion of nuclear or PSH. Sensitivity scenarios include linearly-scaled demand projections, the REF with high (cost-optimal) nuclear, the REF with high (cost-optimal) PSH, and the REF without hydrogen. All scenarios have a 90\% carbon emissions reduction target by 2050. Sub-plots include (A) new and existing generating capacity (MW) per technology, (B) new and existing storage capacity (MWh) for storage technology, (C) average system costs, and (D) total GHG emissions. Costs expressed in 2020 real U.S. dollars (USD) and Indian Rupee (INR) with a conversion rate of 72 INR to USD.}
    \label{fig:DR_alt_techs}
\end{figure*}

In addition to solar, wind, and battery storage, we also consider the cost-optimal expansion of three other commercially available technologies---long-duration green hydrogen storage, pumped storage hydro (PSH), and nuclear---that could play a role in India's future low carbon electricity system. Deploying each of these technologies has substantial uncertainty. Green hydrogen is a relatively new technology, both PSH and nuclear have siting challenges, and all three have substantial uncertainty in costs. However, because of it's relatively abundant storage potential, we include long-duration hydrogen storage as part of our reference scenario. We then test the sensitivity of costs and emissions to expanding PSH and nuclear. 

Long-duration green hydrogen storage has the potential to balance energy demand and supply across the whole year and can help manage the inter-seasonal variability in renewable energy, especially of wind, which generates most of its electricity in the monsoon months. However, our results show that green hydrogen plays only a limited role in India's future low-carbon system. Specifically, hydrogen storage, along with its electrolyzers for hydrogen generation and fuel cells for electricity generation, is deployed only in 2050 and in the reference scenario (labeled \textit{VRE \& ESS mid cost} in Figure~\ref{fig:cost-capacity_generation}B and \textit{Bottom-up demand + H2} in Figure~\ref{fig:DR_alt_techs}B) and high battery cost scenario (\textit{VRE \& ESS mid cost} in Figure~\ref{fig:cost-capacity_generation}B). Notably, when clean energy cost projections are low (\textit{VRE \& ESS low cost} scenario), greater capacity of VRE and battery storage is deployed and no hydrogen storage is built. In the higher clean energy cost scenarios, when hydrogen storage is deployed, new battery energy capacity requirements decrease because hydrogen replaces some short-term balancing services provided by battery storage. Even in the {VRE \& ESS mid and high cost} scenarios, while a large energy capacity of underground hydrogen storage is deployed (1,000 and 2,900~GWh) in 2050, only up to 3.8-11.5~GW of hydrogen power capacity (both electrolyzer and fuel cell) is installed, representing 0.1-0.3\% of total installed generation and storage power capacity. 

If green hydrogen is not considered as a candidate technology (\textit{REF without H2}) because of the uncertainty in its cost, potential, and commercial viability in India, the electricity system relies on greater deployment of both shorter duration battery storage and VRE to meet the 90\% carbon emissions target but with a negligible (below 0.1\%) increase in system cost (Figure~\ref{fig:DR_alt_techs} and Supplementary Table~\ref{stab:alternative_capacity}). The limited need for long-duration (annual balancing) storage is because of the relatively lower inter-seasonal variation in solar generation as well as a substantial correlation between the monthly combined generation from weather-dependent resources (solar, wind, and hydro) and electricity demand (Supplementary Figure~\ref{sfig:vre_monthly_dispatch}). 

Allowing cost-optimal expansion increases nuclear capacity from 13~GW to 66~GW (\textit{REF without H2 + nuclear}), and PSH from 3.1~GW/19~GWh to 120~GW/880~GWh by 2050 (\textit{REF without H2 + PSH}). These capacities are below the Government of India’s 100~GW target for nuclear but above the corresponding target for PSH \cite{press_information_bureau_sustainable_2025, cea_roadmap_2026}. Importantly, expansion of PSH and nuclear avoids long-duration green hydrogen storage and substantially reduces the requirement for battery storage energy capacity (38\% lower than reference) and VRE capacity (8\% lower). At the same time, this expansion has a relatively small effect on total costs in 2050 (less than 2\%) (Figure~\ref{fig:DR_alt_techs} and Supplementary Table~\ref{stab:alternative_capacity}). 

The largest difference in cost comes from demand patterns. If demand patterns remain the same as 2019 (\textit{Linearly-scaled demand + H2}), with no demand shift toward solar hours, system costs are higher by 5.4\% in 2040 and 9.3\% in 2050 (Figure~\ref{fig:DR_alt_techs}C). This scenario also shows in higher demand for green hydrogen storage energy and power capacity because of the greater variability in inter-seasonal net demand. 

In summary, all three clean energy technologies have only a modest effect on costs in India's deeply decarbonized power system while demand response where demand is more aligned with solar hours has a substantial impact. We did not include other emerging technologies like Carbon, Capture, and Storage (CCS) coupled with coal, gas, or biomass power plants, mainly because CCS has not demonstrated commercial viability in India.

\subsection{Cost structure of the electricity system}


\begin{figure*}[htb!]
    \centering
    \includegraphics[scale = .36, trim = {0cm 0cm 0cm 0cm}, clip]{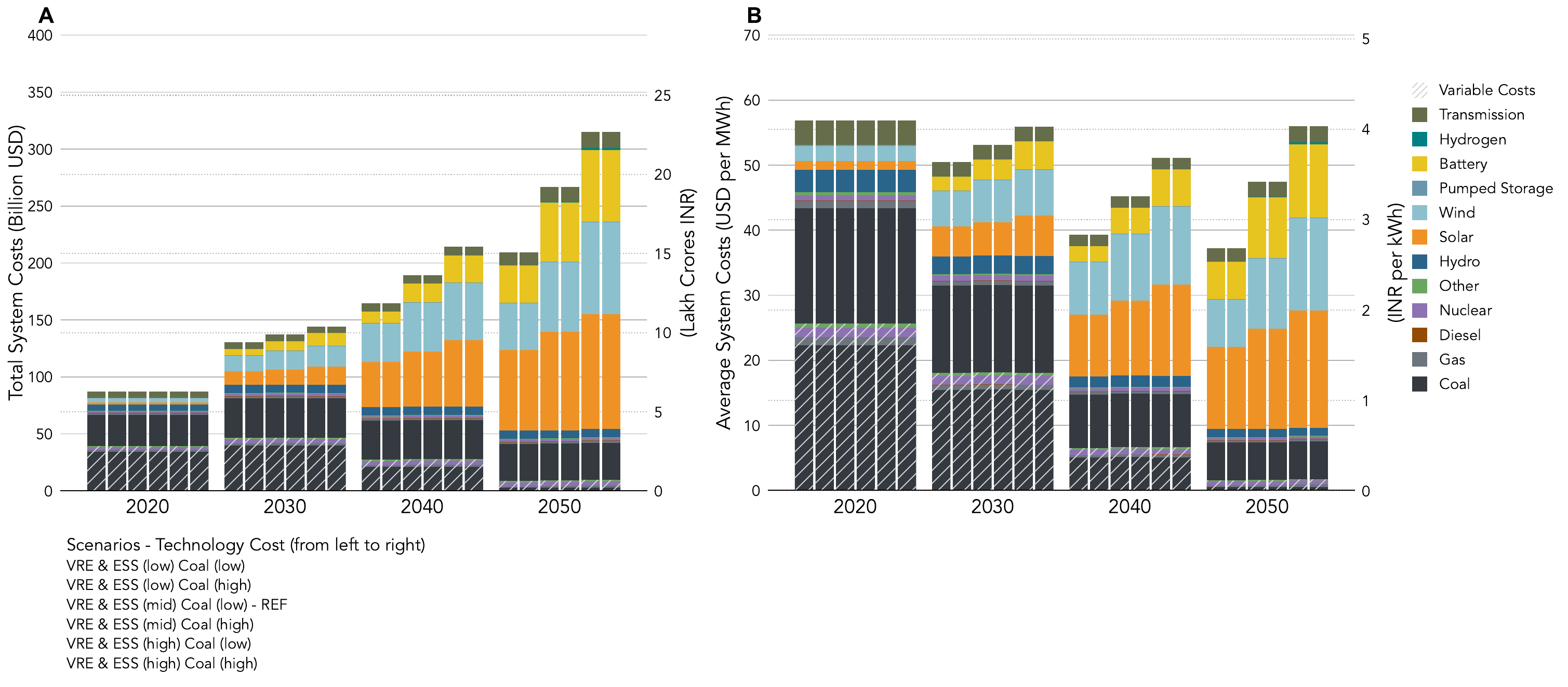}
    \caption{\textbf{Total system costs (A) and average system costs (B) across 
    the technology cost scenarios with clean energy target, 90\% carbon target by 2050, and mid-growth demand projection with bottom-up profiles.} Costs are shown by technology type for the four periods (2020-2050) across six technology cost scenarios. Fixed costs are shown by solid colors and variable costs have hatched lines. Left y axis is in U.S. dollars (USD) and right y axis is in Indian rupees (INR). Costs do not include intrastate or distribution costs. Costs expressed in 2020 real U.S. dollars (USD) and Indian Rupee (INR) with a conversion rate of 72 INR to USD.}
    \label{fig:cost-technology_costs}
\end{figure*}

Technology costs, demand profiles and projections, and clean energy and carbon policies, all drive differences in future total and average system costs. We first focus on the technology cost scenarios, which result in the largest differences in system costs. System costs shown here include only generation, storage, and interstate transmission capital and operational costs and do not include electricity distribution or intrastate transmission and congestion costs.

As electricity demand rises with India's economic growth, total system costs increase over time, more than doubling to USD 165-214 billion (INR 11.5-15.0 lakh crores) by 2040 and increasing 2.4-3.6 times to USD 209-315 billion (INR 14.6-22.1 lakh crores) by 2050 (Figure~\ref{fig:cost-technology_costs}A). However, the average system costs across all future investment periods and technology cost projections including the high-cost scenario (\textit{VRE \& ESS (high) Coal (high)}) remain below the 2020 level of USD 57 per MWh (INR 4.1 per kWh) (Figure~\ref{fig:cost-technology_costs}B). 


Solar, wind, and battery storage costs have a substantial effect on total system costs. Compared to the reference technology cost projections (\textit{VRE \& ESS (mid)} scenarios) under a 90\% carbon target by 2050, high VRE and storage costs (\textit{VRE \& ESS (high)} scenarios) result in 13\% and 18\% greater system costs in 2040 and 2050 compared to mid-cost projections, whereas low VRE and storage costs (\textit{VRE \& ESS (low)} scenarios) have 13\% and 22\% lower system costs, respectively. In contrast, because of its small share in new capacity investments and generation from new power plants, coal costs have limited effect on system costs. 

Even under conservative high-cost projections for VRE and storage (\textit{VRE \& ESS (high)} scenarios), real average system costs under a 90\% carbon target remain below 2020 levels in 2040 and are comparable to 2020 levels in 2050. For VRE and storage low-cost projection, real average system costs continue to fall through the years (31\% lower in 2040 and 35\% lower in 2050 compared to 2020). Results for all additional scenarios including bottom-up and linearly-scaled demand projections and alternative technologies (green hydrogen, pumped storage hydro, and nuclear) are provided in the Supplementary Figures~\ref{sfig:demand-technology_costs}–\ref{sfig:alternative-technology_costs}.

As the share of near-zero variable cost wind and solar resources increases and the share of fossil fuel-based resources decreases in India's electricity system, the share of variable costs also commensurately shrinks (Figure~\ref{fig:cost-technology_costs}). Across the technology cost scenarios, the share of variable costs decreases from 45\% in 2020 to 32-36\% in 2030, 13-17\% in 2040, and 3-4\% in 2050. Total variable costs of coal increases from 2020 to 2030 as coal generation increases but then decreases substantially in 2040 and 2050 as coal generation diminishes. The share of variable costs within just coal's total costs falls from 54\% in 2030 to 38-39\% in 2040 and 8-9\% in 2050, highlighting coal plants increasing role in providing capacity for reliability and decreasing energy generation because of unfavorable economics and climate targets. 


\subsection{State-level generation and interstate electricity trade}

\begin{figure*}[htb!]
    \centering
    \includegraphics[scale = .27, trim = {0cm 4cm 0cm 0cm}, clip]{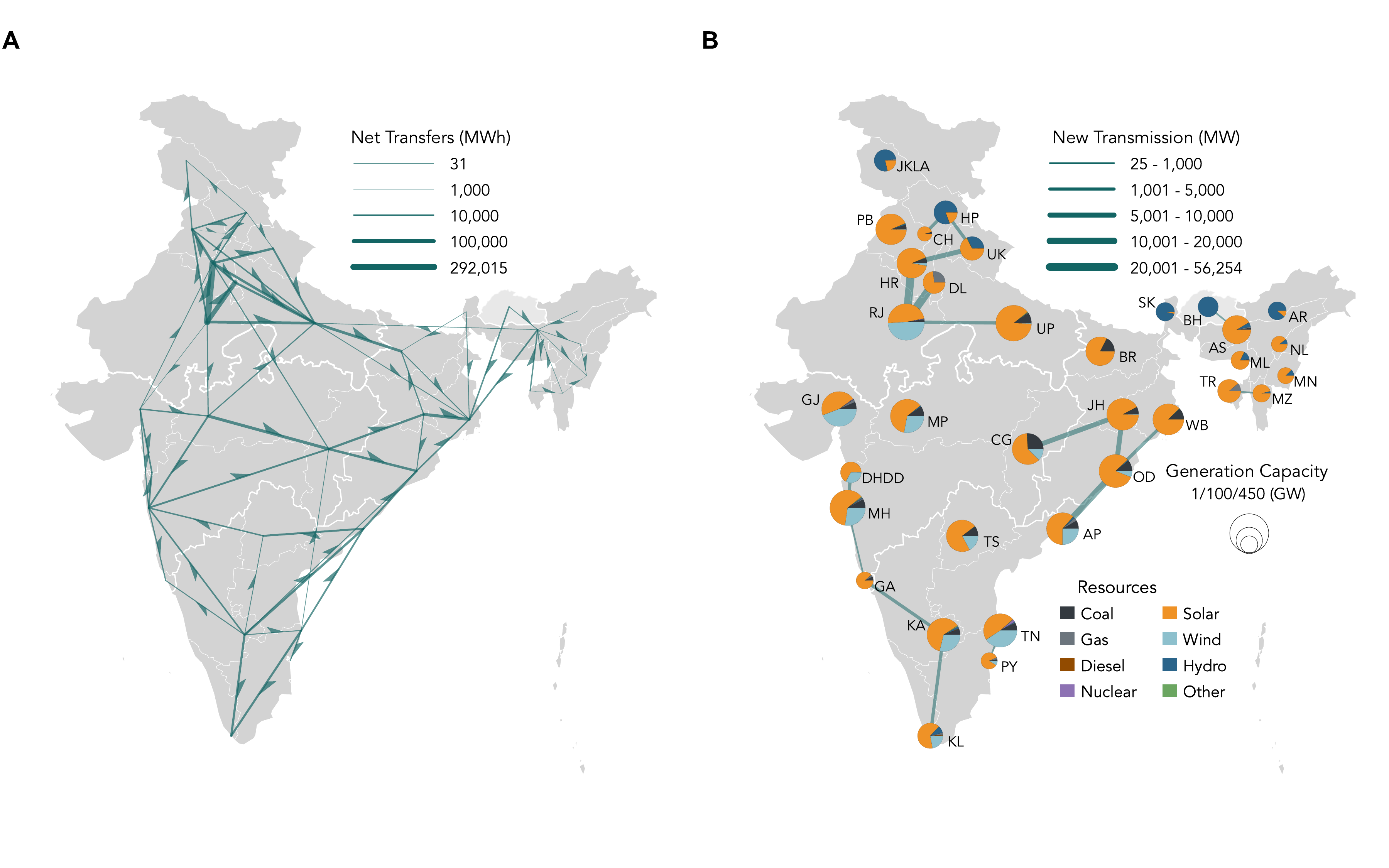}
    \includegraphics[scale = .2855, trim = {0cm 0cm 0cm 0cm}, clip]{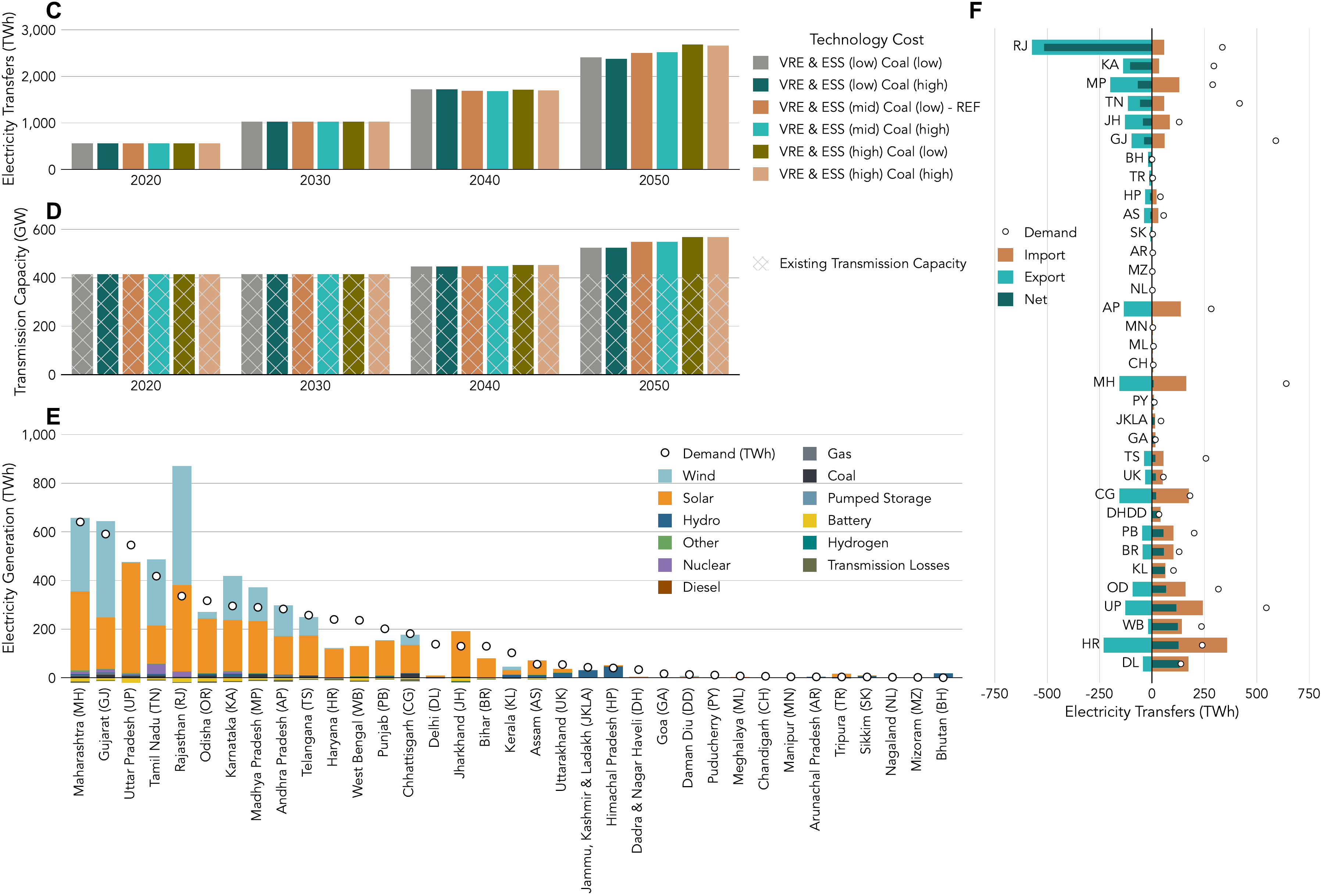}
    \caption{\textbf{Electricity transfers, new transmission capacity, and state import exports.} Interstate net electricity transfers (A) and new interstate transmission transfer capacity and state-wise share of generation sources (B) in 2050 for the reference scenario. Total interstate electricity transfers (C) and existing and new interstate transmission transfer capacity (D) from 2020 to 2050 across variable renewable energy (VRE), energy storage system (ESS), and coal technology cost scenarios. State-wise electricity generation and demand (E) and electricity transfers and demand (F) in 2050 for the reference scenario.}
    \label{fig:cost-transmission}
\end{figure*}

The state-level spatial resolution of our model allows examination of the impact of a more distributed build-out of clean energy across India, reflecting state policies of renewable purchase obligation (RPO) targets (Supplementary Figure~\ref{sfig:zonal_level-RPO}). To the 22 largest states by electricity demand (except the city-state of Delhi), we apply a constraint of requiring 50\% of state-wide electricity demand to be met by wind and solar energy deployed within the state boundaries in 2040 and 2050. Beyond this limit, states with relatively poor renewable resources can import electricity from higher quality out-of-state renewable energy sites if such imports are cost optimal in spite of costs incurred from interstate transmission losses. Results show that applying this constraint does not impose any substantial additional costs in 2040 or 2050 (less than 0.1\%; see Supplementary Figure~\ref{sfig:additional_3-capacity_summary} and Table~\ref{stab:additional_3_capacity}). The state-level RPO target ensures that deployment of VRE is relatively spatially balanced across the country except in states like Delhi, Haryana, West Bengal, Bihar, and Kerala, which either lack high quality VRE resources or border states with high quality renewable resources (Figure~\ref{fig:cost-transmission}B,E). Rajasthan by far has the largest deployment of combined wind and solar capacity, followed by Maharashtra, Gujarat, Tamil Nadu, Uttar Pradesh, and Karnataka. Northern and eastern states like Uttar Pradesh, Haryana, Punjab, West Bengal, Odisha, Jharkhand, and Bihar, lack high quality wind resources and rely mainly on solar resources for energy generation. Out of the large-demand states, notably, the city-state of Delhi has inadequate land-based renewable resources and relies on imports for meeting the majority of its electricity demand in all future periods. 

The shift from fossil-based energy sources to a much more spatially and temporally heterogeneous solar and wind resource-based electricity system leads to increased electricity transfers across India's interstate transmission network over time (Figure~\ref{fig:cost-transmission}A,B and Supplementary Figure~\ref{sfig:zonal-level_tx_exchange_capacity}). Total electricity transfers increase substantially with each period, increasing 3--3.1 times by 2040 and 4.3--4.8 times by 2050 (Figure~\ref{fig:cost-transmission}C and Supplementary Table~\ref{stab:ref-transfers}). Transfers are especially large coming out of Rajasthan as the state exports its high quality renewable generation to other states. With increased electricity transfers, requirements for new transmission transfer capacity also increases but only modestly (26-37\% increase in 2050 compared to 2020) (Figure~\ref{fig:cost-transmission}D). We attribute this surprisingly modest increase, at least in part, to the relatively coarse representation of inter-state electricity transfers in this study, which may understate congestion on the underlying AC transmission network.




State-level imports and exports depend on the differences between the state's electricity demand and generation potential (Figure~\ref{fig:cost-transmission}E) but also the state's location in the overall network where it could be a conduit for electricity transfers across regions (see Supplementary Figures~\ref{sfig:zonal_level-capacity_generation} and \ref{sfig:zonal_level-energy_dispatch}). Rajasthan is the largest net exporter of energy because of its large solar and wind deployment relative to its demand in 2050 (Figure~\ref{fig:cost-transmission}F). The state's exports, 66\% of its own in-state electricity demand, help meet demand in capacity-deficit northern states including Delhi and Haryana. Most other large western and southern states with the exception of Kerala and Telangana are net exporters (Karnataka, Madhya Pradesh, Tamil Nadu, and Gujarat) or close to zero net electricity exchanges (Andhra Pradesh and Maharashtra) (Figure~\ref{fig:cost-transmission}F). Most large northern and eastern states including Delhi, Haryana, West Bengal, Uttar Pradesh, Odisha, Bihar, and Punjab are net importers, mainly because of their relatively poor renewable resources. At the same time, many states experience substantial imports and exports, much larger than their net transfers, highlighting the importance of transmission and electricity transfers to balance the diurnal and seasonal variability of solar and wind across India (see Supplementary Figure~\ref{sfig:zonal_level-transmission_exchange}).  

\section{Discussion and Conclusion}

While there are substantial uncertainties in technology cost and electricity demand projections, examining a wide but feasible space of input parameters allows us to draw robust conclusions about cost-optimal investments under realistic constraints, and their impacts on costs and emissions. Here, we discuss insights that are robust to these uncertainties. 

\subsection{Future system costs decline across technology cost and demand projections while meeting ambitious carbon reduction targets}

India's relatively low average GDP per capita and development challenges makes it important for the country to prioritize electricity affordability while balancing its tradeoffs with climate mitigation. Our results show that across all clean and fossil technology cost and electricity demand projection scenarios, real average system cost i.e., cost of electricity supply, is lower in each future period (2030-2050) compared to 2020 even with a 90\% carbon mitigation target. In other words, even with the most pessimistic assumptions about technological advancements driving down wind, solar PV, and battery storage costs, electricity cost is expected to decline relative to the existing cost (estimated for 2020). While most previous studies \cite{barbar_impact_2023, gulagi_role_2022, rodrigues_indias_2023}, with one exception \cite{bhattacharya_bending_2024}, project declines in real system costs by 2050, our finding is robust across a wide range of technology cost projections, demand growth trajectories, ambitious carbon mitigation targets, reliability requirements, and inter-state transmission losses and congestion.  This is a critical result that allows India's policymakers to adopt ambitious climate goals including clean energy targets without increasing current costs of electricity supply. Realizing these cost reductions makes India’s renewable energy and storage procurement auctions critical for capturing the gains in future technology cost declines.



Declining technology costs drive the large deployment of solar PV, wind, and battery storage. In fact, in the near-term (2030), over 790~GW of clean energy or non-fossil capacity is deployed cost-optimally, exceeding the official 500~GW target, notwithstanding on-the-ground constraints including land availability and spur line interconnections, which we do not capture in this study. In the medium term (2040), system costs for a scenario with a carbon target of 39\% below 2020 levels (in line with a 90\% carbon target by 2050) are similar to one that imposes no carbon target, again suggesting that declining clean technology costs can make an ambitious carbon target cost-effective. However, not having a carbon or a clean energy target in 2040 or 2050 results in only about a 20\% decline in carbon emissions relative to 2020 across both periods assuming the mid-projection for technology costs (Figure~\ref{fig:summary}C). Even with the low VRE and storage cost projection, carbon emissions decline to only 46\% by 2050 compared to 2020 in the absence of a carbon target. Thus, to meet a 90\% carbon emissions reduction goal, which is in line with India's carbon neutrality goal by 2070, climate policies or clean energy targets are essential.


\subsection{Solar, wind, and battery storage dominate new investments while other low-carbon technologies play a limited role}

With its rapidly declining costs and relatively flexible deployment across space and size, solar PV accounts for two-thirds to three-fourths share of new installed generation capacity across all future periods. Wind power's higher capacity factors relative to solar and its improving technology and declining costs makes it the second generation technology to dominate new investments. To meet cost optimal targets in the reference scenario, deployment rates for solar will need to rise three times from 28~GW per year in 2025 to 83~GW per year between 2030-2040, and four times to 106~GW per year between 2040-2050 (Supplementary Figure~\ref{sfig:installation_rate}). Deployment rates for wind will need to scale even more from 2024 levels (4.2~GW per year) by a factor of 6-7 times to 29~GW per year from 2030-2050. In the last 10 years, solar deployment has scaled up substantially, increasing 9 times from 3.1~GW per year in 2015 to 28~GW per year in 2025, whereas wind power deployment has been relatively flat (3.4-4.2~GW per year) (Supplementary Figure~\ref{sfig:installation_rate}). This reflects greater constraints on wind power development, which is not only restricted to limited areas that have high wind resources but also likely faces land and other on-the ground constraints. In contrast, solar resource is much more evenly distributed across the region, offering more flexibility and choices to solar project developers. If wind energy is unable to scale deployment because of on-the-ground constraints, solar deployment may need to increase beyond the optimal capacity estimated in this study to meet India's clean energy and climate goals. 

In addition to solar and wind, chemical battery storage plays a major role, not only in balancing energy supply and demand but also in providing reliability services. The detailed reliability modeling in this study, including a planning reserve margin met by effective load carrying capacity-based contribution from generation and storage assets, drives a substantially higher power and energy capacity requirement for batteries compared to other studies (Supplementary Figure~\ref{sfig:studies_comparison}). This study considered Lithium Ion-based batteries but other storage technologies with a different chemistry that can be cost-competitive could also play a role.


Apart from solar PV, wind, and battery storage, other low carbon technologies considered in this study could also play a role but have only a modest effect on system costs. For instance, quadrupling the current nuclear capacity to 66~GW or increasing the PSH capacity by 50 times from 18.8~GWh to 876~GWh has a relatively small impact (less than 2\%) on total system costs in 2050 under a 90\% carbon emission reduction target. However, this expansion of nuclear and PSH could avoid substantial deployment of both battery storage (38\%) and VRE capacity (8\%). Despite the promising potential of long-duration green hydrogen storage in interseasonal energy balancing, it also surprisingly has only a modest impact on total system costs, especially when agricultural and other demand is shifted to day-time solar hours. This result suggests that expanding generation and storage capacities along multiple axes yield similar costs. In the near to medium term, India could focus on and scale up solar PV, wind, and battery storage deployment while expanding nuclear and PSH to the extent possible under ground realities and in the long term, continue exploring opportunities for other low-carbon technologies like green hydrogen as they become cost-effective over time. For instance, green hydrogen could play a major role in decarbonizing India's industry and sharing this infrastructure with the power sector may result in cost saving opportunities \cite{song_deep_2022, yang_economic_2025}. Further, emerging low-carbon technologies like enhanced geothermal energy that were not considered in this study could also play a role in the future but additional research is needed to assess its cost-competitiveness.

Lastly, technology costs of generation technologies like solar PV, wind, and battery storage are not the only drivers of electricity cost declines under carbon emission reduction targets. Aligning electricity demand with solar hours through demand response programs, especially for the agricultural sector, which is part of our reference scenario, substantially reduces electricity costs (5.1-9.8\% lower in the mid to long term) compared to scenarios where demand profiles remain the same as 2020. Demand response and energy efficiency programs remain critical in limiting cost increases while ensuring reliability.



\subsection{Ambitious but cost-effective clean energy targets}



India is reasonably on track to meet its official target of 500~GW of non-fossil fuel capacity by 2030 \cite{ministry_of_new_and_renewable_energy_india_2025}. This study could inform India's clean energy targets in the medium term (2040). Under the reference scenario that has a 39\% carbon emission reduction target in 2040 relative to 2020 levels, and assumes medium cost decline trajectories of solar PV, wind, and battery storage, low coal costs, and a climate change-impacted but solar-aligned future electricity demand, system costs are similar to a scenario in which no clean energy or carbon targets are applied. Under this scenario, solar and wind capacity in 2040 is 1,080~GW and 480~GW, respectively, which in total is more than three times the 2030 official target of 450~GW. Solar and wind generate 75\% of 2040 electricity demand, whereas renewable energy share (which also includes biomass, hydropower, and WHR) increases to 81\% and total clean energy share, which includes nuclear, rises to 86\%. Power and energy capacities of battery storage are 390~GW and 790~GWh, which are approximately three times the capacity targets in 2030. Our cost-optimal solar PV, wind, and battery storage capacities in 2040 remain within the relatively broad ranges reported in prior studies of India’s power system transition (380–1,700~GW for solar PV, 340–760~GW for wind, and 130–390~GW/540–1,800~GWh for battery storage), see Supplementary Figure~\ref{sfig:studies_comparison}. Over the long term (2050), our estimates for the reference scenario suggest a cost-optimal VRE capacity of 2,900~GW that is 40\% greater than the tentative non-fossil fuel capacity target of 1,800~GW by 2047 \cite{ministry_of_new_and_renewable_energy_india_2025}. Overall, compared with India’s current targets, our study suggests that substantially more ambitious clean energy targets can be achieved cost-effectively without exacerbating electricity affordability challenges.

While clean energy capacity targets can serve as a guiding benchmark for national energy planning, what drives procurement in India is the state-level renewable energy and energy storage portfolio obligations (RPO and ESO), which are shares of renewable energy and battery discharge energy as a percentage of energy demand \cite{piyush_singh_renewable_2022, ajay_tewari_renewable_2023}. Although the southern and western states have substantially greater wind resources compared to northern and eastern states, the relative similarity in solar resource quality across India and low costs of solar PV allows all states with the exception of Delhi to meet at least half of their renewable energy demand from in-state resources across all periods with similar system costs compared to a scenario that imposes no requirement for in-state deployment. This suggests that most states and their distribution utilities can pursue their own procurement and deployment strategies to meet their obligation targets without significantly affecting costs. 

\subsection{Changing cost structures, interest rates, and decreasing capacity factors of coal impact markets and contracts}

Compared to India's coal-dominated electricity system in 2020, the system cost of India's future electricity system is increasingly dominated by fixed costs as technologies like solar, wind, and battery storage, which have high upfront costs and low variable costs, are increasingly deployed (Figure~\ref{fig:cost-technology_costs}). Relying on these technologies can help reduce the electricity sector's exposure to future inflation in fuel costs. However, high interest rates can have a substantial negative impact on the deployment of low-carbon technologies because of the greater upfront capital requirement compared to fossil fuel projects. In addition, technologies with low variable costs will increasingly drive marginal prices toward zero, undermining investment recovery in India's electricity market and requiring market mechanisms to value capacity, flexibility, and reliability \cite{denholm_four_2021}.

Increasing decarbonization of India's electricity system decreases not only the deployment of fossil fuel generators but also their operation or capacity factors, which can affect their financial viability. Our results show that over time but within the lifetimes of power plants, capacity factors of fuel-based generators, especially coal, decrease substantially as solar and wind energy with near-zero marginal costs displace higher marginal cost fuel-based energy generation (Supplementary Figure~\ref{sfig:dispatch_plot}). Yet, coal plants are needed to meet the primary reserve margin and provide reliability services. As a result, fixed costs as a share of total coal power plant costs increase and variable costs decrease. Furthermore, coal plants are required to increasingly cycle up and down to balance the variability in renewable energy generation and electricity demand, which leads to increasing equipment wear and tear, thus incurring additional maintenance costs \cite{usdoe_office_of_electricity_delivery_and_energy_reliability_oe_power_2012, cea_wear_2026}. For coal plants to remain economically viable for reliability and balancing purposes, their increasing costs per unit of electricity generation may need to be increasingly recovered from the fixed component of their compensation or through the ancillary services markets, necessitating changes in contracts and electricity markets. However, we caution that variable cost components in long-term contracts that do not reflect the true marginal costs of coal power plants can distort merit-order dispatch, leading to those coal plants to which such contracts apply being dispatched ahead of lower-cost generators and thereby increasing total system costs and carbon emissions.


\subsection{Detailed studies required for transmission planning, operations, and pricing}


While our study shows substantial increase in interstate electricity trade (4.3--4.8 times in 2050 compared to 2020 levels), we record only a modest increase in interstate transmission transfer capacity. This result is driven by the coarse spatial resolution of our model, which does not capture the power flow constraints on transmission lines, both within and between states. To adequately inform transmission planning, capacity expansion and production cost models need to be coupled closely with transmission models that can simulate power flows in future electricity system scenarios and build new transmission capacity based on congestion. Voltage and frequency stability studies and ``n-1'' contingency studies, which examine the effect of a transmission line or generator failure on the rest of the system, will be needed to assess future transmission plans and operations \cite{cea_manual_2023}. Lastly, improvements in transmission pricing and allocation are essential for optimal transmission utilization and providing appropriate signals for new investments \cite{prayas_energy_group_price_2025}. 

\subsection{Other considerations and future research}

While our study shows that India's power sector can deeply decarbonize with no additional system costs compared to 2020 levels, more research is needed on technical, environmental, and social impacts of India's future decarbonized system. We discuss some of these research considerations below. 

Ensuring reliability in a future decarbonized power system, especially one that will rely mostly on weather-dependent renewable resources, requires a robust analysis of resource adequacy. This study considers only a single weather year in the capacity expansion model but relies on capacity credits derived from 18 years of weather data and the planning reserve margin constraint to identify adequate cost-optimal generation, storage, and transmission capacity that meets future demand. Designing and operating future scenarios of India's electricity system under multiple weather years to account for interannual variability in generation and demand, incorporating the effects of climate change on generation, transmission, and demand, estimating reliability metrics like unserved energy, and examining the effects of extreme events will improve the robustness of the results. 

Decarbonized electricity systems that predominantly rely on inverter-based and low inertia solar, wind, and battery technologies have low system synchronous inertia relative to systems with thermal and hydro generators that have large rotating turbines (Supplementary Figure~\ref{sfig:inertia_requirements} and Table~\ref{stab:gen_inertia}). This is a major concern for system operators \cite{grid-india_grid_2025}. Future studies should examine the reliability risks associated with low-inertia power systems \cite{milano_foundations_2018} and assess the potential and costs of inverter-based technologies for providing synthetic inertia \cite{denholm_inertia_2020}.




Both solar and wind can have land footprints larger than coal and natural gas infrastructure \cite{lovering_land-use_2022}, which could result in substantial trade-offs against other land uses like agriculture and conservation. While previous studies have shown that adequate potential for both solar and wind exist on suitable low-conflict lands \cite{kiesecker_renewable_2019, deshmukh_geospatial_2019, kiesecker_road_2023}, studies have found actual deployment on prime agricultural land \cite{ortiz_artificial_2022} and that plants can have adverse local environmental impacts if sited in water-scarce regions \cite{luo_parched_2018, wri-india_renewable_2021}. More research on existing deployment patterns, potential low-conflict land use pathways, and low water consumption practices for solar technologies could enable planners to set policies and rules that avoid adverse effects of renewable energy siting.


Renewable energy and battery storage manufacturing and deployment could potentially create more jobs than India's coal industry that it aims to replace. \textcite{pai_meeting_2021} show that total energy sector jobs in India are projected to grow under most climate policy and socioeconomic assumptions, but their analysis is limited to direct employment effects \cite{tyagi_indias_2022}. A more comprehensive study examining the impacts of decarbonization pathways on not just direct jobs and wage compensation, but also the indirect effects of employment on the rest of the economy is essential. 



This study shows that India can substantially decarbonize its electricity sector by 2050 without increasing system costs or exacerbating electricity affordability challenges. But several challenges including deployment scale-up, transmission planning, land use tradeoffs, and employment shifts, need to be addressed to meet ambitious carbon mitigation targets. The multi-model framework presented in this study can be useful for examining the economics and environmental impacts of power system decarbonization under uncertainty in other regions of the world. 

\section*{Data Availability}

\emph{GridPath-India} is publicly available in the Dryad repository \cite{terren-serrano_gridpath_2026}. The repository contains the \emph{GridPath} input files for the scenarios analyzed in this investigation, including transmission and project portfolios, operational characteristics, reliability requirements, policy targets, effective load-carrying capability assumptions for variable renewable energy projects, availability factors, and temporal structures necessary to replicate the results. The repository also includes project-level wind and solar capacity factor profiles for multiple technology configurations, along with the India-specific technology cost projections used in this study.

\section*{Code Availability}

The solar and wind site suitability analysis was developed using \emph{MapRE}, publicly available at \url{https://github.com/cetlab-ucsb/mapre}. The \emph{GridPath} open-source power system modeling platform is available at \url{https://github.com/blue-marble/gridpath}. The visualization tools developed to generate the figures in research are publicly available in a GitHub repository at \url{https://github.com/cetlab-ucsb/gridpath_india_viz}.



\section*{Acknowledgements}

G.T.S, R.D., M.M., A.M., and S.P. were partially supported by the World Resource Institute (WRI). Use was made of computational facilities purchased with funds from the National Science Foundation (CNS-1725797) and administered by the Center for Scientific Computing (CSC). The CSC is supported by the California NanoSystems Institute and the Materials Research Science and Engineering Center (MRSEC; NSF DMR 2308708) at UC Santa Barbara.

\section*{Author contributions: CRediT}

G.T.S. and R.D. conceptualized the study. R.D. acquired funding and supervised the project. R.D. and S.P. handled project administration. G.T.S., R.D., S.D., and A.M. developed the methodology. G.T.S., R.D., S.D., M.M, and A.M. developed the software and curated the data. G.T.S., R.D., and M.M. conducted the formal analyses and investigations. G.T.S., R.D., and M.M worked on the visualization. G.T.S., R.D., S.D, M.M., and S.P. drafted the manuscript. G.T.S., R.D., S.D., S.P., and A.G. reviewed and edited the manuscript.

\section*{Competing interests declaration}

The authors declare no competing interests.

\section*{Declaration of generative AI use}

The authors used the AI tool ChatGPT to assist with coding and data processing tasks, as well as editing and improving the clarity of the manuscript during the preparation of this work. After using this tool, the authors reviewed and edited the content as needed and take full responsibility for the content of the publication.

\printbibliography







\clearpage


\section*{Supplementary material}

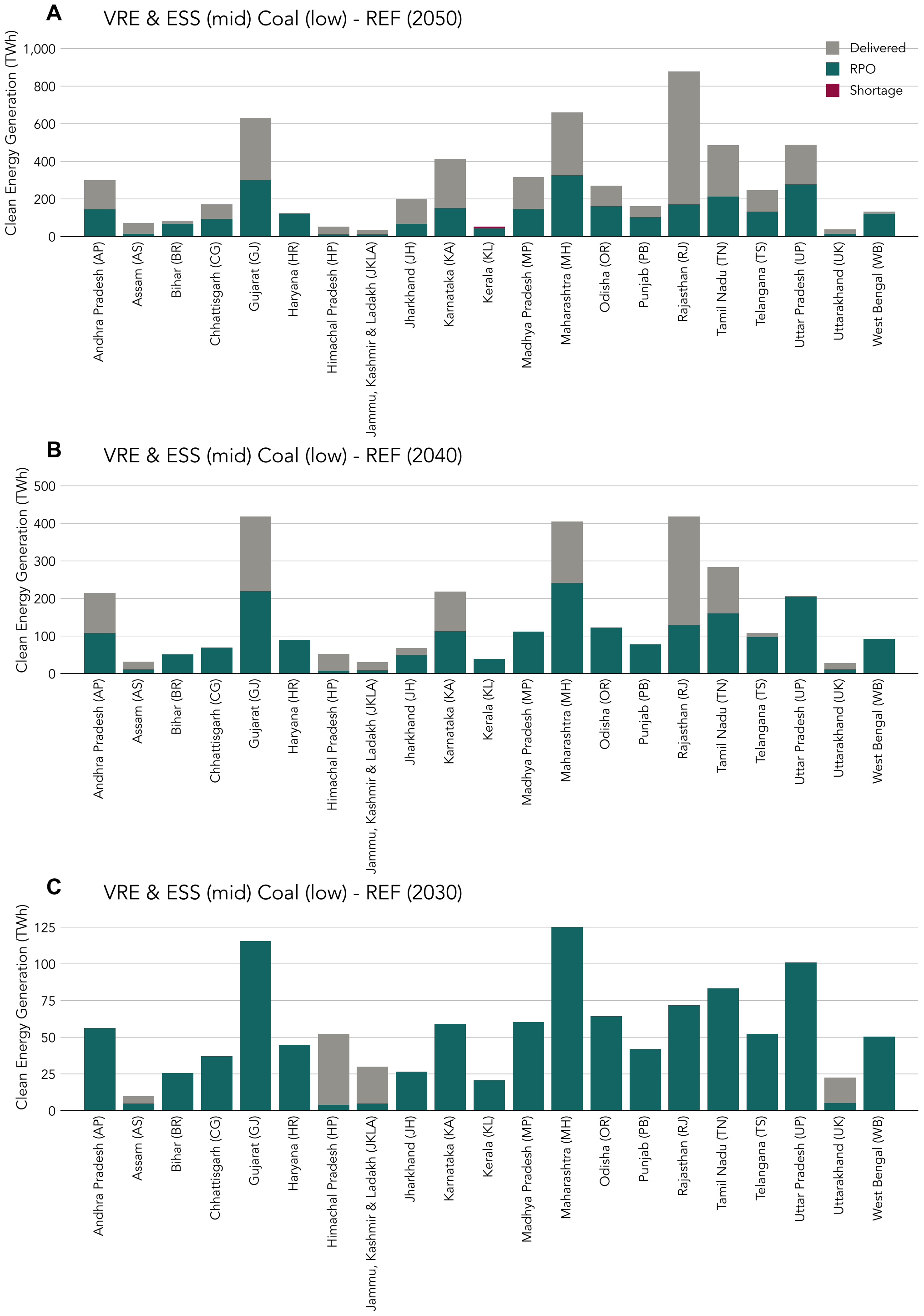


**Supplementary Figure 17. Renewable purchase obligations (RPO)**. State-level RPOs for the reference scenario during the different investment periods 2050 (A), 2040 (B), and 2030 (C). RPOs are only imposed on the states with the largest electricity demand. The RPO is projected to 2050 following the trend given by CEA until 2035. The RPO is the same percentage of the electricity demand for all states, and half for “hilly” states.

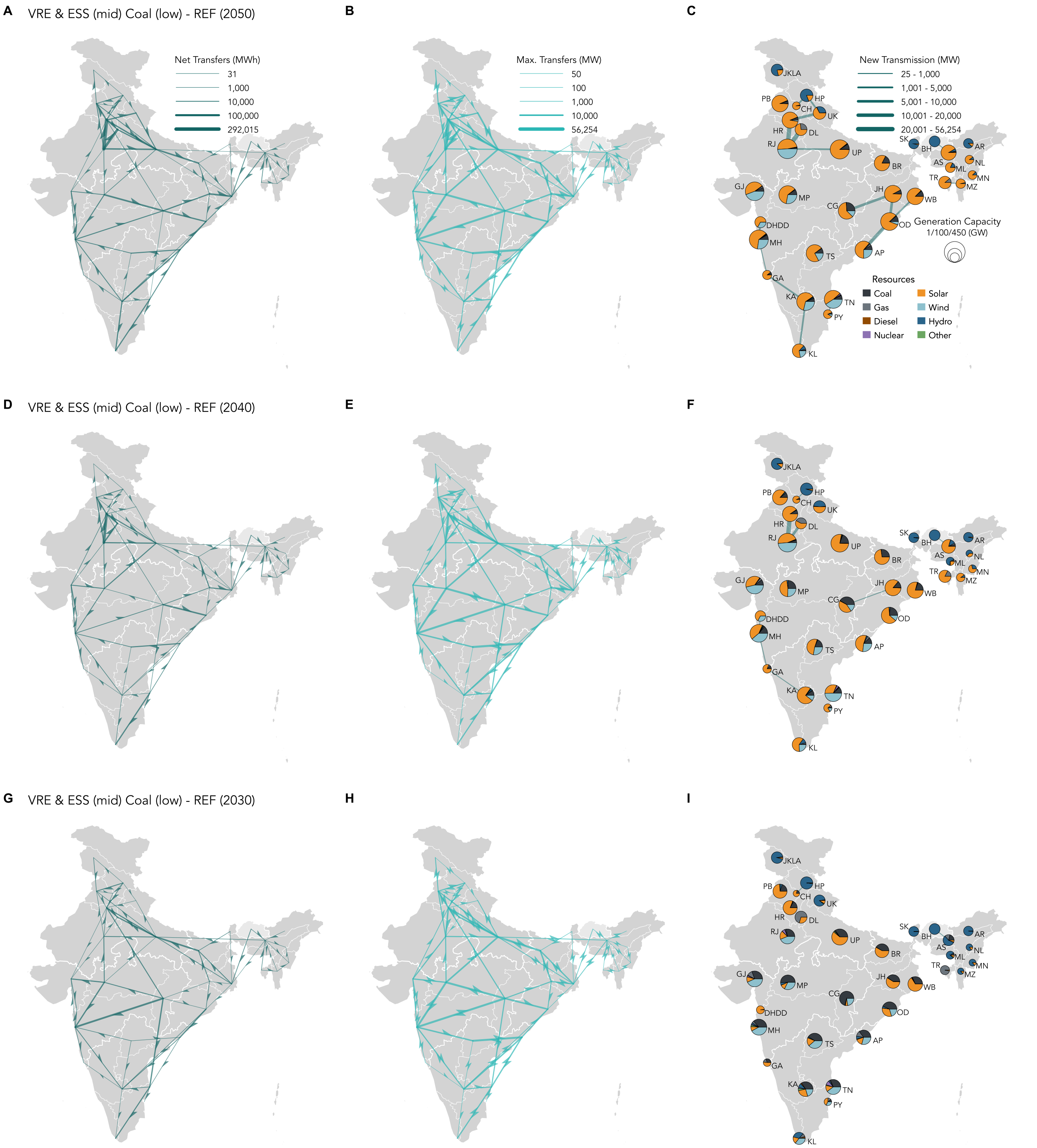


**Supplementary Figure 18. State-level net electricity transfers, maximum electricity transfers, and new transmission and total generation capacity**. Total net electricity exchange (A, D, G) for the different investment periods 2050 (A, B, C), 2040 (D, E, F), and 2030 (G, H, I). The line thickness represents the net imports/exports, and the arrow points towards the net importing state. Maximum instantaneous electricity transfer (B, E, H). The line thickness represents transfer electricity capacity, the arrows show the transfer directions that reach the maximum transfer capacity. New transmission capacity and Total generation capacity per technology (C, F, I). The line thickness represents the new transmission capacity, and the area of the charts represents the total generation capacity.

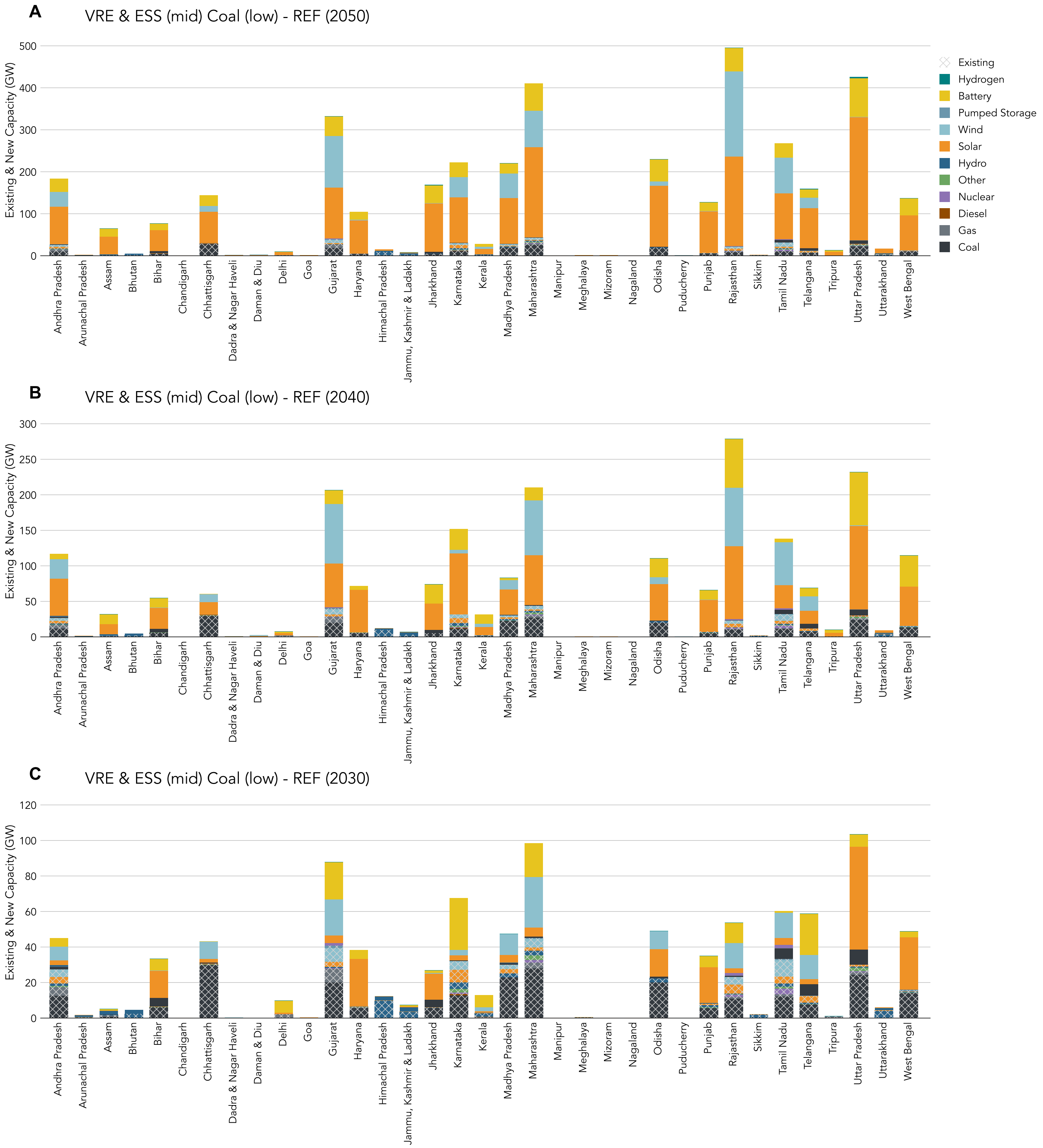


**Supplementary Figure 19. State-level generating capacity per technology**. Existing and new generating capacity on the different investment periods 2050 (A), 2040 (B), and 2030 (C) for the reference scenario. The hatch pattern differentiates existing from new capacity.

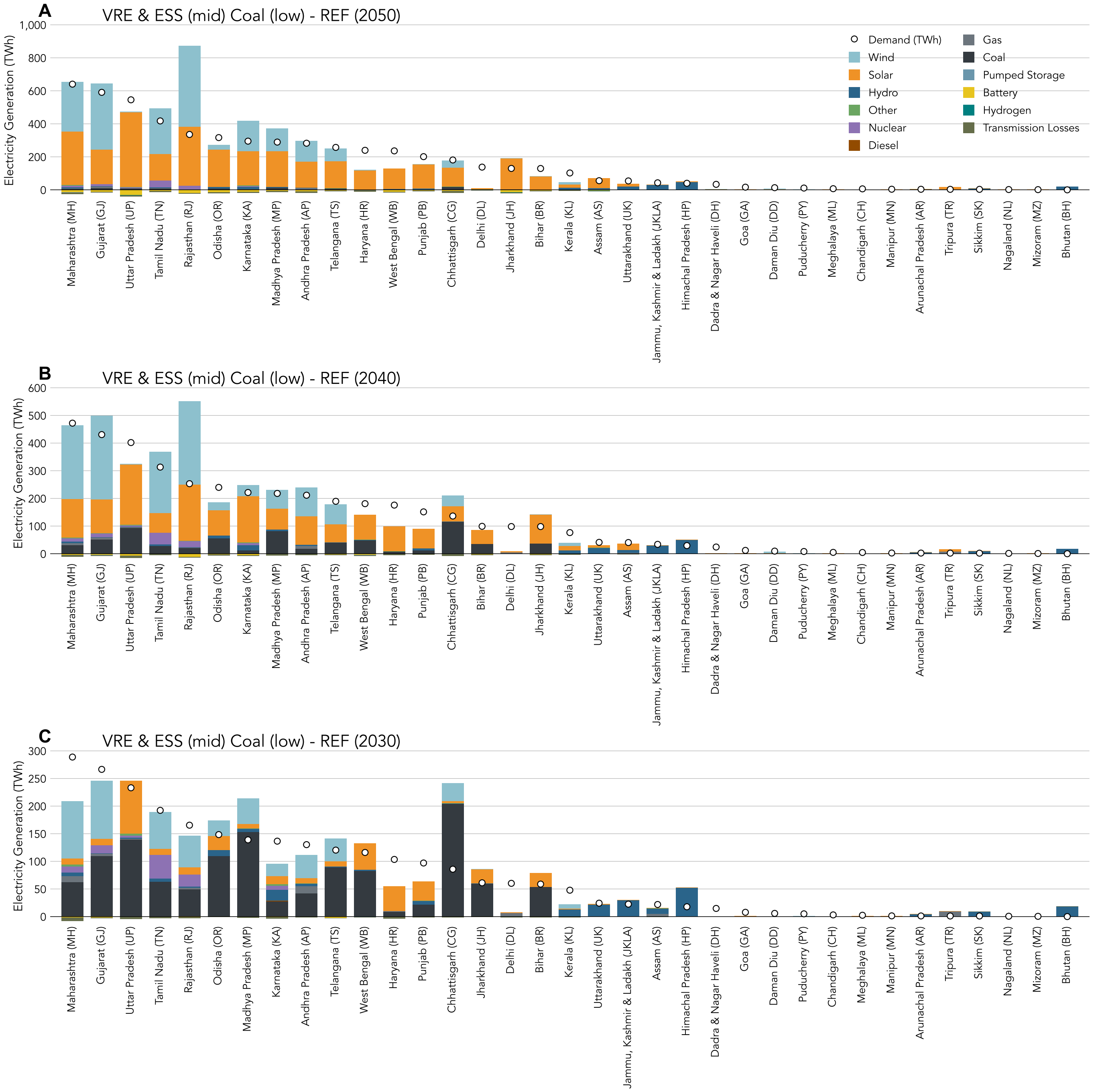


**Supplementary Figure 20. State-level electricity generation per technology**. Electricity generation during the different investment periods, 2050 (A), 2040 (B), and 2030 (C), under the reference scenario. Negative generation represents transmission and storage losses, and the white dot is the total electricity demand in each state.

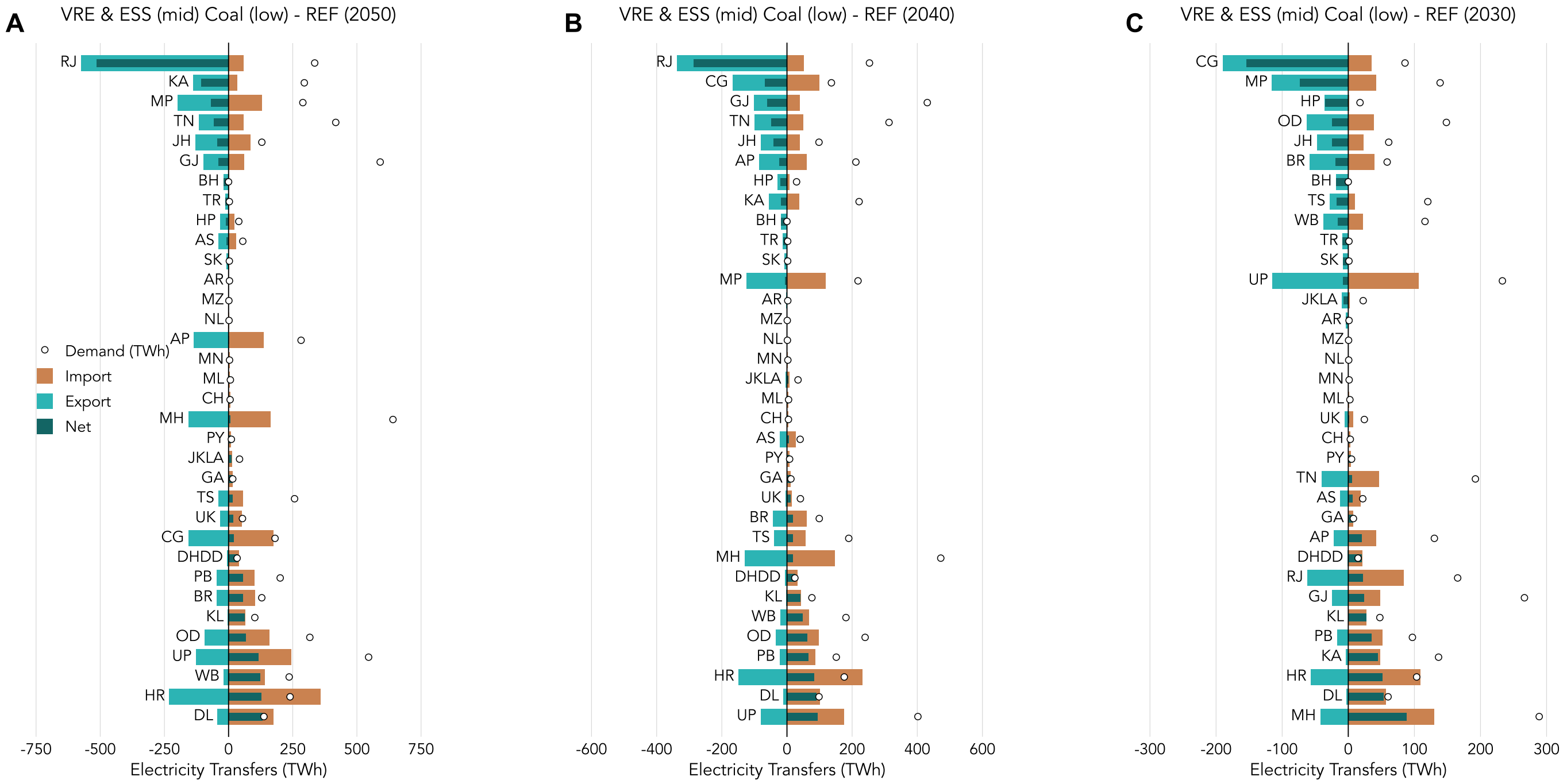


**Supplementary Figure 21. State-level electricity transfers**. Total exports (light green), imports (orange), and net imports or exports for each estate in the reference scenario on the different investment periods: 2050 (A), 2040 (B), and 2030 (C).

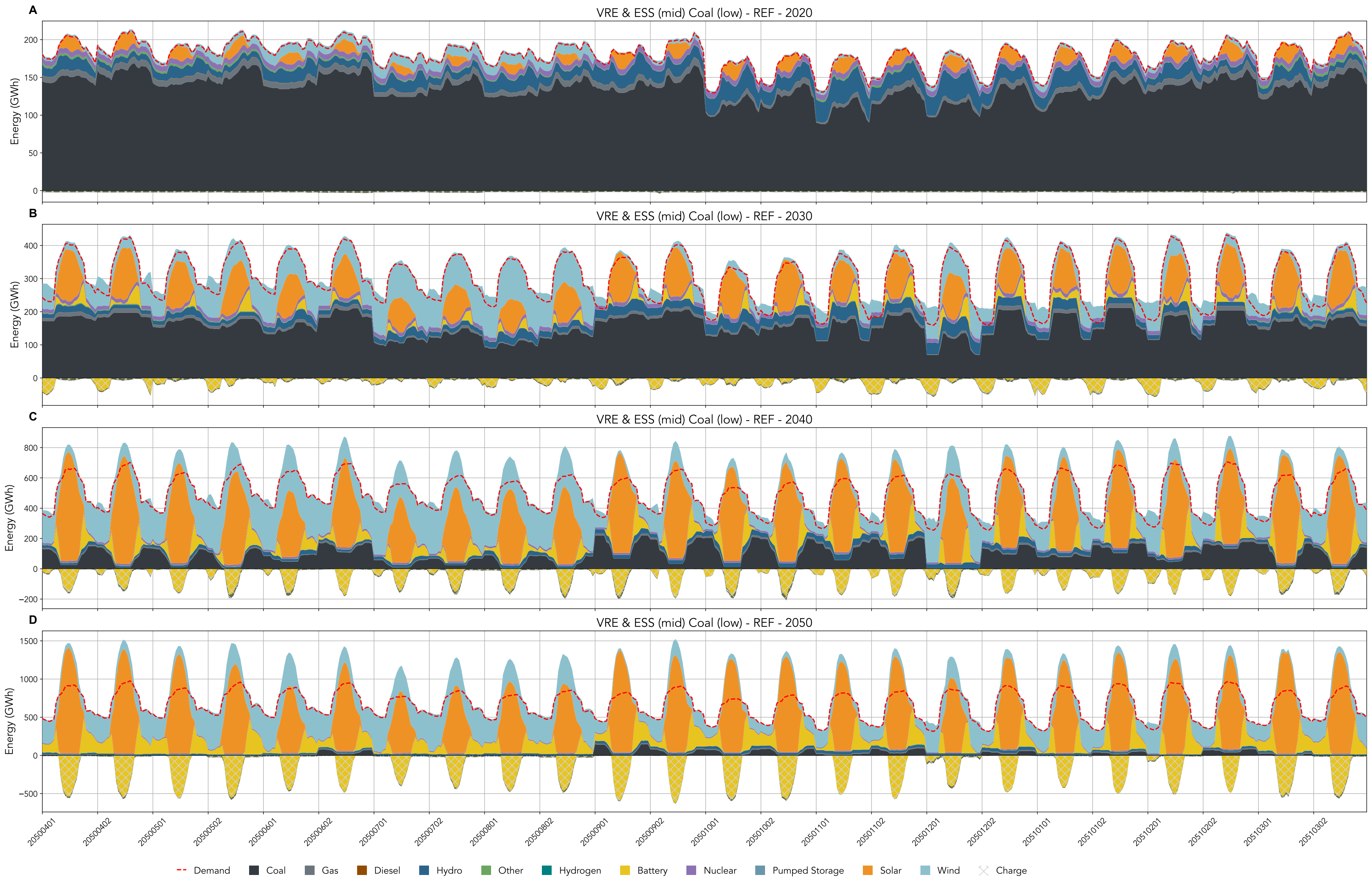


**Supplementary Figure 22. System level power dispatch for the reference scenario (REF) in the different investment period.** Power dispatch on the representative days used in the capacity expansion model. The first day of the month is the median day, and the second day is the peak day. The power dispatch from each technology is aggregated at the system level in period 2020 (A), 2030 (B), 2040 (C), and 2050 (D).

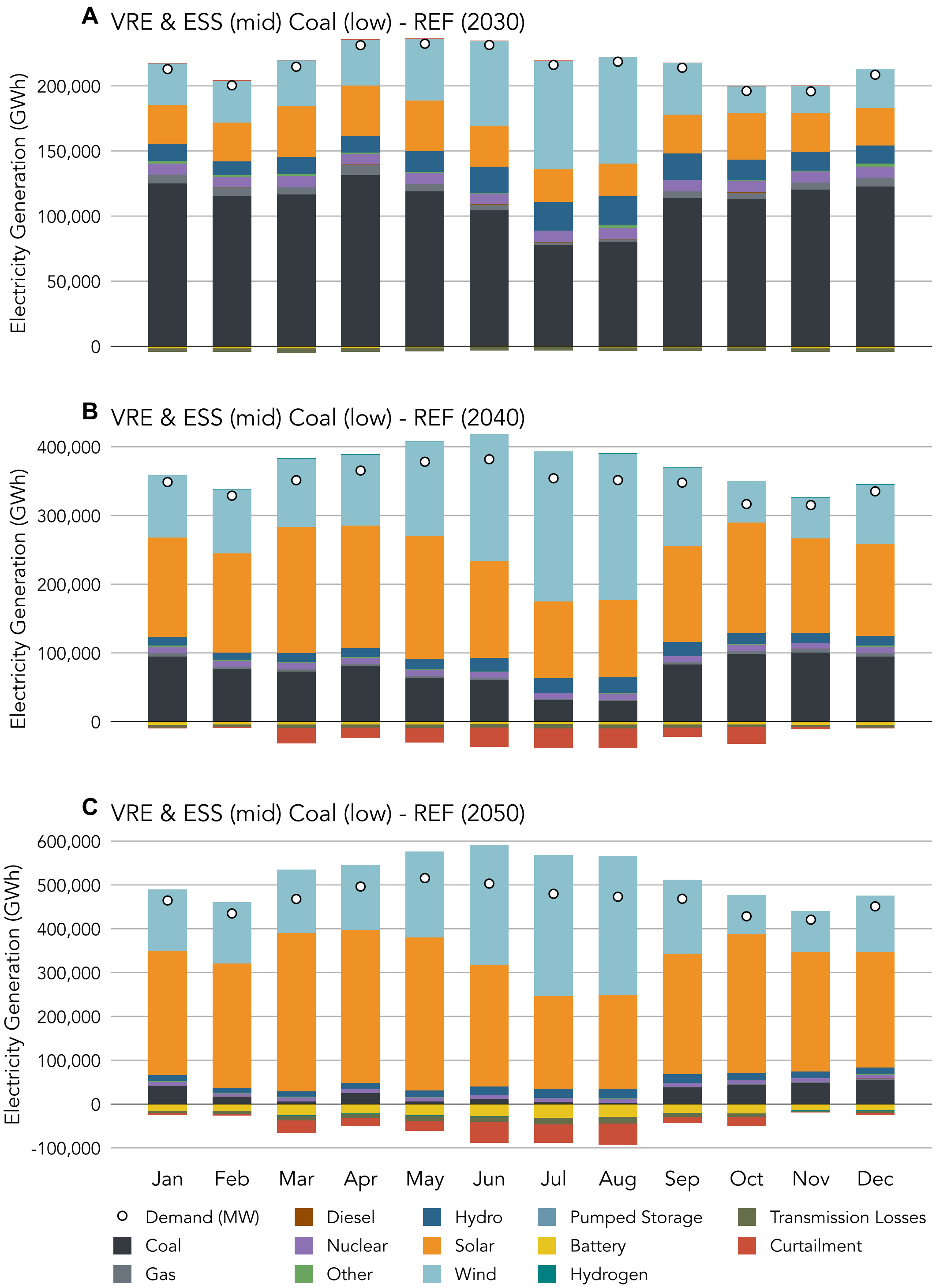


**Supplementary Figure 23. Solar, wind, and hydro (ROR and storage) monthly power dispatch for the reference scenario (REF) in the different investment periods.** The power dispatch is from an 8,760 production cost model, aggregated at the system level for 2030 (A), 2040 (B), and 2050 (C).

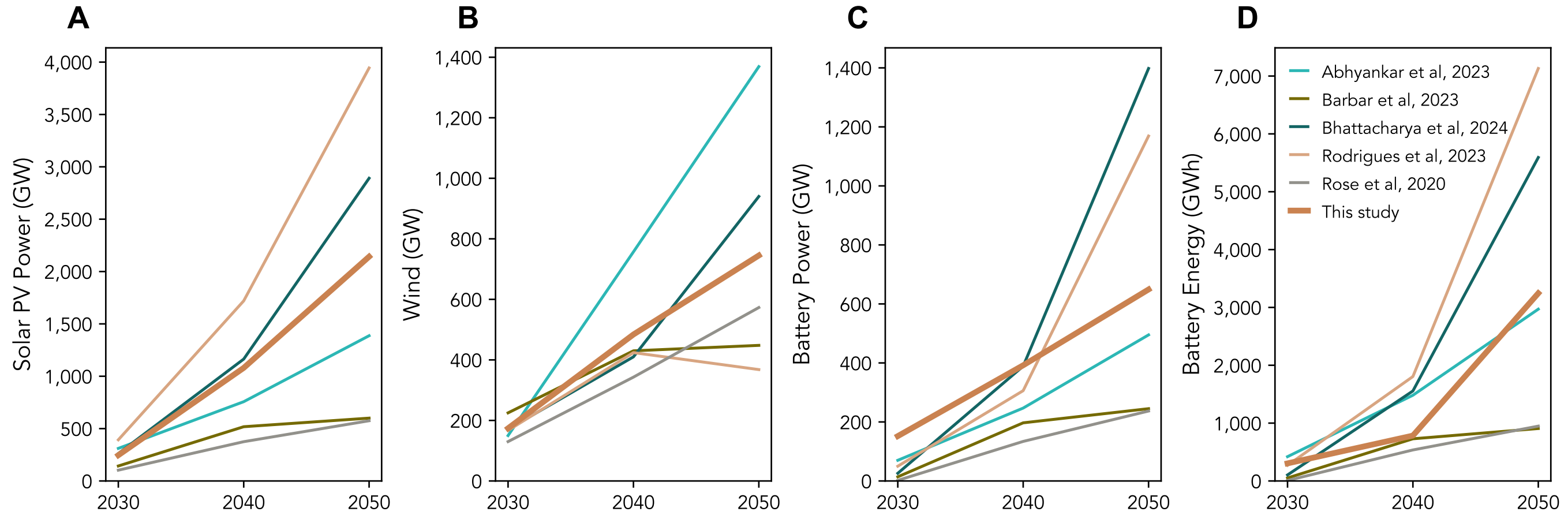


**Supplementary Figure 24. Battery capacity comparison between recent studies.** Panel (A) shows the comparison of battery storage power (GW) capacity between studies. Panel (B) is the comparison of battery storage energy capacity (GWh) in these same studies. The comparison in based on scenarios similar to our reference scenario.

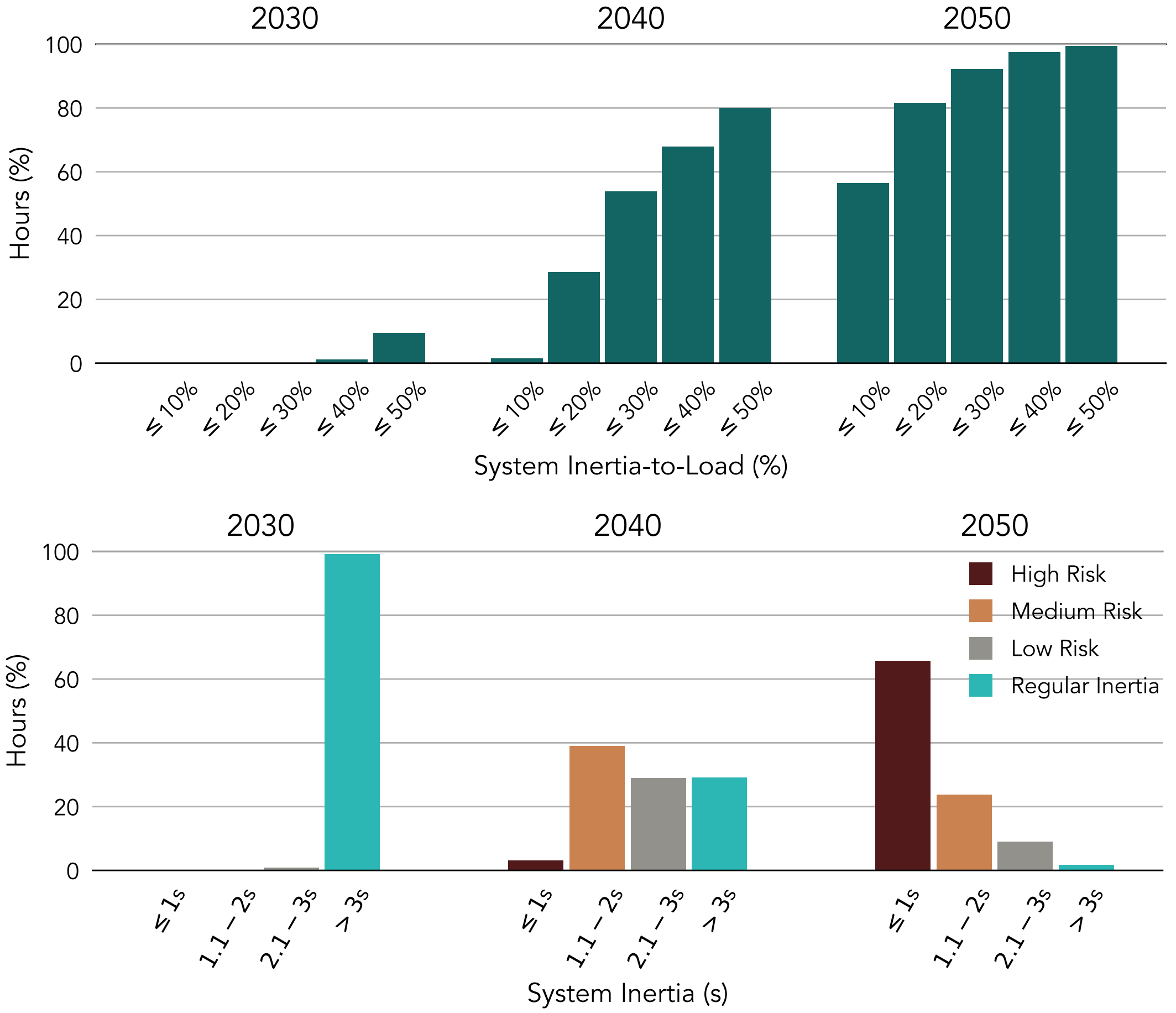


**Supplementary Figure 25. System inertia for the reference scenario.** Percentage of hours in a year that the power system operates with inertia capacity below 10%, 20%, 30%, 40%, and 50% of the load (top). Percentage of hours in a year that the power system operates with inertia resistant to frequency instability duration below 1s, between 1.1-2s and 2.1-3s, and 3s (standard operation).

## Supplementary Tables

**Supplementary Table 1: Literature focused on the energy transition of the electricity sector in India from 2020 (continued on the next page).** We highlight the contribution of each referred study to the timeline, investment periods in capacity expansion, temporal resolution in the production cost, candidate projects spatial resolution, analyzed scenarios, energy sectors considered in the study, aggregated peak and total demand, consideration of operational reliability constraints, findings summary, candidate project technologies, capacity expansion model, and interest rates assumptions on the cost projections.

| Reference | Timeline | Periods | Temporal | Spatial | Scenarios | Sector | Demand (peak) | Reliability | Summary | Technologies | Model | Interest rates |
|---|---|---|---|---|---|---|---|---|---|---|---|---|
| Lu et al., 2020 | 2040 | 1 period | 4 weeks (28 days) with 24 hours. | 5 load zones; 10 solar and wind projects. | 0%; 20%; 40%; 60%; 80% RPS. | Power. | N/A (475GW) | N/A | High VRE shares reduce the electricity cost. The electricity system will rely on wind more than on solar. | Wind, solar, hydro, gas, coal, PSH, battery, offshore wind. | MATLAB | 7% |
| Rose et al., 2020 | 2017-2040 | 41 periods (1 year increments) | 5 days with 7 hours; peak hour per season (5 seasons). | 34 load zones; 146 solar and wind projects. | Reference (mid); Wind (low); Battery (low); PV (low). | Power. | 4190TWh (640GW) | 15% (PRM); 5% demand (reserves). | Generation concentrates in states with good VRE resources, resulting in new transmission requirements. This study does not use Indian costs and has low energy and peak demand. | Wind, solar, hydro, gas, coal, PSH, battery, nuclear. | ReEDS | N/A |
| Gulagi et al., 2022 | 2015-2050 | 7 periods (5-year increments) | 356 days with 24 hours. | 22 load zones; 44 solar and wind projects. | 28-150 EUR/tCO2 from 2020-2050 (Carbon Tax). | Power. | 5921 TWh (1023GW) | N/A | Electricity cost drops to half by 2050. High transmission requirements. The driver for the transition is a carbon tax. | Wind, solar, hydro, gas, coal, PSH, battery, biomass, diesel, H2. | LUT Energy System Transition Model | They use WACC - 11% in 2015, declining steadily to 7% in 2050 (except solar which is 4%) |
| Song et al. 2022 | 2050 | 1 period | 356 days with 24 hours. | 5 load zones; 10 solar and wind projects. | Reference; 100% RPS; 00% RPS with H2 storage; 100% H2 storage and industry demand. | Power and industry. | 8323TWh (N/A) | N/A (spinning and non-spinning reserves). | The integration of H2 storage reduces curtailment and system costs. Similar capacity of solar and wind. | Wind, solar, hydro, gas, coal, PSH, battery. | MATLAB | 7% |
| Barbar et al., 2023 | 2020-2050 | 5 periods (10 year increments) | 140 days with 24 hours. | 5 load zones; 30 solar and wind projects. | Reference; Low storage costs; High AC efficiency; Low gas price. | Power. | 4773TWh (901GW); 4973TWh (901GW); 4199TWh (677GW); 4773TWh (877GW) | N/A | Similar solar and wind capacity with except when low storage cost. Gas's role is marginal. Incentives are necessary to avoid coal capacity. | Wind, solar, hydro, gas, coal, PSH, battery, nuclear. | GenX | 9% |
| Rodrigues et al., 2023 | 2020-2050 | 7 periods (5 years increments) | 12 weeks (84 days) with 24 hours; monthly peak demand week. | 5 load zones;100 solar and wind projects. | Solar and wind capacity constraint; Solar and wind capacity unconstraint; 100% RPS. | Power. | 4985 TWh (700 GW) | 5% demand + 5% wind (reserves); 90% thermal (availbility). | High VRE is the least-cost expansion and coal retirements by 2060. VRE resources max out by 2050. Wind capacity is much lower than solar. | Wind, solar, hydro, gas, coal, PSH, battery, nuclear, H2. | PyPSA | 10% |
| Jain et al., 2023 | 2020-2040 | 5 periods (5 years increments) | 12 days with 24 hours; average hour across days in a month. | 1 load zone. | No constraints; Unit commitment; Minimum stable capacity; Maximum ramp-rate; Start-up and downtime costs; Heat rate curve; Incremental carbon tax; COP26 pledge. | Power. | 3750TWh (543GW) | N/A | Absence of constraints in thermal generators results in VRE overstimation. The transition will take decades. | Wind, solar, hydro, gas, coal, PSH, battery. | The Integrated MARKAL-EFOM System (TIMES) | 10% |

**Supplementary Table 2: (continued) Literature focused on the energy transition of the electricity sector in India from 2020**.

| Reference | Timeline | Periods | Temporal | Spatial | Scenarios | Sector | Demand (peak) | Reliability | Summary | Technologies | Model | Interest rates |
|---|---|---|---|---|---|---|---|---|---|---|---|---|
| Bhattacharya et al., 2024 | 2020-2070 | 11 periods (5-year increments) | 12 days with 24 hours; monthly peak demand. | 1 load zone; 104 solar and wind projects. | Reference; high nuclear; net-zero with COP26 pledge - no new nuclear, net-zero with COP26 pledge - new nuclear. | Power and industry. | 7900TWh (1850GW) | 5% (PRM). | Coal capacity will reach 511GW by 2045 and decline afterwards. Fulfilling the COP26 pledge will increase the cost of electricity. | Wind, solar, hydro, gas, coal, PSH, battery. | The Integrated MARKAL-EFOM System (TIMES) | 9% |
| Abhyankar et al., 2023 | 2020-2050 | 5 periods (1 year increments) | 8760 hours with 24 hours. | 1 load zone; 104 solar and wind projects. | COP26 pledge, 46% (2030), 80%, (2040), and 90% RPS (2050). | Power, transportation, and industry. | 3,850TWh (low electrification); 5150TWh (EV adoption); 6600TWh (high industry electrification) | N/A | 95% coal reduction by 2047. Coal generation is not expected to decline until 2035 due to demand growth. Domestic coal extraction peaks in the mid-2020s. | Wind, solar, hydro, gas, coal, PSH, battery, biomass, offshore wind. | PLEXOS | N/A |
| Rudnick et al., 2022 | 2040 | 1 period (2040) | 8760 hours for 1 year (2040). | 5 zones (regional); Up to 299GW of wind and 721GW of solar. | 40 scenarios - C-CAP (high, low), ST cost (high, low), VRE cost (high, low), NG cost (high, low), and Policy (RPS, Carbon Tax). | Power. | 396GW; 422GW | Storage - 100% reserves, coal and biomass - 40% for spinning reserves. Gas - 50% spinning reserves. VRE and backup diesel generators - assumed to be unable to provide reserves; No PRM. | Wind and solar capacities restricted to 299GW and 721GW respectively; storage at 485GW. Includes explicit CO2 emissions limitations. | Wind, solar, hydro, gas, coal, PSH, battery, nuclear. | MIT-GenX | N/A |
| Lugovoy et al., 2021 | 2050 | 1 period (2050) | 8760 hours for 1 year (2050). | 32 zones (114 wind locations and 60 solar). | 150 scenarios – Technology (PV, onshore wind, PV + wind, PV+onshore+offshore); Balancing (generic, interregional, storage+grid, dd flexibility); Demand (1x, 3x, 5x of 2019); Optimism: low, medium, high. | Power. | 1x: 1300TWh; 2x: 3800TWh; 3x:6400 TWh | N/A | Solar-wind complementarity reduces storage needs; LCOE 30-54USD/kWh. | Wind, solar, hydro, gas, coal, PSH, battery, nuclear. | IDEEA | N/A |
| Sharma et al., 2021 | 2050 | 1 period (2050) | N/A | N/A | 4 scenarios - ref; medium ambition; high ambition; NZ | Economy wide. | NZ non fossil capacity is 2439GW (99% of demand) | N/A | 21 million employment impacted in fossil dependent sectors. Annual power sector costs peak in 2035 at 4.8% of GDP, stabilize at 2.5% by 2050. | Wind, solar, hydro, gas, coal, diesel, PSH, battery, nuclear, biomass, offshore wind. | EPS-India | N/A |
| Singh et al, 2021 | 2020-2070 | 5 periods (2020-2070) | Does not model grid balancing. | N/A | 16 scenarios - CCS availability yes/no; H2 availability yes/no. | Power, transportation, and industry. | 5758TWh | N/A | India needs 10 trillion USD to achieve NZ by 2070. | Coal, hydro, nuclear, solar, wind, other, H2. | GCAM (Global Change Analysis Model) | N/A |

**Supplementary Table 3: Technology cost, policy, and demand scenarios**. The technology cost scenarios consider low/mid/high variable renewable energy (VRE) and energy storage system (ESS) costs, and low/high supercritical coal. The policy scenarios include the enacted policies of a 500 MW clean-energy generation capacity target by 2030 and the state-wise renewable purchase obligation (RPO) projected to 2050, and explore 80%/90%/100% $CO_2$ emission reduction targets from 2020 levels by 2050. The demand scenarios consider low/mid/high demand scenarios from bottom-up or linearly-scaled demand profiles.

| Scenarios | Analysis | VRE Costs | ESS Costs | Coal Costs | C-CAP | Alternative Technology | RPO | Capacity Factors | Demand Projections | Transmission | Clean Target |
|---|---|---|---|---|---|---|---|---|---|---|---|
| | Implications on the energy transition from: different technology cost scenarios; uncertain demand profiles and projections; ambition in the policy targets; considering alternative technologies. | India-specific declining costs trends for VRE technologies (onshore and offshore wind; single-axis, fixed-tilt and rooftop solar) - Low: rapid decline; Mid: expected decline; High: slow decline. | India-specific declining ESS costs trends for batteries, hydrogen, and PSH - Low: rapid decline; Mid: expected decline; High: slow decline. | India-specific static coal capacity expansion cost and fixed state-level fuel costs - Low: expected project cost; High: high project cost. | Carbon emissions reduction target from 2020 levels - 80% by 2050; 90% by 2050; 100% C-CAP by 2050. | Proposed projects of uncertain technologies - Nuclear: nuclear projects in states with existing nuclear capacity; Hydrogen: long-term hydrogen storage projects in states with appropriate geological formations; PSH: proposed PSH projects in states with assessed capacity. | RPO established by the Ministry of Power projected to 2050 - 50% by 2040: decelerating targets trend, static after 2040; 65% by 2050: decelerating targets trend; 75% by 2050: linear trend in targets. | Wind bias correction by 2019 CEA state-level capacity factors - Existing: correction applied to existing projects; New: bias correction applied to existing, new onshore and offshore wind projects. | State-level hourly demand profiles - PIER: PIER v2.0 bottom-up projections synchronized to 2019 profiles; ICED: linearly-scaled 2019 profiles total and peak demand to projections. | Inter-state transmission capacity from ReEDs-India model - Existing: no transmission expansion; New: existing lines expansion. | National clean-energy (solar, wind, hydro, biomass, and WHR) capacity target - 500 MW by 2030. |
| **VRE & ESS (low) Coal (low)** | Technology Cost | Low | Low | Low | 90% by 2050 | Hydrogen | 50% by 2050 | Wind (existing) | PIER v2.0 (mid) | Tx (new) | 500 MW by 2030 |
| **VRE & ESS (low) Coal (high)** | Technology Cost | Low | Low | High | 90% by 2050 | Hydrogen | 50% by 2050 | Wind (existing) | PIER v2.0 (mid) | Tx (new) | 500 MW by 2030 |
| **VRE & ESS (mid) Coal (low)** | Technology Cost | Mid | Mid | Low | 90% by 2050 | Hydrogen | 50% by 2050 | Wind (existing) | PIER v2.0 (mid) | Tx (new) | 500 MW by 2030 |
| **VRE & ESS (mid) Coal (high)** | Technology Cost | Mid | Mid | High | 90% by 2050 | Hydrogen | 50% by 2050 | Wind (existing) | PIER v2.0 (mid) | Tx (new) | 500 MW by 2030 |
| **VRE & ESS (high) Coal (low)** | Technology Cost | High | High | Low | 90% by 2050 | Hydrogen | 50% by 2050 | Wind (existing) | PIER v2.0 (mid) | Tx (new) | 500 MW by 2030 |
| **VRE & ESS (high) Coal (high)** | Technology Cost | High | High | High | 90% by 2050 | Hydrogen | 50% by 2050 | Wind (existing) | PIER v2.0 (mid) | Tx (new) | 500 MW by 2030 |
| **Bottom-up Demand (low)** | Demand | Mid | Mid | Low | 90% by 2050 | Hydrogen | 50% by 2050 | Wind (existing) | PIER v2.0 (low) | Tx (new) | 500 MW by 2030 |
| **Bottom-up Demand (high)** | Demand | Mid | Mid | Low | 90% by 2050 | Hydrogen | 50% by 2050 | Wind (existing) | PIER v2.0 (high) | Tx (new) | 500 MW by 2030 |
| **Linearly-scaled Demand (low)** | Demand | Mid | Mid | Low | 90% by 2050 | Hydrogen | 50% by 2050 | Wind (existing) | 2019 ICED (low) | Tx (new) | 500 MW by 2030 |
| **Linearly-scaled Demand (mid)** | Demand | Mid | Mid | Low | 90% by 2050 | Hydrogen | 50% by 2050 | Wind (existing) | 2019 ICED (mid) | Tx (new) | 500 MW by 2030 |
| **Linearly-scaled Demand (high)** | Demand | Mid | Mid | Low | 90% by 2050 | Hydrogen | 50% by 2050 | Wind (existing) | 2019 ICED (high) | Tx (new) | 500 MW by 2030 |
| **No RPOs, Clean Target or Carbon Target** | Policy | Mid | Mid | Low | - | Hydrogen | - | Wind (existing) | PIER v2.0 (mid) | Tx (new) | - |
| **No Clean Target or Carbon Target** | Policy | Mid | Mid | Low | - | Hydrogen | 50% by 2050 | Wind (existing) | PIER v2.0 (mid) | Tx (new) | - |
| **No Carbon Target** | Policy | Mid | Mid | Low | - | Hydrogen | 50% by 2050 | Wind (existing) | PIER v2.0 (mid) | Tx (new) | 500 MW by 2030 |
| **80% Carbon Target** | Policy | Mid | Mid | Low | 80% by 2050 | Hydrogen | 50% by 2050 | Wind (existing) | PIER v2.0 (mid) | Tx (new) | 500 MW by 2030 |
| **90% Carbon Target** | Policy | Mid | Mid | Low | 90% by 2050 | Hydrogen | 50% by 2050 | Wind (existing) | PIER v2.0 (mid) | Tx (new) | 500 MW by 2030 |
| **100% Carbon Target** | Policy | Mid | Mid | Low | 100% by 2050 | Hydrogen | 50% by 2050 | Wind (existing) | PIER v2.0 (mid) | Tx (new) | 500 MW by 2030 |
| **No RPO, Clean Target or Carbon Target (Linearly-Scaled)** | Policy | Mid | Mid | Low | - | Hydrogen | - | Wind (existing) | 2019 ICED (mid) | Tx (new) | - |
| **No Clean Target or Carbon-Target (Linearly-Scaled)** | Policy | Mid | Mid | Low | - | Hydrogen | 50% by 2050 | Wind (existing) | 2019 ICED (mid) | Tx (new) | - |
| **No Carbon Target (Linearly-Scaled)** | Policy | Mid | Mid | Low | - | Hydrogen | 50% by 2050 | Wind (existing) | 2019 ICED (mid) | Tx (new) | 500 MW by 2030 |
| **80% Carbon Target (Linearly-Scaled)** | Policy | Mid | Mid | Low | 80% by 2050 | Hydrogen | 50% by 2050 | Wind (existing) | 2019 ICED (mid) | Tx (new) | 500 MW by 2030 |
| **90% Carbon Target (Linearly-Scaled)** | Policy | Mid | Mid | Low | 90% by 2050 | Hydrogen | 50% by 2050 | Wind (existing) | 2019 ICED (mid) | Tx (new) | 500 MW by 2030 |
| **100% Carbon Target (Linearly-Scaled)** | Policy | Mid | Mid | Low | 100% by 2050 | Hydrogen | 50% by 2050 | Wind (existing) | 2019 ICED (mid) | Tx (new) | 500 MW by 2030 |
| **No Alternative** | Alternative | Mid | Mid | Low | 90% by 2050 | - | 50% by 2050 | Wind (existing) | PIER v2.0 (mid) | Tx (new) | 500 MW by 2030 |
| **Alternative (PSH)** | Alternative | Mid | Mid | Low | 90% by 2050 | PSH | 50% by 2050 | Wind (existing) | PIER v2.0 (mid) | Tx (new) | 500 MW by 2030 |
| **Alternative (Nuclear)** | Alternative | Mid | Mid | Low | 90% by 2050 | Nuclear | 50% by 2050 | Wind (existing) | PIER v2.0 (mid) | Tx (new) | 500 MW by 2030 |
| **Alternative (H2 & PSH)** | Alternative | Mid | Mid | Low | 90% by 2050 | Hydrogen, and PSH | 50% by 2050 | Wind (existing) | PIER v2.0 (mid) | Tx (new) | 500 MW by 2030 |
| **Alternative (all)** | Alternative | Mid | Mid | Low | 90% by 2050 | Hydrogen, Nuclear, and PSH | 50% by 2050 | Wind (existing) | PIER v2.0 (mid) | Tx (new) | 500 MW by 2030 |

**Supplementary Table 4: Average system costs, greenhouse gas (GHG) emissions, and clean energy generation shares for the clean energy and carbon targets scenarios with bottom-up demand profiles**. The average system cost (USD/MWh), clean energy (%) and GHG emissions are grouped by period and scenario.

| | **Costs (USD/MWh)** | | | | **Clean Energy (%)** | | | | **Emissions ($MtCO_2$)** | | | |
|---|---|---|---|---|---|---|---|---|---|---|---|---|
| **Scenario** | **2020** | **2030** | **2040** | **2050** | **2020** | **2030** | **2040** | **2050** | **2020** | **2030** | **2040** | **2050** |
| 100% Carbon Target | 56.9 | 53.1 | 45.5 | 52.1 | 23.9 | 49.1 | 84.1 | 100.0 | 1119.1 | 1261.6 | 632.0 | 0.0 |
| 80% Carbon Target | 56.9 | 53.1 | 45.1 | 45.7 | 23.9 | 49.1 | 81.6 | 95.6 | 1119.1 | 1261.9 | 740.0 | 220.0 |
| 90% Carbon Target - REF | 56.9 | 53.1 | 45.2 | 47.4 | 23.9 | 49.1 | 82.8 | 97.7 | 1119.1 | 1261.8 | 687.0 | 110.0 |
| No Carbon Target | 56.9 | 53.1 | 44.9 | 43.2 | 23.9 | 49.1 | 78.0 | 82.6 | 1119.1 | 1262.1 | 890.6 | 937.2 |
| No Clean Target or Carbon Target | 56.9 | 49.5 | 45.4 | 43.0 | 23.9 | 72.6 | 77.7 | 82.6 | 1119.1 | 677.2 | 902.6 | 937.2 |

**Supplementary Table 5: Average system costs, greenhouse gas (GHG) emissions, and clean energy generation shares for the different demand scenarios**. The average system cost (USD/MWh), clean energy (%) and GHG emissions are grouped by period and scenario.

| | **Costs (USD/MWh)** | | | | **Clean Energy (%)** | | | | **Emissions ($MtCO_2$)** | | | |
|---|---|---|---|---|---|---|---|---|---|---|---|---|
| **Scenario** | **2020** | **2030** | **2040** | **2050** | **2020** | **2030** | **2040** | **2050** | **2020** | **2030** | **2040** | **2050** |
| VRE & ESS (high) Coal (high) | 56.9 | 55.9 | 51.1 | 56.0 | 23.9 | 49.3 | 82.7 | 97.6 | 1119.1 | 1257.6 | 687.0 | 110.0 |
| VRE & ESS (high) Coal (low) | 56.9 | 55.9 | 51.1 | 56.0 | 23.9 | 49.3 | 82.7 | 97.6 | 1119.1 | 1257.6 | 687.0 | 110.0 |
| VRE & ESS (low) Coal (high) | 56.9 | 50.5 | 39.3 | 37.2 | 23.9 | 49.3 | 83.0 | 97.7 | 1119.1 | 1256.3 | 687.0 | 110.0 |
| VRE & ESS (low) Coal (low) | 56.9 | 50.5 | 39.3 | 37.2 | 23.9 | 49.3 | 83.0 | 97.7 | 1119.1 | 1256.3 | 687.0 | 110.0 |
| VRE & ESS (mid) Coal (high) | 56.9 | 53.1 | 45.2 | 47.4 | 23.9 | 49.1 | 82.8 | 97.7 | 1119.1 | 1261.8 | 687.0 | 110.0 |
| VRE & ESS (mid) Coal (low) - REF | 56.9 | 53.1 | 45.2 | 47.4 | 23.9 | 49.1 | 82.8 | 97.7 | 1119.1 | 1261.8 | 687.0 | 110.0 |

**Supplementary Table 6: Average system costs, greenhouse gas (GHG) emissions, and clean energy generation shares for the different demand scenarios**. The average system cost (USD/MWh), clean energy (%) and GHG emissions are grouped by period and scenario.

| | **Costs (USD/MWh)** | | | | **Clean Energy (%)** | | | | **Emissions ($MtCO_2$)** | | | |
|---|---|---|---|---|---|---|---|---|---|---|---|---|
| **Scenario** | **2020** | **2030** | **2040** | **2050** | **2020** | **2030** | **2040** | **2050** | **2020** | **2030** | **2040** | **2050** |
| Bottom-up Demand (high) | 56.9 | 53.6 | 45.6 | 48.1 | 23.9 | 48.8 | 83.6 | 97.8 | 1119.1 | 1283.2 | 687.0 | 110.0 |
| Bottom-up Demand (low) | 56.9 | 53.1 | 45.2 | 46.7 | 23.9 | 50.1 | 82.1 | 97.6 | 1119.1 | 1221.0 | 687.0 | 110.0 |
| Bottom-up Demand (mid) - REF | 56.9 | 53.1 | 45.2 | 47.4 | 23.9 | 49.1 | 82.8 | 97.7 | 1119.1 | 1261.8 | 687.0 | 110.0 |
| Linearly-scaled Demand (high) | 56.7 | 53.1 | 48.0 | 52.2 | 24.2 | 48.6 | 83.6 | 97.5 | 1111.8 | 1289.6 | 687.0 | 110.0 |
| Linearly-scaled Demand (low) | 56.7 | 52.3 | 47.7 | 51.6 | 24.2 | 50.4 | 82.1 | 97.3 | 1111.8 | 1213.0 | 687.0 | 110.0 |
| Linearly-scaled Demand (mid) | 56.7 | 52.6 | 47.6 | 51.6 | 24.2 | 49.2 | 82.7 | 97.4 | 1111.8 | 1262.4 | 687.0 | 110.0 |

**Supplementary Table 7: Average system costs, greenhouse gas (GHG) emissions, and clean energy generation shares for the clean energy and carbon targets scenarios with linearly-scaled demand profiles**. The average system cost (USD/MWh), clean energy (%) and GHG emissions are grouped by period and scenario.

| | **Costs (USD/MWh)** | | | | **Clean Energy (%)** | | | | **Emissions ($MtCO_2$)** | | | |
|---|---|---|---|---|---|---|---|---|---|---|---|---|
| **Scenario** | **2020** | **2030** | **2040** | **2050** | **2020** | **2030** | **2040** | **2050** | **2020** | **2030** | **2040** | **2050** |
| 100% Carbon Target | 56.7 | 52.5 | 47.9 | 59.3 | 24.2 | 49.2 | 84.3 | 100.0 | 1111.8 | 1261.2 | 632.0 | 0.0 |
| 80% Carbon Target | 56.7 | 52.5 | 47.5 | 49.6 | 24.2 | 49.2 | 81.4 | 95.4 | 1111.8 | 1261.1 | 740.0 | 220.0 |
| 90% Carbon Target | 56.7 | 52.6 | 47.6 | 51.6 | 24.2 | 49.2 | 82.7 | 97.4 | 1111.8 | 1262.4 | 687.0 | 110.0 |
| No Carbon Target | 56.7 | 52.5 | 47.4 | 46.6 | 24.2 | 49.3 | 78.1 | 82.0 | 1111.8 | 1258.4 | 876.0 | 936.3 |
| No Clean Target or Carbon Target | 56.7 | 49.7 | 47.8 | 46.4 | 24.2 | 69.6 | 78.3 | 82.0 | 1111.8 | 756.9 | 867.9 | 938.2 |

**Supplementary Table 8: Average system costs, greenhouse gas (GHG) emissions, and clean energy generation shares for the different alternative technology scenarios**. The average system cost (USD/MWh), clean energy (%) and GHG emissions are grouped by period and scenario.

| | Costs (USD/MWh) | | | | Clean Energy (%) | | | | Emissions ($MtCO_2$) | | | |
|---|---|---|---|---|---|---|---|---|---|---|---|---|
| **Scenario** | **2020** | **2030** | **2040** | **2050** | **2020** | **2030** | **2040** | **2050** | **2020** | **2030** | **2040** | **2050** |
| Cost-optimal H2 - REF | 56.9 | 53.1 | 45.2 | 47.4 | 23.9 | 49.1 | 82.8 | 97.7 | 1119.1 | 1261.8 | 687.0 | 110.0 |
| H2 & high PSH & nuclear | 56.9 | 53.1 | 45.1 | 46.7 | 23.9 | 49.1 | 82.9 | 97.7 | 1119.1 | 1261.7 | 687.0 | 110.0 |
| High PSH | 56.9 | 53.1 | 45.1 | 46.9 | 23.9 | 49.1 | 82.9 | 97.7 | 1119.1 | 1261.8 | 687.0 | 110.0 |
| High nuclear | 56.9 | 53.1 | 45.2 | 47.2 | 23.9 | 49.1 | 82.8 | 97.7 | 1119.1 | 1261.8 | 687.0 | 110.0 |
| No H2 or high PSH or nuclear | 56.9 | 53.1 | 45.2 | 47.4 | 23.9 | 49.1 | 82.8 | 97.7 | 1119.1 | 1261.8 | 687.0 | 110.0 |

**Supplementary Table 9: Capacity expansion in each investment period for the different technology cost scenarios**. Total power (MW) and energy (MWh) capacity grouped by period, scenario, and technology.

| Scenario & Technology | Power Capacity (MW) | | | Energy Capacity (MWh) | | |
|---|---|---|---|---|---|---|
| | 2030 | 2040 | 2050 | 2030 | 2040 | 2050 |
| **VRE & ESS (high) Coal (high)** | **958,064** | **2,304,736** | **3,881,756** | **321,317** | **791,762** | **5,684,934** |
| Coal | 277,966 | 272,754 | 252,834 | 0 | 0 | 0 |
| Gas | 24,505 | 24,563 | 28,501 | 0 | 0 | 0 |
| Diesel | 1,214 | 1,214 | 1,214 | 0 | 0 | 0 |
| Nuclear | 12,680 | 12,680 | 12,680 | 0 | 0 | 0 |
| Other | 10,301 | 10,301 | 10,301 | 0 | 0 | 0 |
| Hydro | 56,271 | 56,271 | 56,271 | 0 | 0 | 0 |
| Solar | 240,449 | 1,015,817 | 2,018,600 | 0 | 0 | 0 |
| Wind | 180,299 | 521,536 | 867,436 | 0 | 0 | 0 |
| Pumped Storage | 3,140 | 3,140 | 3,140 | 18,840 | 18,840 | 18,840 |
| Battery | 151,238 | 386,461 | 619,244 | 302,477 | 772,922 | 2,729,840 |
| Hydrogen | 0 | 0 | 11,537 | 0 | 0 | 2,936,254 |
| **VRE & ESS (high) Coal (low)** | **958,064** | **2,304,736** | **3,881,756** | **321,317** | **791,762** | **5,684,934** |
| Coal | 277,966 | 272,754 | 252,834 | 0 | 0 | 0 |
| Gas | 24,505 | 24,563 | 28,501 | 0 | 0 | 0 |
| Diesel | 1,214 | 1,214 | 1,214 | 0 | 0 | 0 |
| Nuclear | 12,680 | 12,680 | 12,680 | 0 | 0 | 0 |
| Other | 10,301 | 10,301 | 10,301 | 0 | 0 | 0 |
| Hydro | 56,271 | 56,271 | 56,271 | 0 | 0 | 0 |
| Solar | 240,449 | 1,015,817 | 2,018,600 | 0 | 0 | 0 |
| Wind | 180,299 | 521,536 | 867,436 | 0 | 0 | 0 |
| Pumped Storage | 3,140 | 3,140 | 3,140 | 18,840 | 18,840 | 18,840 |
| Battery | 151,238 | 386,461 | 619,244 | 302,477 | 772,922 | 2,729,840 |
| Hydrogen | 0 | 0 | 11,537 | 0 | 0 | 2,936,254 |
| **VRE & ESS (low) Coal (high)** | **958,746** | **2,333,689** | **3,927,468** | **322,681** | **797,891** | **3,650,553** |
| Coal | 277,966 | 272,754 | 252,834 | 0 | 0 | 0 |
| Gas | 24,505 | 24,505 | 24,505 | 0 | 0 | 0 |
| Diesel | 1,214 | 1,214 | 1,214 | 0 | 0 | 0 |
| Nuclear | 12,680 | 12,680 | 12,680 | 0 | 0 | 0 |
| Other | 10,301 | 10,301 | 10,301 | 0 | 0 | 0 |
| Hydro | 56,271 | 56,271 | 56,271 | 0 | 0 | 0 |
| Solar | 240,882 | 1,056,781 | 2,228,260 | 0 | 0 | 0 |
| Wind | 179,866 | 506,519 | 674,706 | 0 | 0 | 0 |
| Pumped Storage | 3,140 | 3,140 | 3,140 | 18,840 | 18,840 | 18,840 |
| Battery | 151,920 | 389,526 | 663,558 | 303,841 | 779,051 | 3,631,713 |
| Hydrogen | 0 | 0 | 0 | 0 | 0 | 0 |
| **VRE & ESS (low) Coal (low)** | **958,746** | **2,333,689** | **3,927,468** | **322,681** | **797,891** | **3,650,553** |
| Coal | 277,966 | 272,754 | 252,834 | 0 | 0 | 0 |
| Gas | 24,505 | 24,505 | 24,505 | 0 | 0 | 0 |
| Diesel | 1,214 | 1,214 | 1,214 | 0 | 0 | 0 |
| Nuclear | 12,680 | 12,680 | 12,680 | 0 | 0 | 0 |
| Other | 10,301 | 10,301 | 10,301 | 0 | 0 | 0 |
| Hydro | 56,271 | 56,271 | 56,271 | 0 | 0 | 0 |
| Solar | 240,882 | 1,056,781 | 2,228,260 | 0 | 0 | 0 |
| Wind | 179,866 | 506,519 | 674,706 | 0 | 0 | 0 |
| Pumped Storage | 3,140 | 3,140 | 3,140 | 18,840 | 18,840 | 18,840 |
| Battery | 151,920 | 389,526 | 663,558 | 303,841 | 779,051 | 3,631,713 |
| Hydrogen | 0 | 0 | 0 | 0 | 0 | 0 |
| **VRE & ESS (mid) Coal (high)** | **959,307** | **2,338,309** | **3,901,227** | **323,803** | **804,186** | **4,262,133** |
| Coal | 277,966 | 272,754 | 252,834 | 0 | 0 | 0 |
| Gas | 24,505 | 24,505 | 24,505 | 0 | 0 | 0 |
| Diesel | 1,214 | 1,214 | 1,214 | 0 | 0 | 0 |
| Nuclear | 12,680 | 12,680 | 12,680 | 0 | 0 | 0 |
| Other | 10,301 | 10,301 | 10,301 | 0 | 0 | 0 |
| Hydro | 56,271 | 56,271 | 56,271 | 0 | 0 | 0 |
| Solar | 246,494 | 1,080,804 | 2,142,208 | 0 | 0 | 0 |
| Wind | 174,254 | 483,968 | 744,769 | 0 | 0 | 0 |
| Pumped Storage | 3,140 | 3,140 | 3,140 | 18,840 | 18,840 | 18,840 |
| Battery | 152,482 | 392,673 | 649,522 | 304,963 | 785,346 | 3,247,951 |
| Hydrogen | 0 | 0 | 3,783 | 0 | 0 | 995,342 |
| **VRE & ESS (mid) Coal (low) - REF** | **959,307** | **2,338,309** | **3,901,227** | **323,803** | **804,186** | **4,262,133** |
| Coal | 277,966 | 272,754 | 252,834 | 0 | 0 | 0 |
| Gas | 24,505 | 24,505 | 24,505 | 0 | 0 | 0 |
| Diesel | 1,214 | 1,214 | 1,214 | 0 | 0 | 0 |
| Nuclear | 12,680 | 12,680 | 12,680 | 0 | 0 | 0 |
| Other | 10,301 | 10,301 | 10,301 | 0 | 0 | 0 |
| Hydro | 56,271 | 56,271 | 56,271 | 0 | 0 | 0 |
| Solar | 246,494 | 1,080,804 | 2,142,208 | 0 | 0 | 0 |
| Wind | 174,254 | 483,968 | 744,769 | 0 | 0 | 0 |
| Pumped Storage | 3,140 | 3,140 | 3,140 | 18,840 | 18,840 | 18,840 |
| Battery | 152,482 | 392,673 | 649,522 | 304,963 | 785,346 | 3,247,951 |
| Hydrogen | 0 | 0 | 3,783 | 0 | 0 | 995,342 |

**Supplementary Table 10: Capacity expansion in each investment period for the different demand scenarios**. Total power (MW) and energy (MWh) capacity grouped by period, scenario, and technology.

| Scenario & Technology | Power Capacity (MW) | | | Energy Capacity (MWh) | | |
|---|---|---|---|---|---|---|
| | 2030 | 2040 | 2050 | 2030 | 2040 | 2050 |
| **Linearly-scaled Demand (high)** | **925,383** | **2,289,642** | **4,185,174** | **186,035** | **1,055,459** | **12,162,080** |
| Coal | 277,966 | 272,754 | 252,834 | 0 | 0 | 0 |
| Gas | 59,466 | 98,815 | 98,815 | 0 | 0 | 0 |
| Diesel | 1,214 | 1,214 | 1,214 | 0 | 0 | 0 |
| Nuclear | 12,680 | 12,680 | 12,680 | 0 | 0 | 0 |
| Other | 10,301 | 10,301 | 10,301 | 0 | 0 | 0 |
| Hydro | 56,271 | 56,271 | 56,271 | 0 | 0 | 0 |
| Solar | 251,287 | 906,868 | 2,085,752 | 0 | 0 | 0 |
| Wind | 169,461 | 656,437 | 1,050,072 | 0 | 0 | 0 |
| Pumped Storage | 3,140 | 3,140 | 3,140 | 18,840 | 18,840 | 18,840 |
| Battery | 83,597 | 271,163 | 584,495 | 167,195 | 1,036,619 | 3,618,213 |
| Hydrogen | 0 | 0 | 29,601 | 0 | 0 | 8,525,027 |
| **Linearly-scaled Demand (low)** | **910,268** | **2,054,967** | **3,713,556** | **186,316** | **806,086** | **10,227,564** |
| Coal | 277,966 | 272,754 | 252,834 | 0 | 0 | 0 |
| Gas | 44,210 | 83,258 | 83,258 | 0 | 0 | 0 |
| Diesel | 1,214 | 1,214 | 1,214 | 0 | 0 | 0 |
| Nuclear | 12,680 | 12,680 | 12,680 | 0 | 0 | 0 |
| Other | 10,301 | 10,301 | 10,301 | 0 | 0 | 0 |
| Hydro | 56,271 | 56,271 | 56,271 | 0 | 0 | 0 |
| Solar | 244,377 | 777,310 | 1,842,157 | 0 | 0 | 0 |
| Wind | 176,372 | 587,285 | 930,142 | 0 | 0 | 0 |
| Pumped Storage | 3,140 | 3,140 | 3,140 | 18,840 | 18,840 | 18,840 |
| Battery | 83,738 | 250,755 | 496,776 | 167,476 | 787,246 | 3,070,870 |
| Hydrogen | 0 | 0 | 24,784 | 0 | 0 | 7,137,853 |
| **Linearly-scaled Demand (mid)** | **917,939** | **2,148,506** | **3,895,636** | **189,986** | **916,416** | **10,144,802** |
| Coal | 277,966 | 272,754 | 252,834 | 0 | 0 | 0 |
| Gas | 50,046 | 87,679 | 87,679 | 0 | 0 | 0 |
| Diesel | 1,214 | 1,214 | 1,214 | 0 | 0 | 0 |
| Nuclear | 12,680 | 12,680 | 12,680 | 0 | 0 | 0 |
| Other | 10,301 | 10,301 | 10,301 | 0 | 0 | 0 |
| Hydro | 56,271 | 56,271 | 56,271 | 0 | 0 | 0 |
| Solar | 249,274 | 830,256 | 1,930,705 | 0 | 0 | 0 |
| Wind | 171,474 | 616,636 | 982,709 | 0 | 0 | 0 |
| Pumped Storage | 3,140 | 3,140 | 3,140 | 18,840 | 18,840 | 18,840 |
| Battery | 85,573 | 257,576 | 534,373 | 171,146 | 897,576 | 3,291,447 |
| Hydrogen | 0 | 0 | 23,731 | 0 | 0 | 6,834,515 |
| **Bottom-up Demand (high)** | **964,383** | **2,477,636** | **4,158,133** | **318,199** | **840,293** | **5,708,398** |
| Coal | 278,007 | 272,795 | 252,875 | 0 | 0 | 0 |
| Gas | 32,343 | 32,343 | 32,343 | 0 | 0 | 0 |
| Diesel | 1,214 | 1,214 | 1,214 | 0 | 0 | 0 |
| Nuclear | 12,680 | 12,680 | 12,680 | 0 | 0 | 0 |
| Other | 10,301 | 10,301 | 10,301 | 0 | 0 | 0 |
| Hydro | 56,271 | 56,271 | 56,271 | 0 | 0 | 0 |
| Solar | 230,776 | 1,120,462 | 2,316,402 | 0 | 0 | 0 |
| Wind | 189,972 | 563,682 | 798,561 | 0 | 0 | 0 |
| Pumped Storage | 3,140 | 3,140 | 3,140 | 18,840 | 18,840 | 18,840 |
| Battery | 149,680 | 404,750 | 667,176 | 299,359 | 821,453 | 3,626,559 |
| Hydrogen | 0 | 0 | 7,171 | 0 | 0 | 2,062,999 |
| **Bottom-up Demand (low)** | **955,903** | **2,261,158** | **3,711,824** | **316,996** | **820,264** | **3,728,420** |
| Coal | 277,966 | 272,754 | 252,834 | 0 | 0 | 0 |
| Gas | 24,505 | 24,505 | 24,505 | 0 | 0 | 0 |
| Diesel | 1,214 | 1,214 | 1,214 | 0 | 0 | 0 |
| Nuclear | 12,680 | 12,680 | 12,680 | 0 | 0 | 0 |
| Other | 10,301 | 10,301 | 10,301 | 0 | 0 | 0 |
| Hydro | 56,271 | 56,271 | 56,271 | 0 | 0 | 0 |
| Solar | 245,448 | 1,067,608 | 1,987,431 | 0 | 0 | 0 |
| Wind | 175,300 | 411,974 | 722,932 | 0 | 0 | 0 |
| Pumped Storage | 3,140 | 3,140 | 3,140 | 18,840 | 18,840 | 18,840 |
| Battery | 149,078 | 400,712 | 636,842 | 298,156 | 801,424 | 2,779,916 |
| Hydrogen | 0 | 0 | 3,673 | 0 | 0 | 929,664 |

**Supplementary Table 11: Capacity expansion in each investment period for the different policy scenarios with bottom-up demand profiles**. Total power (MW) and energy (MWh) capacity grouped by period, scenario, and technology.

| Scenario & Technology | Capacity (MW) | | | Capacity (MWh) | | |
|---|---|---|---|---|---|---|
| | 2030 | 2040 | 2050 | 2030 | 2040 | 2050 |
| **100% Carbon Target** | **959,301** | **2,371,716** | **4,283,179** | **323,791** | **797,185** | **20,176,046** |
| Coal | 277,966 | 272,754 | 252,834 | 0 | 0 | 0 |
| Gas | 24,505 | 24,505 | 24,505 | 0 | 0 | 0 |
| Diesel | 1,214 | 1,214 | 1,214 | 0 | 0 | 0 |
| Nuclear | 12,680 | 12,680 | 12,680 | 0 | 0 | 0 |
| Other | 10,301 | 10,301 | 10,301 | 0 | 0 | 0 |
| Hydro | 56,271 | 56,271 | 56,271 | 0 | 0 | 0 |
| Solar | 246,182 | 1,095,711 | 2,510,752 | 0 | 0 | 0 |
| Wind | 174,566 | 505,968 | 641,887 | 0 | 0 | 0 |
| Pumped Storage | 3,140 | 3,140 | 3,140 | 18,840 | 18,840 | 18,840 |
| Battery | 152,475 | 389,173 | 715,446 | 304,951 | 778,345 | 4,562,120 |
| Hydrogen | 0 | 0 | 54,150 | 0 | 0 | 15,595,086 |
| **No Carbon Target** | **959,152** | **2,207,507** | **3,259,294** | **323,494** | **816,507** | **1,501,189** |
| Coal | 277,966 | 272,754 | 281,448 | 0 | 0 | 0 |
| Gas | 24,505 | 24,505 | 49,569 | 0 | 0 | 0 |
| Diesel | 1,214 | 1,214 | 1,214 | 0 | 0 | 0 |
| Nuclear | 12,680 | 12,680 | 12,680 | 0 | 0 | 0 |
| Other | 10,301 | 10,301 | 10,301 | 0 | 0 | 0 |
| Hydro | 56,271 | 56,271 | 56,271 | 0 | 0 | 0 |
| Solar | 246,549 | 993,149 | 1,572,602 | 0 | 0 | 0 |
| Wind | 174,199 | 434,661 | 646,638 | 0 | 0 | 0 |
| Pumped Storage | 3,140 | 3,140 | 3,140 | 18,840 | 18,840 | 18,840 |
| Battery | 152,327 | 398,833 | 625,431 | 304,654 | 797,667 | 1,482,349 |
| Hydrogen | 0 | 0 | 0 | 0 | 0 | 0 |
| **80% Carbon Target** | **959,296** | **2,307,378** | **3,749,267** | **323,780** | **810,097** | **2,502,289** |
| Coal | 277,966 | 272,754 | 252,834 | 0 | 0 | 0 |
| Gas | 24,505 | 24,505 | 24,505 | 0 | 0 | 0 |
| Diesel | 1,214 | 1,214 | 1,214 | 0 | 0 | 0 |
| Nuclear | 12,680 | 12,680 | 12,680 | 0 | 0 | 0 |
| Other | 10,301 | 10,301 | 10,301 | 0 | 0 | 0 |
| Hydro | 56,271 | 56,271 | 56,271 | 0 | 0 | 0 |
| Solar | 246,506 | 1,066,951 | 1,929,999 | 0 | 0 | 0 |
| Wind | 174,243 | 463,935 | 811,818 | 0 | 0 | 0 |
| Pumped Storage | 3,140 | 3,140 | 3,140 | 18,840 | 18,840 | 18,840 |
| Battery | 152,470 | 395,629 | 646,506 | 304,940 | 791,257 | 2,483,449 |
| Hydrogen | 0 | 0 | 0 | 0 | 0 | 0 |
| **No Clean Target or Carbon Target** | **1,199,944** | **2,204,117** | **3,261,899** | **231,186** | **818,170** | **1,507,748** |
| Coal | 277,966 | 272,754 | 281,790 | 0 | 0 | 0 |
| Gas | 24,505 | 24,505 | 48,461 | 0 | 0 | 0 |
| Diesel | 1,214 | 1,214 | 1,214 | 0 | 0 | 0 |
| Nuclear | 12,680 | 12,680 | 12,680 | 0 | 0 | 0 |
| Other | 10,301 | 10,301 | 10,301 | 0 | 0 | 0 |
| Hydro | 56,271 | 56,271 | 56,271 | 0 | 0 | 0 |
| Solar | 492,980 | 995,938 | 1,575,864 | 0 | 0 | 0 |
| Wind | 214,714 | 427,649 | 645,351 | 0 | 0 | 0 |
| Pumped Storage | 3,140 | 3,140 | 3,140 | 18,840 | 18,840 | 18,840 |
| Battery | 106,173 | 399,665 | 626,828 | 212,346 | 799,330 | 1,488,908 |
| Hydrogen | 0 | 0 | 0 | 0 | 0 | 0 |

**Supplementary Table 12: Capacity expansion in each investment period for the different policy scenarios with linearly-scaled demand profiles**. Total power (MW) and energy (MWh) capacity grouped by period, scenario, and technology.

| Scenario & Technology | Power Capacity (MW) | | | Energy Capacity (MWh) | | |
|---|---|---|---|---|---|---|
| | 2030 | 2040 | 2050 | 2030 | 2040 | 2050 |
| **100% Carbon Target** | **918,007** | **2,218,969** | **4,741,021** | **188,703** | **1,177,077** | **76,132,350** |
| Coal | 277,966 | 272,754 | 252,834 | 0 | 0 | 0 |
| Gas | 50,756 | 57,471 | 57,471 | 0 | 0 | 0 |
| Diesel | 1,214 | 1,214 | 1,214 | 0 | 0 | 0 |
| Nuclear | 12,680 | 12,680 | 12,680 | 0 | 0 | 0 |
| Other | 10,301 | 10,301 | 10,301 | 0 | 0 | 0 |
| Hydro | 56,271 | 56,271 | 56,271 | 0 | 0 | 0 |
| Solar | 249,275 | 904,156 | 2,632,962 | 0 | 0 | 0 |
| Wind | 171,474 | 612,462 | 651,190 | 0 | 0 | 0 |
| Pumped Storage | 3,140 | 3,140 | 3,140 | 18,840 | 18,840 | 18,840 |
| Battery | 84,932 | 288,521 | 816,373 | 169,863 | 1,158,237 | 5,096,700 |
| Hydrogen | 0 | 0 | 246,586 | 0 | 0 | 71,016,810 |
| **No Carbon Target** | **918,714** | **2,035,944** | **2,971,645** | **178,791** | **732,144** | **1,495,311** |
| Coal | 282,790 | 284,595 | 285,686 | 0 | 0 | 0 |
| Gas | 51,594 | 99,781 | 194,252 | 0 | 0 | 0 |
| Diesel | 1,214 | 1,214 | 1,214 | 0 | 0 | 0 |
| Nuclear | 12,680 | 12,680 | 12,680 | 0 | 0 | 0 |
| Other | 10,301 | 10,301 | 10,301 | 0 | 0 | 0 |
| Hydro | 56,271 | 56,271 | 56,271 | 0 | 0 | 0 |
| Solar | 249,274 | 752,133 | 1,224,573 | 0 | 0 | 0 |
| Wind | 171,474 | 572,455 | 823,500 | 0 | 0 | 0 |
| Pumped Storage | 3,140 | 3,140 | 3,140 | 18,840 | 18,840 | 18,840 |
| Battery | 79,976 | 243,375 | 360,029 | 159,951 | 713,304 | 1,476,471 |
| Hydrogen | 0 | 0 | 0 | 0 | 0 | 0 |
| **80% Carbon Target** | **917,842** | **2,109,445** | **3,682,999** | **187,126** | **834,214** | **11,871,916** |
| Coal | 277,966 | 272,754 | 252,834 | 0 | 0 | 0 |
| Gas | 51,379 | 98,590 | 102,466 | 0 | 0 | 0 |
| Diesel | 1,214 | 1,214 | 1,214 | 0 | 0 | 0 |
| Nuclear | 12,680 | 12,680 | 12,680 | 0 | 0 | 0 |
| Other | 10,301 | 10,301 | 10,301 | 0 | 0 | 0 |
| Hydro | 56,271 | 56,271 | 56,271 | 0 | 0 | 0 |
| Solar | 249,274 | 803,282 | 1,818,612 | 0 | 0 | 0 |
| Wind | 171,474 | 602,150 | 908,680 | 0 | 0 | 0 |
| Pumped Storage | 3,140 | 3,140 | 3,140 | 18,840 | 18,840 | 18,840 |
| Battery | 84,143 | 249,064 | 485,824 | 168,286 | 815,374 | 2,931,377 |
| Hydrogen | 0 | 0 | 30,978 | 0 | 0 | 8,921,700 |
| **90% Carbon Target** | **917,939** | **2,148,506** | **3,895,636** | **189,986** | **916,416** | **10,144,802** |
| Coal | 277,966 | 272,754 | 252,834 | 0 | 0 | 0 |
| Gas | 50,046 | 87,679 | 87,679 | 0 | 0 | 0 |
| Diesel | 1,214 | 1,214 | 1,214 | 0 | 0 | 0 |
| Nuclear | 12,680 | 12,680 | 12,680 | 0 | 0 | 0 |
| Other | 10,301 | 10,301 | 10,301 | 0 | 0 | 0 |
| Hydro | 56,271 | 56,271 | 56,271 | 0 | 0 | 0 |
| Solar | 249,274 | 830,256 | 1,930,705 | 0 | 0 | 0 |
| Wind | 171,474 | 616,636 | 982,709 | 0 | 0 | 0 |
| Pumped Storage | 3,140 | 3,140 | 3,140 | 18,840 | 18,840 | 18,840 |
| Battery | 85,573 | 257,576 | 534,373 | 171,146 | 897,576 | 3,291,447 |
| Hydrogen | 0 | 0 | 23,731 | 0 | 0 | 6,834,515 |
| **No Clean Target or Carbon Target** | **1,076,660** | **2,037,466** | **2,967,956** | **189,274** | **731,544** | **1,499,090** |
| Coal | 277,966 | 285,882 | 288,036 | 0 | 0 | 0 |
| Gas | 24,505 | 99,331 | 190,701 | 0 | 0 | 0 |
| Diesel | 1,214 | 1,214 | 1,214 | 0 | 0 | 0 |
| Nuclear | 12,680 | 12,680 | 12,680 | 0 | 0 | 0 |
| Other | 10,301 | 10,301 | 10,301 | 0 | 0 | 0 |
| Hydro | 56,271 | 56,271 | 56,271 | 0 | 0 | 0 |
| Solar | 336,270 | 751,764 | 1,218,576 | 0 | 0 | 0 |
| Wind | 269,097 | 575,389 | 825,503 | 0 | 0 | 0 |
| Pumped Storage | 3,140 | 3,140 | 3,140 | 18,840 | 18,840 | 18,840 |
| Battery | 85,217 | 241,495 | 361,535 | 170,434 | 712,704 | 1,480,250 |
| Hydrogen | 0 | 0 | 0 | 0 | 0 | 0 |

**Supplementary Table 13: Capacity expansion in each investment period for the different alternative technology scenarios**. Total power (MW) and energy (MWh) capacity grouped by period, scenario, and technology.

| Scenario & Technology | Power Capacity (MW) | | | Energy Capacity (MWh) | | |
|---|---|---|---|---|---|---|
| | 2030 | 2040 | 2050 | 2030 | 2040 | 2050 |
| **H2 & high PSH & nuclear** | **959,254** | **2,360,729** | **3,672,856** | **323,697** | **1,080,430** | **2,898,080** |
| Coal | 277,966 | 272,754 | 252,834 | 0 | 0 | 0 |
| Gas | 24,505 | 24,505 | 24,505 | 0 | 0 | 0 |
| Diesel | 1,214 | 1,214 | 1,214 | 0 | 0 | 0 |
| Nuclear | 12,680 | 12,680 | 65,657 | 0 | 0 | 0 |
| Other | 10,301 | 10,301 | 10,301 | 0 | 0 | 0 |
| Hydro | 56,271 | 56,271 | 56,271 | 0 | 0 | 0 |
| Solar | 246,488 | 1,148,608 | 1,960,006 | 0 | 0 | 0 |
| Wind | 174,260 | 433,165 | 683,656 | 0 | 0 | 0 |
| Pumped Storage | 3,140 | 54,400 | 120,293 | 18,840 | 386,766 | 876,240 |
| Battery | 152,428 | 346,832 | 498,120 | 304,857 | 693,664 | 2,021,840 |
| Hydrogen | 0 | 0 | 0 | 0 | 0 | 0 |
| **High nuclear** | **959,226** | **2,338,706** | **3,646,141** | **323,640** | **804,019** | **2,729,122** |
| Coal | 277,966 | 272,754 | 252,834 | 0 | 0 | 0 |
| Gas | 24,505 | 24,505 | 24,505 | 0 | 0 | 0 |
| Diesel | 1,214 | 1,214 | 1,214 | 0 | 0 | 0 |
| Nuclear | 12,680 | 12,680 | 70,180 | 0 | 0 | 0 |
| Other | 10,301 | 10,301 | 10,301 | 0 | 0 | 0 |
| Hydro | 56,271 | 56,271 | 56,271 | 0 | 0 | 0 |
| Solar | 246,498 | 1,081,496 | 1,920,662 | 0 | 0 | 0 |
| Wind | 174,251 | 483,756 | 697,290 | 0 | 0 | 0 |
| Pumped Storage | 3,140 | 3,140 | 3,140 | 18,840 | 18,840 | 18,840 |
| Battery | 152,400 | 392,589 | 609,745 | 304,800 | 785,179 | 2,710,282 |
| Hydrogen | 0 | 0 | 0 | 0 | 0 | 0 |
| **High PSH** | **959,320** | **2,361,521** | **3,911,645** | **323,830** | **1,082,120** | **3,391,774** |
| Coal | 277,966 | 272,754 | 252,834 | 0 | 0 | 0 |
| Gas | 24,505 | 24,505 | 24,505 | 0 | 0 | 0 |
| Diesel | 1,214 | 1,214 | 1,214 | 0 | 0 | 0 |
| Nuclear | 12,680 | 12,680 | 12,680 | 0 | 0 | 0 |
| Other | 10,301 | 10,301 | 10,301 | 0 | 0 | 0 |
| Hydro | 56,271 | 56,271 | 56,271 | 0 | 0 | 0 |
| Solar | 246,490 | 1,149,578 | 2,161,048 | 0 | 0 | 0 |
| Wind | 174,259 | 432,546 | 735,525 | 0 | 0 | 0 |
| Pumped Storage | 3,140 | 54,632 | 118,521 | 18,840 | 388,039 | 876,240 |
| Battery | 152,495 | 347,040 | 538,747 | 304,990 | 694,081 | 2,515,534 |
| Hydrogen | 0 | 0 | 0 | 0 | 0 | 0 |
| **No H2 or high PSH or nuclear** | **959,307** | **2,338,545** | **3,901,341** | **323,803** | **804,243** | **3,294,448** |
| Coal | 277,966 | 272,754 | 252,834 | 0 | 0 | 0 |
| Gas | 24,505 | 24,505 | 24,505 | 0 | 0 | 0 |
| Diesel | 1,214 | 1,214 | 1,214 | 0 | 0 | 0 |
| Nuclear | 12,680 | 12,680 | 12,680 | 0 | 0 | 0 |
| Other | 10,301 | 10,301 | 10,301 | 0 | 0 | 0 |
| Hydro | 56,271 | 56,271 | 56,271 | 0 | 0 | 0 |
| Solar | 246,494 | 1,081,046 | 2,142,290 | 0 | 0 | 0 |
| Wind | 174,254 | 483,933 | 745,174 | 0 | 0 | 0 |
| Pumped Storage | 3,140 | 3,140 | 3,140 | 18,840 | 18,840 | 18,840 |
| Battery | 152,482 | 392,701 | 652,932 | 304,963 | 785,403 | 3,275,608 |
| Hydrogen | 0 | 0 | 0 | 0 | 0 | 0 |

**Supplementary Table 14: Capacity expansion in each investment period to analyze the effect of a 90% carbon target**. Total power (MW) and energy (MWh) capacity grouped by period, scenario, and technology.

| Scenario & Technology | Power Capacity (MW) 2030 | 2040 | 2050 | Energy Capacity (MWh) 2030 | 2040 | 2050 |
|---|---|---|---|---|---|---|
| **VRE & ESS (low) Coal (low) & No Carbon Target** | **958,500** | **2,283,316** | **3,506,840** | **322,190** | **797,967** | **2,302,701** |
| Coal | 277,966 | 272,754 | 252,834 | 0 | 0 | 0 |
| Gas | 24,505 | 24,505 | 25,580 | 0 | 0 | 0 |
| Diesel | 1,214 | 1,214 | 1,214 | 0 | 0 | 0 |
| Nuclear | 12,680 | 12,680 | 12,680 | 0 | 0 | 0 |
| Other | 10,301 | 10,301 | 10,301 | 0 | 0 | 0 |
| Hydro | 56,271 | 56,271 | 56,271 | 0 | 0 | 0 |
| Solar | 240,882 | 1,028,204 | 1,821,647 | 0 | 0 | 0 |
| Wind | 179,866 | 484,684 | 658,595 | 0 | 0 | 0 |
| Pumped Storage | 3,140 | 3,140 | 3,140 | 18,840 | 18,840 | 18,840 |
| Battery | 151,675 | 389,563 | 664,579 | 303,350 | 779,127 | 2,283,861 |
| Hydrogen | 0 | 0 | 0 | 0 | 0 | 0 |
| **VRE & ESS (mid) Coal (low) & No Carbon Target** | **959,152** | **2,207,507** | **3,259,294** | **323,494** | **816,507** | **1,501,189** |
| Coal | 277,966 | 272,754 | 281,448 | 0 | 0 | 0 |
| Gas | 24,505 | 24,505 | 49,569 | 0 | 0 | 0 |
| Diesel | 1,214 | 1,214 | 1,214 | 0 | 0 | 0 |
| Nuclear | 12,680 | 12,680 | 12,680 | 0 | 0 | 0 |
| Other | 10,301 | 10,301 | 10,301 | 0 | 0 | 0 |
| Hydro | 56,271 | 56,271 | 56,271 | 0 | 0 | 0 |
| Solar | 246,549 | 993,149 | 1,572,602 | 0 | 0 | 0 |
| Wind | 174,199 | 434,661 | 646,638 | 0 | 0 | 0 |
| Pumped Storage | 3,140 | 3,140 | 3,140 | 18,840 | 18,840 | 18,840 |
| Battery | 152,327 | 398,833 | 625,431 | 304,654 | 797,667 | 1,482,349 |
| Hydrogen | 0 | 0 | 0 | 0 | 0 | 0 |

**Supplementary Table 15: Capacity expansion in each investment period to analyze the effect of Renewable Purchase Obligations (RPO)**. Total power (MW) and energy (MWh) capacity grouped by period, scenario, and technology.

| Scenario & Technology | Power Capacity (MW) 2030 | 2040 | 2050 | Energy Capacity (MWh) 2030 | 2040 | 2050 |
|---|---|---|---|---|---|---|
| **VRE & ESS (mid) Coal (low) & No RPO & 90% Carbon Target** | **937,838** | **2,329,887** | **3,895,792** | **280,865** | **804,994** | **4,183,347** |
| Coal | 277,966 | 272,754 | 252,834 | 0 | 0 | 0 |
| Gas | 24,505 | 24,505 | 24,505 | 0 | 0 | 0 |
| Diesel | 1,214 | 1,214 | 1,214 | 0 | 0 | 0 |
| Nuclear | 12,680 | 12,680 | 12,680 | 0 | 0 | 0 |
| Other | 10,301 | 10,301 | 10,301 | 0 | 0 | 0 |
| Hydro | 56,271 | 56,271 | 56,271 | 0 | 0 | 0 |
| Solar | 237,102 | 1,075,578 | 2,145,284 | 0 | 0 | 0 |
| Wind | 183,647 | 480,368 | 735,400 | 0 | 0 | 0 |
| Pumped Storage | 3,140 | 3,140 | 3,140 | 18,840 | 18,840 | 18,840 |
| Battery | 131,013 | 393,077 | 650,715 | 262,025 | 786,154 | 3,262,775 |
| Hydrogen | 0 | 0 | 3,449 | 0 | 0 | 901,731 |

**Supplementary Table 16: Capacity expansion in each investment period to analyze the effect of an 8% Planning Reserve Margin (PRM).** Total power (MW) and energy (MWh) capacity grouped by period, scenario, and technology.

| Scenario & Technology | Power Capacity (MW) 2030 | 2040 | 2050 | Energy Capacity (MWh) 2030 | 2040 | 2050 |
|---|---|---|---|---|---|---|
| **VRE & ESS (mid) Coal (low) & No PRM & 90% Carbon Target** | **872,975** | **2,168,088** | **3,825,597** | **155,248** | **652,381** | **4,196,540** |
| Coal | 278,157 | 272,944 | 253,024 | 0 | 0 | 0 |
| Gas | 28,677 | 28,677 | 28,677 | 0 | 0 | 0 |
| Diesel | 1,214 | 1,214 | 1,214 | 0 | 0 | 0 |
| Nuclear | 12,680 | 12,680 | 12,680 | 0 | 0 | 0 |
| Other | 10,301 | 10,301 | 10,301 | 0 | 0 | 0 |
| Hydro | 56,271 | 56,271 | 56,271 | 0 | 0 | 0 |
| Solar | 244,234 | 1,033,460 | 2,136,166 | 0 | 0 | 0 |
| Wind | 176,514 | 523,211 | 756,251 | 0 | 0 | 0 |
| Pumped Storage | 3,140 | 3,140 | 3,140 | 18,840 | 18,840 | 18,840 |
| Battery | 61,787 | 226,191 | 564,183 | 136,408 | 633,541 | 3,191,809 |
| Hydrogen | 0 | 0 | 3,690 | 0 | 0 | 985,890 |

**Supplementary Table 17: Capacity expansion in each investment period to analyze the combined effect of no new gas and coal and the linearly-scaled demand.** Total power (MW) and energy (MWh) capacity grouped by period, scenario, and technology.

| Scenario & Technology | Power Capacity (MW) | | | Energy Capacity (MWh) | | |
|---|---|---|---|---|---|---|
| | 2030 | 2040 | 2050 | 2030 | 2040 | 2050 |
| **No New Coal and Gas & ALT** | **917,334** | **2,206,412** | **3,972,869** | **672,985** | **4,627,043** | **20,345,372** |
| Coal | 282,310 | 277,098 | 257,178 | 0 | 0 | 0 |
| Gas | 24,505 | 24,505 | 24,505 | 0 | 0 | 0 |
| Diesel | 1,214 | 1,214 | 1,214 | 0 | 0 | 0 |
| Nuclear | 12,680 | 12,680 | 12,680 | 0 | 0 | 0 |
| Other | 10,301 | 10,301 | 10,301 | 0 | 0 | 0 |
| Hydro | 56,271 | 56,271 | 56,271 | 0 | 0 | 0 |
| Solar | 249,345 | 912,430 | 2,080,414 | 0 | 0 | 0 |
| Wind | 171,404 | 586,698 | 877,324 | 0 | 0 | 0 |
| Pumped Storage | 3,140 | 3,140 | 3,140 | 18,840 | 18,840 | 18,840 |
| Battery | 94,476 | 310,386 | 592,354 | 188,952 | 1,448,809 | 3,769,995 |
| Hydrogen | 11,689 | 11,689 | 57,488 | 465,193 | 3,159,394 | 16,556,536 |
| **No New Coal and Gas & No Pipeline Coal & ALT** | **910,184** | **2,309,759** | **3,977,453** | **9,365,636** | **11,964,395** | **25,540,141** |
| Coal | 241,451 | 236,239 | 216,319 | 0 | 0 | 0 |
| Gas | 24,505 | 24,505 | 24,505 | 0 | 0 | 0 |
| Diesel | 1,214 | 1,214 | 1,214 | 0 | 0 | 0 |
| Nuclear | 12,680 | 12,680 | 12,680 | 0 | 0 | 0 |
| Other | 10,301 | 10,301 | 10,301 | 0 | 0 | 0 |
| Hydro | 56,271 | 56,271 | 56,271 | 0 | 0 | 0 |
| Solar | 249,360 | 1,028,089 | 2,126,985 | 0 | 0 | 0 |
| Wind | 171,389 | 580,854 | 831,473 | 0 | 0 | 0 |
| Pumped Storage | 3,140 | 3,140 | 3,140 | 18,840 | 18,840 | 18,840 |
| Battery | 102,960 | 319,552 | 619,830 | 306,186 | 1,720,829 | 3,997,445 |
| Hydrogen | 36,914 | 36,914 | 74,736 | 9,040,610 | 10,224,726 | 21,523,856 |
| **No New Gas & ALT** | **920,537** | **2,193,538** | **3,971,156** | **470,168** | **3,005,629** | **19,460,993** |
| Coal | 295,382 | 290,169 | 270,249 | 0 | 0 | 0 |
| Gas | 24,505 | 24,505 | 24,505 | 0 | 0 | 0 |
| Diesel | 1,214 | 1,214 | 1,214 | 0 | 0 | 0 |
| Nuclear | 12,680 | 12,680 | 12,680 | 0 | 0 | 0 |
| Other | 10,301 | 10,301 | 10,301 | 0 | 0 | 0 |
| Hydro | 56,271 | 56,271 | 56,271 | 0 | 0 | 0 |
| Solar | 249,326 | 887,896 | 2,056,071 | 0 | 0 | 0 |
| Wind | 171,423 | 597,418 | 900,728 | 0 | 0 | 0 |
| Pumped Storage | 3,140 | 3,140 | 3,140 | 18,840 | 18,840 | 18,840 |
| Battery | 90,672 | 304,320 | 581,258 | 181,344 | 1,366,885 | 3,677,076 |
| Hydrogen | 5,625 | 5,625 | 54,740 | 269,984 | 1,619,904 | 15,765,077 |
| **Linearly-scaled Demand (mid) - ALT** | **917,939** | **2,148,506** | **3,895,636** | **189,986** | **916,416** | **10,144,802** |
| Coal | 277,966 | 272,754 | 252,834 | 0 | 0 | 0 |
| Gas | 50,046 | 87,679 | 87,679 | 0 | 0 | 0 |
| Diesel | 1,214 | 1,214 | 1,214 | 0 | 0 | 0 |
| Nuclear | 12,680 | 12,680 | 12,680 | 0 | 0 | 0 |
| Other | 10,301 | 10,301 | 10,301 | 0 | 0 | 0 |
| Hydro | 56,271 | 56,271 | 56,271 | 0 | 0 | 0 |
| Solar | 249,274 | 830,256 | 1,930,705 | 0 | 0 | 0 |
| Wind | 171,474 | 616,636 | 982,709 | 0 | 0 | 0 |
| Pumped Storage | 3,140 | 3,140 | 3,140 | 18,840 | 18,840 | 18,840 |
| Battery | 85,573 | 257,576 | 534,373 | 171,146 | 897,576 | 3,291,447 |
| Hydrogen | 0 | 0 | 23,731 | 0 | 0 | 6,834,515 |

**Supplementary Table 18: VRE curtailment, unserved energy (USE), and energy storage obligation (ESO) in the production cost model simulations for the technology cost scenarios**. The unserved energy requirement by CEA is 0.05%.

| Scenario | Demand (TWh) | | | Curtailment (GWh) | | | Curtailment (%) | | | USE (GWh) | | | USE (%) | | | ESO (%) | | |
|---|---|---|---|---|---|---|---|---|---|---|---|---|---|---|---|---|---|---|
| | 2030 | 2040 | 2050 | 2030 | 2040 | 2050 | 2030 | 2040 | 2050 | 2030 | 2040 | 2050 | 2030 | 2040 | 2050 | 2030 | 2040 | 2050 |
| VRE & ESS (low) Coal (low) | 2571 | 4174 | 5603 | 845 | 132 | 231 | 0 | 4.3 | 6.3 | 0 | 179942 | 351239 | 0.033 | 0.003 | 0.004 | 3.3 | 7 | 17.1 |
| VRE & ESS (mid) Coal (low) - REF | 2571 | 4174 | 5603 | 744 | 147 | 192 | 0 | 4.5 | 4.4 | 0 | 186164 | 248938 | 0.029 | 0.004 | 0.003 | 3.3 | 7 | 26.7 |
| VRE & ESS (high) Coal (low) | 2571 | 4174 | 5603 | 761 | 193 | 129 | 0 | 3.8 | 6.6 | 0 | 160685 | 370635 | 0.03 | 0.005 | 0.002 | 3.3 | 6.8 | 26.5 |

**Supplementary Table 19: Transmission corridors net, maximum, and minimum transfers for the reference scenario.** The line name is represented by the abbreviation of the source load zone (from) and the abbreviation of the sink load zone (to). A positive sign in the net, maximum, and minimum transfers represent the original source-sink direction (from-to), a negative sign represent sink-source direction (to-from). The table shows the net (TWh), maximum (MW), and minimum (MW) transfers for each line in the different investment periods of the reference scenario (REF). The lines from DH and DD are combined for simplification.

| Line | Max. transfer (MW) | | | Min. transfer (MW) | | | Net transfers (TWh) | | |
|---|---|---|---|---|---|---|---|---|---|
| | 2030 | 2040 | 2050 | 2030 | 2040 | 2050 | 2030 | 2040 | 2050 |
| AP-KA | 11,177 | 11,177 | 11,177 | 0 | -11,177 | -11,177 | 11,799 | 22 | -40,362 |
| AP-OR | 5,012 | 5,600 | 16,665 | -5,600 | -5,600 | -16,665 | -32,517 | 4,816 | 28,822 |
| AP-TN | 8,815 | 9,075 | 9,075 | -4,213 | -9,075 | -9,075 | 6,170 | -10,299 | -17,460 |
| AP-TS | 9,039 | 20,933 | 20,933 | -20,141 | -20,746 | -20,933 | -5,557 | 27,341 | 25,002 |
| AR-AS | 1,066 | 1,050 | 1,038 | -140 | -374 | -668 | 3,680 | 2,380 | 1,400 |
| AS-BH | 0 | 883 | 883 | -894 | -894 | -894 | -6,149 | -5,294 | -3,055 |
| AS-ML | 306 | 702 | 1,036 | -41 | 0 | -675 | 1,292 | 3,705 | 3,850 |
| AS-MN | 163 | 1,250 | 1,250 | -11 | -1,187 | -1,250 | 219 | 1,031 | 705 |
| AS-MZ | 0 | 171 | 307 | -22 | -358 | -700 | -7 | -398 | -1,260 |
| AS-NL | 100 | 1,375 | 1,488 | -25 | -994 | -1,187 | 324 | 1,426 | 2,217 |
| AS-TR | 0 | 2,300 | 2,300 | -1,038 | -2,300 | -2,300 | -8,205 | -12,056 | -11,986 |
| AS-UP | 3,000 | 3,000 | 3,000 | 0 | -3,000 | -3,000 | 7,280 | 3,726 | 2,085 |
| AS-WB | 2,774 | 3,656 | 3,656 | -2,737 | -3,656 | -3,656 | 2,043 | 4,412 | 17,218 |
| BH-WB | 2,100 | 2,100 | 2,100 | 0 | 0 | 0 | 12,077 | 12,853 | 15,003 |
| BR-JH | 3,416 | 9,394 | 9,394 | -9,394 | -9,394 | -9,394 | -14,770 | -26,896 | -9,992 |
| BR-SK | 0 | 1,075 | 1,294 | -1,294 | -1,294 | -1,294 | -7,195 | -3,889 | -1,685 |
| BR-UP | 18,022 | 18,022 | 18,022 | -4,492 | -18,022 | -18,022 | 57,709 | 7,390 | -50,244 |
| BR-WB | 1,231 | 7,824 | 7,824 | -7,824 | -7,824 | -7,824 | -16,659 | 5,195 | 6,749 |
| CG-HR | 6,000 | 6,000 | 6,000 | -6,000 | -6,000 | -6,000 | 21,263 | 8,428 | -5,306 |
| CG-JH | 5,508 | 5,508 | 9,337 | -5,508 | -5,508 | -9,337 | -2,250 | 11,178 | 37,047 |
| CG-MH | 19,070 | 19,070 | 19,070 | 0 | -19,070 | -19,070 | 75,967 | -12,616 | -31,017 |
| CG-MP | 18,235 | 18,707 | 18,707 | 0 | -18,707 | -18,707 | 16,912 | 4,517 | -55,208 |
| CG-OR | 12,173 | 12,173 | 12,173 | -12,173 | -12,173 | -12,173 | 4,175 | 50,128 | 39,768 |
| CG-TN | 6,000 | 6,000 | 6,000 | 0 | -6,000 | -6,000 | 38,420 | 4,630 | -5,457 |
| CH-HP | 0 | 156 | 440 | -440 | -619 | -1,157 | -3,140 | -3,880 | -5,249 |
| CH-HR | 154 | 154 | 154 | -154 | -154 | -154 | 260 | -387 | -402 |
| CH-PB | 0 | 100 | 100 | -100 | -100 | -100 | -34 | -24 | -50 |
| DHDD-GJ | 684 | 1,406 | 1,323 | -1,749 | -1,749 | -1,749 | -9,736 | -17,233 | -15,308 |
| DHDD-MH | 914 | 1,120 | 1,120 | -1,120 | -1,120 | -2,111 | -10,470 | -8,831 | -19,721 |
| DL-HR | 5,072 | 7,143 | 8,953 | -8,953 | -8,953 | -8,953 | -22,116 | -42,118 | -21,109 |
| DL-RJ | 208 | 6,141 | 208 | -208 | -6,141 | -22,543 | -644 | -46,220 | -139,584 |
| DL-UP | 0 | 9,160 | 9,160 | -9,160 | -9,160 | -9,160 | -31,131 | -1,947 | 29,679 |
| GA-KA | 200 | 239 | 865 | -200 | -239 | -1,041 | -85 | -1,588 | -6,269 |
| GA-MH | 0 | 345 | 403 | -1,341 | -1,382 | -1,382 | -7,216 | -9,874 | -8,859 |
| GJ-HR | 2,500 | 2,500 | 2,500 | -2,500 | -2,500 | -2,500 | -573 | 6,961 | 4,311 |
| GJ-MH | 4,355 | 4,355 | 4,355 | -4,355 | -4,355 | -4,355 | -3,052 | 11,553 | 8,569 |
| GJ-MP | 0 | 7,450 | 7,450 | -7,450 | -7,450 | -7,450 | -29,765 | 19,763 | 12,209 |
| GJ-RJ | 6,034 | 6,034 | 6,034 | -6,034 | -6,034 | -6,034 | -211 | 5,636 | -1,459 |
| HP-JKLA | 517 | 517 | 517 | 0 | -517 | -517 | 1,242 | 1,436 | 2,246 |
| HP-PB | 5,452 | 7,763 | 7,504 | -1,252 | -6,455 | -7,713 | 19,279 | 12,068 | 14,569 |
| HP-UK | 486 | 486 | 1,218 | -486 | -486 | -1,218 | 81 | 1,768 | 2,030 |
| HR-HP | 1,067 | 3,477 | 3,477 | -3,477 | -3,477 | -3,477 | -10,900 | -1,704 | 12,849 |
| HR-PB | 9,429 | 11,014 | 11,014 | -6,579 | -11,014 | -11,014 | -5,337 | 31,868 | 44,856 |
| HR-RJ | 11,334 | 24,472 | 43,168 | -11,334 | -24,472 | -56,254 | -7,030 | -177,183 | -291,936 |
| HR-UK | 1,362 | 1,362 | 8,149 | -1,362 | -1,362 | -8,149 | -2,751 | 5,227 | 39,602 |
| HR-UP | 0 | 10,205 | 10,205 | -10,205 | -10,205 | -10,205 | -27,264 | 31,696 | 45,298 |
| JH-OR | 3,812 | 3,812 | 6,234 | -3,812 | -3,812 | -6,234 | 8,315 | 2,699 | 7,110 |
| JH-WB | 5,620 | 15,913 | 15,913 | -6,516 | -15,913 | -15,913 | -824 | 21,084 | 62,089 |
| JKLA-PB | 3,949 | 2,956 | 2,714 | -2,333 | -3,330 | -4,942 | 8,620 | -2,370 | -10,005 |
| KA-KL | 1,165 | 1,165 | 3,323 | -1,165 | -1,165 | -3,323 | 1,895 | 5,828 | 31,350 |
| KA-MH | 4,032 | 6,012 | 6,012 | -6,012 | -6,012 | -6,012 | -13,645 | 9,661 | 16,937 |
| KA-OR | 0 | 2,000 | 2,000 | -2,000 | -2,000 | -2,000 | -7,308 | 4,367 | 5,203 |
| KA-TN | 4,563 | 4,563 | 4,563 | -4,563 | -4,563 | -4,563 | -9,019 | -2,933 | 1,413 |
| KA-TS | 0 | 917 | 917 | -917 | -917 | -917 | -4,599 | 890 | 3,405 |
| KL-TN | 0 | 6,394 | 6,394 | -6,394 | -6,394 | -6,394 | -25,668 | -34,628 | -30,449 |
| MH-TS | 9,800 | 9,800 | 9,800 | -9,800 | -9,800 | -9,800 | -7,751 | -11,416 | -12,858 |
| MN-MZ | 0 | 50 | 50 | -50 | -50 | -50 | -380 | -280 | -180 |
| MN-NL | 44 | 1,144 | 1,059 | -31 | -1,200 | -1,200 | 20 | -399 | -1,427 |
| MP-MH | 17,154 | 17,154 | 17,154 | -35 | -17,154 | -17,154 | 38,534 | 17,526 | 30,844 |
| MP-RJ | 5,750 | 5,750 | 5,750 | -5,750 | -5,750 | -5,750 | 18,351 | -11,941 | -12,049 |
| MP-UP | 12,250 | 12,250 | 12,250 | -12,250 | -12,250 | -12,250 | 3,187 | 24,925 | 6,051 |
| MZ-TR | 0 | 100 | 100 | -100 | -100 | -100 | -553 | -538 | -525 |
| OR-WB | 1,100 | 1,100 | 3,321 | -1,100 | -1,100 | -3,321 | -2,542 | 736 | 14,693 |
| PB-RJ | 5,000 | 5,000 | 5,000 | -5,000 | -5,000 | -5,000 | -4,977 | -25,639 | -19,190 |
| PB-UP | 0 | 6,594 | 6,594 | -6,594 | -6,594 | -6,594 | -7,465 | 1,925 | 13,891 |
| PY-TN | 0 | 0 | 222 | -732 | -1,258 | -1,434 | -4,181 | -7,333 | -9,784 |
| RJ-UP | 6,327 | 6,327 | 6,327 | -6,327 | -6,327 | -6,327 | -16,982 | 29,490 | 49,876 |
| SK-WB | 1,250 | 1,250 | 1,250 | 0 | -1,096 | -1,250 | 945 | 3,155 | 4,413 |
| UP-UK | 3,848 | 4,710 | 12,405 | -3,615 | -3,258 | -12,699 | 4,067 | 4,741 | -22,924 |
| UP-WB | 0 | 3,000 | 3,000 | -3,000 | -3,000 | -3,000 | -10,619 | 838 | 3,489 |

**Supplementary Table 20: Technology-specific inertia constant ranges.** Min and max inertia ($H(s)$) used to characterize synchronous generator contributions to system inertia.

| Technology | Min. (s) | Max. (s) | Notes |
|---|---|---|---|
| **Subcritical Coal Large** | 6 | 9 | Big steam turbines have large generators high inertia. |
| **Subcritical Coal Small** | 4 | 7 | Smaller units slightly lower inertia than large coal |
| **Supercritical Coal** | 5 | 8 | Similar to large coal but a bit lower due to design optimizations |
| **CCGT** | 3 | 6 | Medium inertia |
| **CT** | 2 | 4 | Aeroderivative frame CTs generally lower inertia |
| **Diesel** | 1.5 | 3 | Medium speed diesel gensets relatively low inertia per MVA. |
| **Hydro ROR** | 3 | 6 | Depends on head and runner often comparable to or slightly higher than CCGT |
| **Hydro Storage** | 3 | 6 | Conventional reservoir hydro similar to ROR |
| **Biomass** | 3 | 6 | Steam cycle with synchronous generator similar to small coal |
| **WHR** | 2 | 4 | Smaller turbogenerators driven by waste heat lower MVA so modest contribution per unit |
| **Nuclear** | 5 | 9 | Large steam turbines with big generators among the highest inertia per unit. |
| **PSH** | 3 | 7 | Pumped storage units operating as generators or condensers |

**Supplementary Table 21: Hierarchical organization of the national grid in India.** The "National Load Despatch Center" regulates the multiple interconnected power grids, operated by the "Regional Load Despatch Centers" in charge of managing the interactions between the "State Load Despatch Centers".

| Regional Load Despatch Centers | State | Abbreviation |
|---|---|---|
| Northern | Chandigarh | CH |
| | Delhi | DL |
| | Haryana | HR |
| | Himachal Pradesh | HP |
| | Jammu, Kashmir & Ladakh | JKLH |
| | Punjab | PJ |
| | Rajasthan | RJ |
| | Uttar Pradesh | UP |
| | Uttarakhand | UK |
| Northeastern | Arunachal Pradesh | AR |
| | Assam | AS |
| | Manipur | MN |
| | Meghalaya | ML |
| | Mizoram | MZ |
| | Nagaland | NL |
| | Tripura | TR |
| Western | Dadra & Nagar Haveli | DH |
| | Daman & Diu | DD |
| | Chhattisgarh | CG |
| | Goa | GA |
| | Gujarat | GJ |
| | Madhya Pradesh | MP |
| | Maharashtra | MH |
| Southern | Andhra Pradesh | AP |
| | Karnataka | KA |
| | Kerala | KL |
| | Puducherry | PY |
| | Tamil Nadu | TN |
| | Telangana | TG |
| Eastern | Bihar | BR |
| | Jharkhand | JH |
| | Odisha | OR |
| | Sikkim | SK |
| | West Bengal | WB |
| Interconnection | Bhutan | BH |

**Supplementary Table 22: Candidate projects, available capacity, and Renewable Purchase Obligations (RPO).** States with a candidate project by technology have a mark, the type, or total available capacity. States with an RPO target have a mark. We assume all new coal projects use supercritial technology. The available capacity for PSH projects is in GW/GWh. The available capacity for solar PV, onshore and offshore wind is in GW. The states are ordered by alphabetically by dispatch center.

| **State** | **Battery** | **CCGT** | **CT** | **Coal** | **Hydrogen** | **Nuclear** | **PSH** | **Solar** | **Wind** | **Offshore** | **RPO** |
|---|---|---|---|---|---|---|---|---|---|---|---|
| Delhi | ✓ | ✓ | ✓ | | Cavern | | | | | | |
| Haryana | ✓ | ✓ | ✓ | ✓ | Tank | | | 69.9 | 2.5 | | ✓ |
| Himachal Pradesh | | | | | | | | 2 | | | ✓ |
| Jammu, Kashmir & Ladakh | ✓ | ✓ | ✓ | | Cavern | | | 0.4 | | | ✓ |
| Punjab | ✓ | | | ✓ | Cavern | | | 86.9 | 356.1 | | ✓ |
| Rajasthan | ✓ | ✓ | ✓ | ✓ | Cavern | ✓ | 15.7/94.1 | 819.6 | 8,902 | | ✓ |
| Uttar Pradesh | ✓ | ✓ | ✓ | ✓ | Cavern | ✓ | 16.5/99 | 268.7 | 1.5 | | ✓ |
| Uttarakhand | | | | | | | 1/6 | 9.5 | | | ✓ |
| Arunachal Pradesh | | | | | | | 2.2/13.3 | 10.3 | | | |
| Assam | ✓ | | | ✓ | Cavern | | | 88.9 | | | ✓ |
| Manipur | | | | | | | | 0.7 | | | |
| Meghalaya | | | | | | | | 6.9 | | | |
| Mizoram | | | | | | | | 1.6 | | | |
| Nagaland | | | | | | | | 1.3 | | | |
| Tripura | ✓ | ✓ | ✓ | | Cavern | | | 16.2 | | | |
| Dadra & Nagar Haveli | | | | | | | | 0.1 | 0.7 | | |
| Daman & Diu | | | | | | | | 1.2 | 1.2 | 0.7 | |
| Chhattisgarh | ✓ | | | ✓ | Tank | | 9/54 | 241.7 | 15.9 | | ✓ |
| Goa | | | | | | | | 1.5 | | | |
| Gujarat | ✓ | ✓ | ✓ | ✓ | Cavern | ✓ | 6.2/37.2 | 351.2 | 330.5 | 8.4 | ✓ |
| Madhya Pradesh | ✓ | | | ✓ | Cavern | | 6.9/41.3 | 602.9 | 206.6 | | ✓ |
| Maharashtra | ✓ | ✓ | ✓ | ✓ | Tank | ✓ | 25.6/153.4 | 546.6 | 296.8 | | ✓ |
| Andhra Pradesh | ✓ | ✓ | ✓ | ✓ | Tank | ✓ | 26.4/158.3 | 231.3 | 191.8 | 11.4 | ✓ |
| Karnataka | ✓ | | | ✓ | Tank | ✓ | 4.3/25.6 | 359.8 | 359.5 | | ✓ |
| Kerala | ✓ | ✓ | ✓ | | Tank | | | 7.8 | 4.7 | | ✓ |
| Puducherry | | | | | | | | 0.1 | 76 | 0.1 | |
| Tamil Nadu | ✓ | ✓ | ✓ | ✓ | Tank | ✓ | 8.3/49.8 | 155.2 | 109.9 | 11.1 | ✓ |
| Telangana | ✓ | | | ✓ | Cavern | | 8.4/50.1 | 223.2 | 113.8 | | ✓ |
| Bihar | ✓ | | | ✓ | Cavern | | | 45.1 | 0.1 | | ✓ |
| Jharkhand | ✓ | | | ✓ | Cavern | | 1.5/9 | 111 | 0.4 | | ✓ |
| Odisha | ✓ | ✓ | ✓ | ✓ | Cavern | | 6.4/3.8 | 183.4 | 26.6 | 5.8 | ✓ |
| West Bengal | ✓ | ✓ | ✓ | ✓ | Cavern | | 4.7/2.8 | 70.6 | 3.1 | | ✓ |

**Supplementary Table 23: Technology heat rates and $CO_2$ emissions**. We consider different heat rates per technology, and different $CO_2$ emission rates per fuel. The electricity generation technologies for gas are Combustion Turbines (CT) and Combined Cycle Gas Turbines. Technology for diesel is assumed to be Internal Combustion (IC) engines.

| **Technology** | **Fuel** | **Heat Rate (MMBtu/MWh)** | **Emissions Rate ($tCO_2$/MMBtu)** |
|---|---|---|---|
| Supercritical (new) | Coal | 10.04 | 0.096 |
| CCGT (new) | Gas | 8.05 | 0.052 |
| CT (new) | Gas | 12.62 | 0.052 |
| Subcritical (small) | Coal | 10.89 | 0.096 |
| Subcritical (large) | Coal | 10.04 | 0.096 |
| CCGT | Gas | 8.05 | 0.052 |
| IC | Diesel | 11.48 | 0.073 |
| WHR | WHR | 12.3 | 0 |
| Biomass | Biomass | 12.3 | 0 |

**Supplementary Table 24: Pipeline project capacity.** States with hydro run-of-river (ROR), supercritical coal, or nuclear capacity on the pipeline. The capacity is in GW, and the states are grouped by dispatch centers.

| State | Hydro | Coal | Nuclear |
|---|---|---|---|
| Himachal Pradesh | 2.22 | | |
| Jammu, Kashmir & Ladakh | 2.56 | | |
| Punjab | 0.21 | | |
| Rajasthan | | | 1.4 |
| Uttar Pradesh | | 8.58 | |
| Uttarakhand | 1.39 | | |
| Arunachal Pradesh | 1.15 | | |
| Assam | 2 | | |
| Chhattisgarh | | 0.8 | |
| Gujarat | | | 1.4 |
| Madhya Pradesh | 0.4 | 0.8 | |
| Maharashtra | | 0.66 | |
| Andhra Pradesh | 0.96 | 1.6 | |
| Karnataka | | 0.37 | |
| Kerala | 0.1 | | |
| Tamil Nadu | | 6.07 | 2 |
| Telangana | | 6.68 | |
| Bihar | | 4.87 | |
| Jharkhand | | 4.38 | |
| Odisha | | 1.05 | |
| West Bengal | 0.12 | | |
| Bhutan | 2.22 | | |

**Supplementary Table 25: Historical distribution and transmission losses.** States are grouped by regional dispatch centers. NE includes Arunachal Pradesh, Manipur, Meghalaya, Mizoram and Nagaland. Gujarat includes Daman & Diu and Dadra & Nagar Haveli. The minium losses are limited to the 7%.

| State | 2006 | 2007 | 2008 | 2009 | 2010 | 2011 | 2012 | 2013 | 2014 | 2015 | 2016 | 2017 | 2018 | 2019 | 2020 | 2021 | 2022 | 2023 |
|---|---|---|---|---|---|---|---|---|---|---|---|---|---|---|---|---|---|---|
| Delhi | 48.6 | 33.0 | 28.7 | 22.2 | 22.1 | 20.0 | 19.3 | 22.1 | 19.9 | 21.5 | 19.6 | 19.4 | 17.4 | 12.6 | 12.4 | 13.5 | 12.6 | 11.6 |
| Haryana | 30.5 | 33.3 | 32.8 | 30.7 | 31.0 | 29.7 | 28.6 | 36.0 | 35.8 | 34.0 | 27.4 | 33.7 | 28.2 | 22.8 | 22.6 | 23.8 | 19.5 | 18.8 |
| Himachal Pradesh | 23.6 | 19.8 | 17.0 | 15.5 | 20.5 | 22.2 | 18.6 | 19.1 | 21.0 | 20.8 | 18.8 | 22.4 | 21.8 | 14.3 | 14.3 | 14.4 | 13.7 | 14.2 |
| Jammu, Kashmir & Ladakh | 44.9 | 52.0 | 55.7 | 58.0 | 67.4 | 63.3 | 61.8 | 56.6 | 54.7 | 53.1 | 50.1 | 47.0 | 45.5 | 42.2 | 37.2 | 43.9 | 44.2 | 50.9 |
| Punjab | 31.6 | 25.1 | 22.9 | 23.1 | 23.4 | 24.9 | 23.1 | 20.3 | 20.6 | 18.5 | 18.6 | 17.6 | 17.4 | 14.8 | 16.0 | 14.8 | 14.5 | 14.2 |
| Rajasthan | 39.9 | 35.6 | 34.7 | 31.5 | 30.0 | 27.9 | 27.9 | 24.9 | 26.9 | 27.5 | 29.1 | 29.7 | 28.7 | 24.8 | 24.3 | 27.6 | 28.2 | 20.3 |
| Uttar Pradesh | 32.6 | 33.5 | 28.6 | 30.9 | 33.1 | 34.0 | 32.4 | 26.9 | 29.1 | 27.2 | 24.5 | 23.6 | 23.2 | 21.2 | 22.4 | 25.1 | 21.8 | 20.9 |
| Uttarakhand | 36.0 | 34.5 | 35.7 | 41.8 | 25.3 | 30.0 | 28.7 | 26.9 | 21.8 | 24.5 | 25.6 | 25.4 | 25.0 | 22.2 | 20.1 | 22.0 | 21.3 | 18.4 |
| Assam | 40.3 | 33.7 | 38.6 | 37.6 | 32.8 | 34.2 | 33.5 | 30.7 | 31.1 | 27.6 | 25.2 | 25.0 | 24.6 | 24.0 | 27.9 | 28.8 | 25.5 | 17.6 |
| NE | 49.7 | 57.8 | 48.5 | 45.4 | 45.1 | 40.4 | 39.0 | 33.4 | 33.9 | 36.9 | 36.0 | 34.2 | 32.9 | 35.4 | 33.0 | 31.6 | 30.0 | 25.9 |
| Chhattisgarh | 31.1 | 31.7 | 29.8 | 26.4 | 18.6 | 15.1 | 16.5 | 28.8 | 28.4 | 29.2 | 30.8 | 25.6 | 24.0 | 25.1 | 25.8 | 25.6 | 23.7 | 22.2 |
| Goa | 19.7 | 20.9 | 21.2 | 17.1 | 17.0 | 15.3 | 12.4 | 13.3 | 12.7 | 15.0 | 18.0 | 18.0 | 17.2 | 16.2 | 10.9 | 15.9 | 8.1 | 7.1 |
| Gujarat | 27.9 | 24.9 | 25.6 | 23.5 | 22.0 | 18.6 | 21.1 | 17.9 | 17.4 | 19.0 | 18.9 | 18.1 | 17.6 | 19.6 | 20.6 | 20.0 | 17.9 | 14.3 |
| Madhya Pradesh | 40.1 | 39.2 | 35.6 | 38.5 | 38.3 | 37.6 | 34.5 | 31.4 | 31.5 | 32.3 | 28.6 | 26.8 | 26.8 | 27.6 | 27.3 | 27.6 | 24.7 | 22.4 |
| Maharashtra | 31.6 | 31.6 | 29.8 | 23.9 | 25.2 | 20.7 | 20.0 | 21.8 | 21.8 | 20.4 | 19.9 | 18.2 | 17.9 | 19.5 | 21.2 | 19.6 | 17.4 | 14.7 |
| Andhra Pradesh | 20.1 | 18.7 | 22.4 | 19.6 | 18.4 | 16.6 | 17.5 | 19.3 | 20.1 | 17.9 | 16.2 | 16.2 | 16.1 | 17.6 | 17.0 | 16.0 | 15.4 | 14.2 |
| Karnataka | 29.8 | 25.9 | 18.9 | 17.0 | 18.8 | 17.3 | 12.7 | 11.1 | 10.2 | 11.5 | 10.5 | 13.4 | 13.2 | 16.0 | 17.2 | 16.7 | 16.4 | 16.3 |
| Kerala | 23.5 | 19.1 | 17.8 | 13.2 | 19.6 | 18.3 | 17.2 | 17.7 | 15.0 | 15.4 | 11.6 | 18.1 | 15.8 | 16.4 | 14.8 | 13.5 | 13.6 | 13.5 |
| Tamil Nadu | 18.7 | 19.5 | 18.3 | 18.0 | 18.2 | 13.4 | 16.3 | 14.5 | 11.0 | 11.2 | 10.9 | 12.7 | 16.0 | 19.0 | 17.3 | 13.3 | 13.4 | 15.2 |
| Telangana | - | - | - | - | - | - | - | - | - | 15.7 | 16.5 | 18.7 | 18.3 | 14.9 | 15.3 | 15.0 | 14.4 | 12.2 |
| Bihar | 44.0 | 50.7 | 48.8 | 46.4 | 43.6 | 50.8 | 50.9 | 49.4 | 47.3 | 46.3 | 42.8 | 30.3 | 30.3 | 32.5 | 31.1 | 31.3 | 27.3 | 25.8 |
| Jharkhand | 26.8 | 26.2 | 23.2 | 24.3 | 22.2 | 17.1 | 14.3 | 13.3 | 13.1 | 17.2 | 17.0 | 16.3 | 16.3 | 18.4 | 16.7 | 21.0 | 22.7 | 23.0 |
| Odisha | 45.6 | 43.3 | 39.4 | 42.7 | 37.0 | 42.5 | 44.6 | 39.8 | 38.9 | 42.0 | 39.1 | 38.2 | 38.1 | 34.9 | 22.6 | 24.4 | 22.9 | 21.3 |
| Sikkim | 10.7 | 26.9 | 36.8 | 38.8 | 39.0 | 33.7 | 31.1 | 28.1 | 23.1 | 25.0 | 26.5 | 23.8 | 35.5 | 32.7 | 34.3 | 33.9 | 31.9 | 27.7 |
| West Bengal | 24.8 | 23.6 | 21.3 | 16.8 | 18.3 | 22.4 | 23.2 | 24.1 | 24.0 | 24.7 | 22.2 | 22.4 | 22.2 | 18.3 | 16.3 | 17.5 | 16.9 | 18.2 |

**Supplementary Table 26: Linear projections of distribution and transmission losses.** States are grouped by regional dispatch centers. NE includes Arunachal Pradesh, Manipur, Meghalaya, Mizoram and Nagaland. Gujarat includes Daman & Diu and Dadra & Nagar Haveli. The minium losses are limited to the 7%.

| State | 2024 | 2025 | 2026 | 2027 | 2028 | 2029 | 2030 | 2031 | 2032 | 2033 | 2034 | 2035 | 2036 | 2037 | 2038 | 2039 | 2040 | 2041 | 2042 | 2043 | 2044 | 2045 | 2046 | 2047 | 2048 | 2049 | 2050 | 2051 |
|---|---|---|---|---|---|---|---|---|---|---|---|---|---|---|---|---|---|---|---|---|---|---|---|---|---|---|---|---|
| Delhi | 11.3 | 10.9 | 10.6 | 10.3 | 10.0 | 9.7 | 9.4 | 9.1 | 8.8 | 8.6 | 8.3 | 8.1 | 7.8 | 7.6 | 7.4 | 7.1 | 7.0 | 7.0 | 7.0 | 7.0 | 7.0 | 7.0 | 7.0 | 7.0 | 7.0 | 7.0 | 7.0 | 7.0 |
| Haryana | 18.3 | 17.9 | 17.4 | 17.0 | 16.6 | 16.1 | 15.7 | 15.3 | 15.0 | 14.6 | 14.2 | 13.9 | 13.5 | 13.2 | 12.8 | 12.5 | 12.2 | 11.9 | 11.6 | 11.3 | 11.0 | 10.7 | 10.5 | 10.2 | 10.0 | 9.7 | 9.5 | 9.2 |
| Himachal Pradesh | 13.9 | 13.7 | 13.4 | 13.2 | 12.9 | 12.7 | 12.4 | 12.2 | 12.0 | 11.7 | 11.5 | 11.3 | 11.1 | 10.9 | 10.7 | 10.5 | 10.3 | 10.1 | 9.9 | 9.7 | 9.5 | 9.3 | 9.2 | 9.0 | 8.8 | 8.6 | 8.5 | 8.3 |
| Jammu, Kashmir & Ladakh | 50.1 | 49.2 | 48.4 | 47.6 | 46.8 | 46.0 | 45.2 | 44.5 | 43.7 | 43.0 | 42.2 | 41.5 | 40.8 | 40.2 | 39.5 | 38.8 | 38.2 | 37.5 | 36.9 | 36.3 | 35.7 | 35.1 | 34.5 | 33.9 | 33.3 | 32.7 | 32.2 | 31.7 |
| Punjab | 13.8 | 13.4 | 13.0 | 12.6 | 12.2 | 11.9 | 11.5 | 11.2 | 10.8 | 10.5 | 10.2 | 9.9 | 9.6 | 9.3 | 9.0 | 8.7 | 8.5 | 8.2 | 8.0 | 7.7 | 7.5 | 7.3 | 7.1 | 7.0 | 7.0 | 7.0 | 7.0 | 7.0 |
| Rajasthan | 19.8 | 19.4 | 18.9 | 18.5 | 18.1 | 17.7 | 17.3 | 16.9 | 16.5 | 16.1 | 15.8 | 15.4 | 15.0 | 14.7 | 14.4 | 14.0 | 13.7 | 13.4 | 13.1 | 12.8 | 12.5 | 12.2 | 12.0 | 11.7 | 11.4 | 11.2 | 10.9 | 10.7 |
| Uttar Pradesh | 20.3 | 19.7 | 19.1 | 18.6 | 18.1 | 17.6 | 17.1 | 16.6 | 16.1 | 15.6 | 15.2 | 14.8 | 14.3 | 13.9 | 13.5 | 13.1 | 12.8 | 12.4 | 12.1 | 11.7 | 11.4 | 11.1 | 10.7 | 10.4 | 10.1 | 9.8 | 9.6 | 9.3 |
| Uttarakhand | 17.9 | 17.3 | 16.8 | 16.3 | 15.8 | 15.3 | 14.9 | 14.4 | 14.0 | 13.6 | 13.2 | 12.8 | 12.4 | 12.0 | 11.7 | 11.3 | 11.0 | 10.6 | 10.3 | 10.0 | 9.7 | 9.4 | 9.1 | 8.9 | 8.6 | 8.3 | 8.1 | 7.9 |
| Assam | 17.1 | 16.6 | 16.1 | 15.6 | 15.1 | 14.7 | 14.2 | 13.8 | 13.4 | 13.0 | 12.6 | 12.2 | 11.9 | 11.5 | 11.2 | 10.8 | 10.5 | 10.2 | 9.9 | 9.6 | 9.3 | 9.0 | 8.7 | 8.5 | 8.2 | 8.0 | 7.7 | 7.5 |
| NE | 25.1 | 24.4 | 23.6 | 22.9 | 22.2 | 21.6 | 20.9 | 20.3 | 19.7 | 19.1 | 18.5 | 18.0 | 17.4 | 16.9 | 16.4 | 15.9 | 15.4 | 15.0 | 14.5 | 14.1 | 13.7 | 13.3 | 12.9 | 12.5 | 12.1 | 11.7 | 11.4 | 11.0 |
| Chhattisgarh | 22.1 | 21.9 | 21.8 | 21.7 | 21.5 | 21.4 | 21.3 | 21.1 | 21.0 | 20.8 | 20.7 | 20.6 | 20.4 | 20.3 | 20.2 | 20.1 | 19.9 | 19.8 | 19.7 | 19.6 | 19.4 | 19.3 | 19.2 | 19.1 | 18.9 | 18.8 | 18.7 | 18.6 |
| Goa | 7.0 | 7.0 | 7.0 | 7.0 | 7.0 | 7.0 | 7.0 | 7.0 | 7.0 | 7.0 | 7.0 | 7.0 | 7.0 | 7.0 | 7.0 | 7.0 | 7.0 | 7.0 | 7.0 | 7.0 | 7.0 | 7.0 | 7.0 | 7.0 | 7.0 | 7.0 | 7.0 | 7.0 |
| Gujarat | 14.0 | 13.6 | 13.3 | 13.0 | 12.6 | 12.3 | 12.0 | 11.7 | 11.4 | 11.2 | 10.9 | 10.6 | 10.4 | 10.1 | 9.9 | 9.6 | 9.4 | 9.2 | 8.9 | 8.7 | 8.5 | 8.3 | 8.1 | 7.9 | 7.7 | 7.5 | 7.3 | 7.2 |
| Madhya Pradesh | 21.7 | 21.1 | 20.4 | 19.8 | 19.2 | 18.6 | 18.1 | 17.5 | 17.0 | 16.5 | 16.0 | 15.5 | 15.1 | 14.6 | 14.2 | 13.8 | 13.3 | 12.9 | 12.5 | 12.2 | 11.8 | 11.5 | 11.1 | 10.8 | 10.5 | 10.1 | 9.8 | 9.5 |
| Maharashtra | 14.3 | 13.9 | 13.5 | 13.1 | 12.7 | 12.3 | 11.9 | 11.6 | 11.2 | 10.9 | 10.5 | 10.2 | 9.9 | 9.6 | 9.3 | 9.1 | 8.8 | 8.5 | 8.3 | 8.0 | 7.8 | 7.5 | 7.3 | 7.1 | 7.0 | 7.0 | 7.0 | 7.0 |
| Andhra Pradesh | 14.0 | 13.7 | 13.5 | 13.3 | 13.1 | 12.9 | 12.7 | 12.5 | 12.3 | 12.1 | 11.9 | 11.7 | 11.5 | 11.3 | 11.1 | 11.0 | 10.8 | 10.6 | 10.4 | 10.3 | 10.1 | 10.0 | 9.8 | 9.6 | 9.5 | 9.3 | 9.2 | 9.0 |
| Karnataka | 15.9 | 15.4 | 15.0 | 14.5 | 14.1 | 13.7 | 13.3 | 12.9 | 12.6 | 12.2 | 11.9 | 11.5 | 11.2 | 10.9 | 10.6 | 10.3 | 10.0 | 9.7 | 9.4 | 9.1 | 8.9 | 8.6 | 8.4 | 8.1 | 7.9 | 7.7 | 7.5 | 7.3 |
| Kerala | 13.2 | 12.9 | 12.6 | 12.4 | 12.1 | 11.8 | 11.5 | 11.3 | 11.0 | 10.8 | 10.6 | 10.3 | 10.1 | 9.9 | 9.7 | 9.4 | 9.2 | 9.0 | 8.8 | 8.6 | 8.4 | 8.3 | 8.1 | 7.9 | 7.7 | 7.5 | 7.4 | 7.2 |
| Tamil Nadu | 15.0 | 14.8 | 14.5 | 14.3 | 14.1 | 13.9 | 13.7 | 13.5 | 13.3 | 13.1 | 12.9 | 12.7 | 12.5 | 12.3 | 12.1 | 11.9 | 11.8 | 11.6 | 11.4 | 11.2 | 11.1 | 10.9 | 10.7 | 10.6 | 10.4 | 10.3 | 10.1 | 9.9 |
| Telangana | 11.9 | 11.5 | 11.2 | 10.9 | 10.5 | 10.2 | 9.9 | 9.7 | 9.4 | 9.1 | 8.8 | 8.6 | 8.3 | 8.1 | 7.9 | 7.6 | 7.4 | 7.2 | 7.0 | 7.0 | 7.0 | 7.0 | 7.0 | 7.0 | 7.0 | 7.0 | 7.0 | 7.0 |
| Bihar | 25.0 | 24.3 | 23.6 | 22.9 | 22.2 | 21.5 | 20.9 | 20.2 | 19.6 | 19.0 | 18.5 | 17.9 | 17.4 | 16.9 | 16.3 | 15.9 | 15.4 | 14.9 | 14.5 | 14.0 | 13.6 | 13.2 | 12.8 | 12.4 | 12.1 | 11.7 | 11.3 | 11.0 |
| Jharkhand | 22.7 | 22.4 | 22.1 | 21.8 | 21.5 | 21.2 | 21.0 | 20.7 | 20.4 | 20.1 | 19.9 | 19.6 | 19.4 | 19.1 | 18.9 | 18.6 | 18.4 | 18.1 | 17.9 | 17.7 | 17.5 | 17.2 | 17.0 | 16.8 | 16.6 | 16.3 | 16.1 | 15.9 |
| Odisha | 20.7 | 20.1 | 19.5 | 18.9 | 18.3 | 17.8 | 17.2 | 16.7 | 16.2 | 15.7 | 15.2 | 14.8 | 14.3 | 13.9 | 13.5 | 13.1 | 12.7 | 12.3 | 11.9 | 11.6 | 11.2 | 10.9 | 10.6 | 10.3 | 10.0 | 9.7 | 9.4 | 9.1 |
| Sikkim | 27.6 | 27.5 | 27.3 | 27.2 | 27.0 | 26.9 | 26.8 | 26.6 | 26.5 | 26.4 | 26.2 | 26.1 | 26.0 | 25.9 | 25.7 | 25.6 | 25.5 | 25.3 | 25.2 | 25.1 | 25.0 | 24.8 | 24.7 | 24.6 | 24.5 | 24.3 | 24.2 | 24.1 |
| West Bengal | 18.0 | 17.7 | 17.5 | 17.3 | 17.0 | 16.8 | 16.6 | 16.3 | 16.1 | 15.9 | 15.7 | 15.5 | 15.2 | 15.0 | 14.8 | 14.6 | 14.4 | 14.2 | 14.0 | 13.8 | 13.7 | 13.5 | 13.3 | 13.1 | 12.9 | 12.7 | 12.6 | 12.4 |

**Supplementary Table 27: Specified capacity (MW) per load zone for each technology**. Initial capacity in FY20 included in *GridPath-India*. We considered multiple coal technologies (supercritical, or subcritical small or large); all coal technologies are grouped for simplification.

| Load Zone | Biomass | CCGT | Coal | Diesel | PSH | Hydro (ROR) | Hydro (storage) | Nuclear | Solar PV | WHR | Wind | Total |
|---|---|---|---|---|---|---|---|---|---|---|---|---|
| **AP** | 484 | 4899 | 13519 | | | 50 | 1560 | | 3610 | | 4090 | **28211** |
| **AR** | | | | | | 515 | | | 5 | | | **520** |
| **AS** | 2 | 644 | 883 | | | 100 | 250 | | 40 | 38 | | **1957** |
| **BH** | | | | | | 2326 | | | | | | **2326** |
| **BR** | 125 | | 6406 | | | | | | 150 | | | **6680** |
| **CH** | | | 14 | | | | | | | | | **14** |
| **CG** | 275 | | 30679 | 100 | | | 120 | | 230 | | | **31404** |
| **DH** | | | 8 | | | | | | 5 | | | **13** |
| **DD** | | | | | | | | | 19 | | | **19** |
| **DL** | | 2208 | | | | | | | | | | **2208** |
| **GA** | | 48 | 84 | | | | | | 3 | | | **135** |
| **GJ** | 77 | 7551 | 20277 | | 1440 | 250 | 300 | 440 | 2945 | | 7540 | **40820** |
| **HR** | 241 | 432 | 5704 | | | | | | 250 | | | **6627** |
| **HP** | 9 | | 33 | | | 6694 | 3115 | | 32 | | | **9883** |
| **JKLA** | | 175 | | | | 3449 | | | 18 | | | **3642** |
| **JH** | 4 | | 5843 | | | 130 | 40 | | 35 | | | **6052** |
| **KA** | 1887 | | 13451 | 880 | | 742 | 2902 | 880 | 7275 | | 4790 | **32807** |
| **KL** | | 534 | 62 | 234 | | 149 | 1708 | | 140 | | 60 | **2886** |
| **MP** | 107 | | 23838 | | | 345 | 1890 | | 2255 | | 2515 | **30950** |
| **MH** | 2584 | 3207 | 28623 | | 400 | 494 | 2153 | 1400 | 1800 | 60 | 5000 | **45722** |
| **MN** | | | | | | | 105 | | 3 | | | **108** |
| **ML** | 14 | | 39 | | | 286 | 36 | | | | | **375** |
| **MZ** | | | | | | | 60 | | | | | **60** |
| **NL** | | | | | | | 75 | | | | | **75** |
| **OR** | 59 | | 20123 | | | | 2143 | | | | | **22325** |
| **PY** | | 33 | 25 | | | | | | 5 | | | **63** |
| **PB** | 473 | | 6364 | | | 496 | 600 | | 945 | | | **8879** |
| **RJ** | 121 | 1023 | 11635 | | | | 411 | 1280 | 5135 | | 4295 | **23900** |
| **SK** | | | | | | 2169 | | | | | | **2169** |
| **TN** | 1013 | 1027 | 15504 | | 400 | 240 | 1538 | 3440 | 3915 | | 9300 | **36377** |
| **TS** | 160 | | 7936 | | | 654 | 146 | | 3620 | | 125 | **12641** |
| **TR** | | 1132 | | | | | | | 8 | | | **1140** |
| **UP** | 2117 | 1493 | 25121 | | | 72 | 430 | 440 | 1095 | | | **30768** |
| **UK** | 130 | | 165 | | | 2558 | 1198 | | 315 | | | **4367** |
| **WB** | 320 | 100 | 14897 | | 900 | 378 | 63 | | 110 | | | **16768** |
| **Total** | **10203** | **24505** | **251234** | **1214** | **3140** | **22098** | **20842** | **7880** | **33963** | **98** | **37715** | **412892** |

**Supplementary Table 28: Specified transmission corridors capacity and characteristics**. The line name represents the abbreviation of the source load zone (from) and the abbreviation of the sink load zone (to). The table shows the specified maximum and minimum capacity (MW) for each transmission correction (from-to), and transmission losses estimated for each corridor. The corridors distance are approximated by the distance between the major city in each load zone.

| Line | Capacity (MW) | Loss Factor (%) |
|---|---|---|
| AP-KA | 11,177 | 5.01 |
| AP-OR | 5,600 | 2.43 |
| AP-TN | 9,075 | 3.77 |
| AP-TS | 20,933 | 3.18 |
| AR-AS | 12,731 | 1.32 |
| AS-BH | 200 | 1.65 |
| AS-MN | 1,250 | 1.66 |
| AS-ML | 2,912 | 0.42 |
| AS-MZ | 700 | 1.80 |
| AS-NL | 1,708 | 1.24 |
| AS-TR | 2,300 | 1.64 |
| AS-UP | 3,000 | 6.69 |
| AS-WB | 3,656 | 3.27 |
| BH-WB | 2,100 | 3.32 |
| BR-JH | 9,394 | 2.06 |
| BR-SK | 1,294 | 2.48 |
| BR-UP | 18,022 | 2.71 |
| BR-WB | 7,824 | 2.94 |
| CH-HR | 154 | 1.64 |
| CH-HP | 440 | 0.34 |
| CH-PB | 100 | 0.56 |
| CG-HR | 6,000 | 5.65 |
| CG-JH | 5,508 | 3.11 |
| CG-MP | 18,707 | 3.84 |
| CG-MH | 19,070 | 5.87 |
| CG-OR | 12,173 | 2.80 |
| CG-TN | 6,000 | 5.71 |
| DH-GJ | 1,749 | 1.93 |
| DH-MH | 1,034 | 0.83 |
| DD-GJ | 1,406 | 1.81 |
| DD-MH | 1,120 | 0.93 |
| DL-HR | 8,953 | 0.14 |
| DL-RJ | 208 | 1.47 |
| DL-UP | 9,160 | 2.58 |
| GA-KA | 200 | 3.03 |
| GA-MH | 1,382 | 2.61 |
| GJ-HR | 2,500 | 4.75 |
| GJ-MP | 7,450 | 2.09 |
| GJ-MH | 4,355 | 2.74 |
| GJ-RJ | 6,034 | 3.35 |
| HR-HP | 3,477 | 1.86 |
| HR-PB | 11,014 | 1.93 |
| HR-RJ | 11,334 | 1.40 |
| HR-UP | 10,205 | 2.48 |
| HR-UK | 1,362 | 1.40 |
| HP-JKLA | 517 | 2.49 |
| HP-PB | 8,405 | 0.79 |
| HP-UK | 486 | 0.73 |
| JKLA-PB | 15,646 | 2.28 |
| JH-OR | 3,812 | 1.76 |
| JH-WB | 15,913 | 1.40 |
| KA-KL | 1,165 | 3.13 |
| KA-MH | 6,012 | 5.25 |
| KA-OR | 2,000 | 7.43 |
| KA-TN | 4,563 | 1.81 |
| KA-TS | 917 | 3.10 |
| KL-TN | 6,394 | 3.89 |
| MP-MH | 17,154 | 3.16 |
| MP-RJ | 5,750 | 2.89 |
| MP-UP | 12,250 | 4.29 |
| MH-TS | 9,800 | 3.86 |
| MN-MZ | 50 | 1.07 |
| MN-NL | 1,200 | 0.78 |
| MZ-TR | 100 | 0.91 |
| OR-WB | 1,100 | 2.27 |
| PY-TN | 1,434 | 0.86 |
| PB-RJ | 5,000 | 2.77 |
| PB-UP | 6,594 | 4.17 |
| RJ-UP | 6,327 | 3.17 |
| SK-WB | 1,250 | 3.29 |
| UP-UK | 12,699 | 2.99 |
| UP-WB | 3,000 | 5.52 |

# Supplementary Notes

## 1 Solar PV and Wind Zones

The potential VRE zones were selected using the Multi-criteria analysis for planning Renewable Energy (*MapRE*). *MapRE* was developed at the Lawrence Berkeley National Laboratory (LBNL) and is maintained at the University of California Santa Barbara (UCSB). This tool identifies appropriate zones for utility-scale solar and wind technologies in developing countries.

The Solar PV and onshore wind zoning analysis was conducted using a combination of *ArcGIS* and *ArcPy* [27]. The zoning process is split up into the following stages.

**Site Suitability Analysis.** In the first stage, we identify suitable sites for renewable energy deployment. Some basic assumptions are made, such as a land use efficiency of 30%, as well as an average capacity factor (20% for both solar and wind). In order to constrain where renewable projects can be built, many geographic features are excluded from potential renewable sites – such as protected areas, urban zones, water bodies, steep slopes, and infrastructure buffers. Each of these features is added as layers that include processing criteria that determine how each exclusion should be applied spatially.

We process resource rasters (solar irradiation for solar PV and wind speed for wind turbines), with a threshold placed to only keep locations where the resource available meets this threshold. The resource rasters are processed against the geographic feature constraints, creating a map where suitable areas meet both the minimum resource threshold and avoid all exclusion criteria. Areas smaller than a minimum threshold (2 $km^2$ in this case) are filtered out.

**Project Area.** This stage takes the suitable sites from the site suitability analysis and creates discrete project areas. Whereas the sites in the previous stage are provided as continuous coverage across suitable terrain, we require renewable development to occur in specific project boundaries. To accomplish this, we overlay a "fishnet grid" system on the suitable areas. This "fishnet grid" serves as a spatial framework for dividing large continuous suitable areas into manageable project-sized parcels. In this process, areas above certain size thresholds are intersected with the "fishnet grid" to create smaller parcels, while areas below minimum thresholds are aggregated with neighboring suitable areas to create viable project sizes.

This spatial process ensures that the resulting project areas are neither too large for single development projects nor too small to be economically viable. The configurable parameters in this stage are: grid cell size (5 km), minimum contiguous area thresholds (2 $km^2$), and maximum area limits (25 $km^2$).

**Zones Clustering.** The final stage groups individual project areas into larger development zones. We configured the maximum area of the largest project to 100 $km^2$ (for solar and for wind). This clustering process sums generation capacity and energy production values while averaging cost and performance metrics, creating zones that represent logical units for transmission planning and large-scale development. To avoid clustering that causes zones to be in multiple states, we do not allow zones to exist across state boundaries.

We obtained 566 solar PV candidate zones and 311 onshore wind candidate zones. However, these projects include zones located in the Andaman and Nicobar Islands, where the power grid is not connected to mainland India and thus not included in our proposed *GridPath-India* model. After removing the candidate zones in the Andaman and Nicobar Islands, we obtain the 556 solar PV and 300 onshore wind candidate zones in our *GridPath-India* model.

**Offshore Wind Zones (REZoning).** The offshore wind candidate sites were selected using the Renewable Energy Zoning tool (*REZoning*) [34]. *REZoning* is an online-based tool developed by the World Bank Group, Energy Sector Management Assistance Program (ESMAP), and UCSB. The website provides a platform to identify suitable zones for VRE projects and allows importing area shapefiles. This tool enables multi-criteria filtering based on topographic limitations, land-use constraints, and system performance.

We define a 25 km resolution spatial grid. The distance of the candidate sites to a port is $0.5 \geq d_{\text{port}} \leq 50$ km, to an anchorage is $d_{\text{anchorage}} \leq 100$ km, and to a UNESCO World Heritage Site is $d_{\text{UNESCO}} \leq 500$ km. The bathymetry (i.e., distance to the ocean floor) must be $d_{\text{depth}} \geq -1$ km. In addition, we discarded protected areas, inertial areas, coral reefs, and wetlands. The candidate sites resulting from this analysis are 46.

## 2 Electricity System Model Formulation

*GridPath* is written in *Python* using the optimization package *Pyomo* [3], and the resulting mixed-integer linear program is solved with the commercial solver *Gurobi* [49].

The inputs to *GridPath* include projected hourly electricity demand for each investment period, existing and candidate generator capacities, hourly capacity factors for weather-dependent generators (biomass, wind, solar, and hydro ROR), monthly hydropower budgets (hydro storage), candidate generator capacities in the pipeline, and transmission infrastructure capacities. The required techno-economic parameters include generators' fixed O&M costs, variable O&M costs (which include fuel costs in this project), unit investment costs of expanding the transmission lines, heat rates, capital costs per technology, lifetime, and $CO_2$ emission rates. The proposed policy parameters include capacity group requirements for specifying clean energy deployment targets, $CO_2$ emission limits (i.e., carbon caps), and renewable energy purchase obligations (share of electricity demand to be met by renewable energy). The system reliability parameters include planning reserve margin (additional available generation or storage capacity above peak demand as a percentage of peak demand), generators' availability factors, firm and variable generators' capacity credits (available capacity during peak demand), and projections (state and regional level) of hourly regulation reserves and spinning reserve provisions.

The outputs from *GridPath* are new generator, storage, and transmission project capacities by investment period, as well as hourly electricity dispatch, curtailment, unserved energy, and transmission losses, exports and imports among the states, variable O&M costs, reserve provisions, and $CO_2$ emissions.

| **Set** | **Index** | **Description** |
|---|---|---|
| $\Pi$ | $\pi$ | Years or periods when investment decisions are made. |
| $\Delta$ | $\delta$ | Horizons over which storage/hydro is balanced. |
| $\Delta^\pi \subset \Delta$ | $\delta^\pi$ | Horizons within an investment period $\pi$. |
| T | $\tau$ | Timepoints when operational decisions are made. |
| $T^\pi \subset T$ | $\tau$ | Operational timepoints in investment period $\pi$. |
| $T^\delta \subset T$ | $\tau^\delta$ | Operational timepoints in a sampled day $\delta$. |
| | $\pi^\tau$ | The year in which timepoint $\tau$ occurs. |
| | $\delta^\tau$ | The sampled day during which timepoint $\tau$ occurs. |
| | $hr^\tau$ | Number of hours in timepoint $\tau$. |
| Z | $z$ | Load zones (or nodes). |
| ZF | $zf$ | "From" load zones (or nodes) for transmission lines. |
| ZT | $zt$ | "To" load zones (or nodes) for transmission lines. |
| BAF | $baf$ | Balancing areas for frequency regulation reserves. |
| BAS | $bas$ | Balancing areas for spinning reserves. |
| PZ | $pz$ | Planning reserve margin zones. |
| RZ | $rz$ | Renewable or clean energy target zones. |
| CZ | $cz$ | Carbon CAP zones. |
| CG | $cg$ | Capacity groups. |
| $\Gamma$ | $\gamma$ | All generator and storage projects. |
| $\Gamma^\epsilon \subset \Gamma$ | $\gamma^\epsilon$ | Specified (including existing) generator and storage projects. |
| $\Gamma^\nu \subset \Gamma$ | $\gamma^\nu$ | Candidate generator and storage projects. |
| $\Gamma^\mu \subset \Gamma$ | $\gamma^\mu$ | Must-run generators (e.g., nuclear, biomass, waste-heat-recovery). |
| $\Gamma^\phi \subset \Gamma$ | $\gamma^\phi$ | Dispatchable generators (e.g., CCGT, CT, Coal). |
| $\Gamma^\omega \subset \Gamma$ | $\gamma^\omega$ | Variable generators (e.g., wind, solar, hydro run-of-river). |
| $\Gamma^\sigma \subset \Gamma$ | $\gamma^\sigma$ | Storage projects (e.g., battery, hydrogen, pumped hydro). |
| $\Gamma^\psi \subset \Gamma^\sigma$ | $\gamma^\psi$ | Hydrogen storage projects. |
| $\Gamma^\chi \subset \Gamma$ | $\gamma^\chi$ | Hydro storage projects. |
| $\Gamma^z \subset \Gamma$ | $\gamma^z$ | Projects associated with a load zone. |
| $\Gamma^{rz} \subset \Gamma$ | $\gamma^{rz}$ | Projects associated with a renewable/clean energy target zone. |
| $\Gamma^{cz} \subset \Gamma$ | $\gamma^{cz}$ | Projects associated with a carbon CAP zone. |
| $\Gamma^{cg} \subset \Gamma$ | $\gamma^{cg}$ | Projects associated with a capacity group. |
| $\Lambda$ | $\lambda$ | All transmission lines. |
| $\Lambda^\epsilon \subset \Lambda$ | $\lambda^\epsilon$ | Specified (including existing) transmission lines. |
| $\Lambda^\nu \subset \Lambda$ | $\lambda^\nu$ | Candidate transmission lines. |

| Variable | Description |
|---|---|
| $CB_{\gamma^{\nu},\pi} \geq 0$ | Power capacity to install at each candidate project in each investment period. |
| $CE_{\gamma\subset\Gamma^{\nu}\cap\Gamma^{\sigma},\pi} \geq 0$ | Energy capacity (i.e. power $\times$ duration) to install at candidate storage project in each investment period. |
| $B_{\gamma^{\pi}} \geq 0$ | Power capacity at each project in each investment period. |
| $E_{\gamma\subset\Gamma^{\sigma},\pi} \geq 0$ | Energy capacity (i.e. power $\times$ duration) to at each storage project in each investment period. |
| $U_{\gamma^{\phi},\tau} \geq 0$ | Commitment level of each dispatchable generator in each timepoint. |
| $V_{\gamma^{\phi},\tau} \geq 0$ | Startup of a dispatchable generator at a timepoint. |
| $W_{\gamma^{\phi},\tau} \geq 0$ | Shutdown of a dispatchable generator at a timepoint. |
| $O_{\gamma,\tau} \geq 0$ | Power output from each generator in each timepoint. |
| $C_{\gamma^{\sigma},\tau} \geq 0$ | Charging of each storage project in each timepoint. |
| $D_{\gamma^{\sigma},\tau} \geq 0$ | Discharging from each storage project in each timepoint. |
| $F_{\gamma^{\sigma},\tau} \geq 0$ | Energy available in storage at the start of a timepoint. |
| $CUR_{\gamma\subset\Gamma^{\omega}\cap\Gamma^{\chi},\tau} \geq 0$ | Output curtailment of each variable or hydro project in each timepoint. |
| $Freq_{\gamma\subset\Gamma^{\phi}\cap\Gamma^{\sigma}\cap\Gamma^{\chi},\tau} \geq 0$ | Frequency regulation provision from each dispatchable, storage, or hydro project in each timepoint. |
| $Spin_{\gamma\subset\Gamma^{\phi}\cap\Gamma^{\sigma}\cap\Gamma^{\chi},\tau} \geq 0$ | Spinning reserve provision from each dispatchable, storage, or hydro project in each timepoint. |
| $UE_{z,\tau} \geq 0$ | Unserved energy at each zone in each timepoint. |
| $OE_{z,\tau} \geq 0$ | Overgeneration energy at each zone in each timepoint. |
| $CL_{\lambda^{\nu},\pi} \geq 0$ | Transmission line capacity to install at each candidate transmission line in each investment period. |
| $T_max_cap_{\lambda,\tau}$ | Transmission maximum capacity in each timepoint. |
| $T_min_cap_{\lambda,\tau}$ | Transmission minimum capacity in each timepoint. |
| $O_{\lambda,\tau}$ | Transmission flow on a transmission line in each timepoint. |
| $T_loss_{\lambda,zf,\tau} \geq 0$ | Transmission line loss at the originating (from) load zone in each timepoint. |
| $T_loss_{\lambda,zt,\tau} \geq 0$ | Transmission line loss at the destination (to) load zone in each timepoint. |

| Parameter | Description |
|---|---|
| $D_{z,\tau}$ | Electricity demand or load at load zone in timepoint $\tau$. |
| $\Pi^{f}$ | First investment period. |
| $disc_factor$ | Discount factor derived from the discount rate to calculate net present value. |
| $n_years_rep_{\pi}$ | Number of years represented by each investment period. |
| $tp_weight_{\tau}$ | Weight of each timepoint based on number of timepoints or hours it represents in a year. |
| $B_max_{\gamma^{\nu},\pi}$ | Maximum cumulative power capacity installed at each candidate project in each investment period. |
| $E_max_{\gamma\subset\Gamma^{\nu}\cap\Gamma^{\sigma},\pi}$ | Maximum energy capacity (i.e. power $\times$ duration) installed at candidate storage project in each investment period. |
| $SB_{\gamma^{\epsilon},\pi}$ | Specified power capacity (including existing) installed at each project in each investment period. |
| $SE_{\gamma\subset\Gamma^{\epsilon}\cap\Gamma^{\sigma},\pi}$ | Specified energy capacity (i.e. power $\times$ duration) installed at storage project in each investment period. |
| $MSL_{\gamma^{\phi},\tau}$ | Minimum Stable Level of each dispatchable generator in each timepoint. |
| $MDT_{\gamma^{\phi}}$ | Minimum Down Time of each dispatchable generator. |
| $MUT_{\gamma^{\phi}}$ | Minimum Up Time of each dispatchable generator. |
| $Charging_Eff_{\gamma^{\sigma}}$ | Charging efficiency of storage project. |
| $Discharging_Eff_{\gamma^{\sigma}}$ | Discharging efficiency of storage project. |
| $SL_max_{\lambda^{\epsilon},\pi}$ | Maximum transmission capacity at a specified transmission line in each investment period. |
| $SL_min_{\lambda^{\epsilon},\pi}$ | Minimum transmission capacity at a specified transmission line in each investment period. |
| $SL_{\lambda^{\epsilon},\pi}$ | Specified transmission line capacity in each investment period. |
| $T_loss_factor_{\lambda}$ | Transmission loss factor for transmission line. |

**Capacity Constraints**

Candidate project capacity cannot exceed pre-defined limits.

$$\sum_{\pi=\pi^f}^{\pi} CB_{\gamma,\pi} \leq CB_max_{\gamma,\pi} \qquad \forall \gamma \in \Gamma^{\nu}$$

$$\sum_{\pi=\pi^f}^{\pi} CE_{\gamma,\pi} \leq CE_max_{\gamma,\pi} \qquad \forall \gamma \in \Gamma^{\sigma} \cap \Gamma^{\nu}$$

$$\sum_{\pi=\pi^f}^{\pi} CL_{\lambda,\pi} \leq CL_max_{\lambda,\pi} \qquad \forall \lambda \in \Lambda^{\nu}$$

Total project power and energy capacity at each project is a sum of selected candidate projects and specified projects

$$B_{\gamma,\tau} = \begin{cases} SB_{\gamma,\pi} & \forall \gamma \in \Gamma^{\epsilon} \\ \sum_{\pi=\pi^f}^{\pi} CB_{\gamma,\pi} & \forall \gamma \in \Gamma^{\nu} \end{cases}$$

$$E_{\gamma^{\phi},\tau} = \begin{cases} SE_{\gamma,\pi} & \forall \gamma \in \Gamma^{\phi} \cap \Gamma^{\epsilon} \\ \sum_{\pi=\pi^f}^{\pi} CE_{\gamma,\pi} & \forall \gamma \in \Gamma^{\phi} \cap \Gamma^{\nu} \end{cases}$$

Transmission max capacity is either specified or is the new built capacity

$$T_max_cap_{\lambda,\tau} = \begin{cases} SL_max_{\lambda,\pi} & \forall \lambda \in \Lambda^{\epsilon} \\ \sum_{\pi=\pi^f}^{\pi} CL_{\lambda,\pi} & \forall \lambda \in \Lambda^{\nu} \end{cases}$$

Transmission min capacity is either specified or is the negative of the new built capacity

$$T_min_cap_{\lambda,\tau} = \begin{cases} SL_min_{\lambda,\pi} & \forall \lambda \in \Lambda^{\epsilon} \\ -\sum_{\pi=\pi^f}^{\pi} CL_{\lambda,\pi} & \forall \lambda \in \Lambda^{\nu} \end{cases}$$

| **Project Operational Constraints** |
|---|
| Dispatchable generator capacity commitment cannot exceed available generator capacity. $$U_{\gamma,\tau} \leq B_{\gamma,\pi} \times availability_factor_{\gamma,\pi} \qquad \forall \gamma \in \Gamma^{\phi}$$ |
| Power output from dispatchable generators must exceed or equal a minimum loading level. $$O_{\gamma,\tau} \geq U_{\gamma,\tau} \times MSL_{\gamma} \qquad \forall \gamma \in \Gamma^{\phi}$$ |
| A dispatchable generator must be committed for its Minimum Up Time. $$\sum_{\theta=\tau}^{\tau+MUT_{\gamma}-1} U_{\gamma,\tau} \geq MUT_{\gamma} \times V_{\gamma,\theta} \qquad \forall \gamma \in \Gamma^{\phi}$$ |
| A dispatchable generator must be down for its Minimum Down Time. $$\sum_{\theta=\tau}^{\tau+MDT_{\gamma}-1} (1 - U_{\gamma,\tau}) \geq MDT_{\gamma} \times W_{\gamma,\theta} \qquad \forall \gamma \in \Gamma^{\phi}$$ |
| A dispatchable generator cannot startup and shutdown on the same timepoint. $$V_{\gamma,\tau} - W_{\gamma,\tau} = U_{\gamma,\tau} - U_{\gamma,\tau-1} \qquad \forall \gamma \in \Gamma^{\phi}$$ |
| Power output plus frequency and spinning reserves from dispatchable generators cannot exceed committed capacity. $$O_{\gamma,\tau} + Freq_{\gamma,\tau} + Spin_{\gamma,\tau} \leq U_{\gamma,\tau} \qquad \forall \gamma \in \Gamma^{\phi}$$ |
| Power output of must-run generator must equal their available capacity. $$O_{\gamma,\tau} = B_{\gamma,\pi} \times availability_factor_{\gamma,\pi} \qquad \forall \gamma \in \Gamma^{\mu}$$ |
| Power output of variable generators cannot exceed their available capacity in a timepoint. $$O_{\gamma,\tau} = B_{\gamma,\pi} \times capacity_factor_{\gamma,\tau} - CUR_{\gamma,\tau} \qquad \forall \gamma \in \Gamma^{\omega}$$ |

| |
|---|
| Net power output plus frequency and spinning reserves from storage cannot exceed storage power capacity. $$D_{\gamma,\tau} - C_{\gamma,\tau} + Freq_{\gamma,\tau} + Spin_{\gamma,\tau} \leq B_{\gamma,\pi} \qquad \forall \gamma \in \Gamma^{\sigma}$$ |
| The energy available in storage cannot exceed the installed energy capacity. $$F_{\gamma,\tau} \leq E_{\gamma,\pi} \qquad \forall \gamma \in \Gamma^{\sigma}$$ |
| Frequency regulation and spinning reserves from storage cannot exceed the energy available in storage at each timepoint. $$Freq_{\gamma,\tau} + Spin_{\gamma,\tau} \leq F_{\gamma,\tau} - \frac{D_{\gamma,\tau}}{Discharging_Eff_{\gamma}} + C_{\gamma,\tau} \times Charging_Eff_{\gamma} \qquad \forall \gamma \in \Gamma^{\sigma}$$ |
| The energy available in storage must equal the energy available in the previous timepoint minus the net power output from the previous timepoint. The last timepoint of each horizon is defined as the previous timepoint for timepoint 1 of that horizon. $$F_{\gamma,\tau} = F_{\gamma,\tau-1} - \frac{D_{\gamma,\tau-1}}{Discharging_Eff_{\gamma}} + C_{\gamma,\tau-1} \times Charging_Eff_{\gamma} \qquad \forall \gamma \in \Gamma^{\sigma}/\Gamma^{\psi}$$ |
| The energy available in storage must equal the weighted energy available in the previous timepoint minus the weighted net power output from the previous timepoint. The last timepoint of each horizon is defined as the previous timepoint for timepoint 1 of that horizon. $$F_{\gamma,\tau} = F_{\gamma,\tau-1} - \frac{tp_weight_{\tau} \times D_{\gamma,\tau-1}}{Discharging_Eff_{\gamma}} + tp_weight_{\tau} \times C_{\gamma,\tau-1} \times Charging_Eff_{\gamma} \qquad \forall \gamma \in \Gamma^{\psi}$$ |
| Power output plus frequency regulation reserves from hydro generators cannot exceed available capacity. $$O_{\gamma,\tau} + Freq_{\gamma,\tau} + Spin_{\gamma,\tau} \leq maximum_power_fraction_{\gamma,\delta} \times B_{\gamma,\pi} \qquad \forall \gamma \in \Gamma^{\chi}$$ |
| Average power output from hydro projects plus curtailment during each horizon must equal the average fraction of available capacity during that horizon. $$\sum_{\tau \in T^{\delta}} ((O_{\gamma,\tau} + CUR_{\gamma,\tau}) \times hr_{\tau}) = \sum_{\tau \in T^{\delta}} (average_power_fraction_{\gamma,\delta} \times B_{\gamma,\pi} \times hr_{\tau}) \qquad \forall \gamma \in \Gamma^{\chi}$$ |

| **Transmission Operational Constraints** |
|---|
| Transmission flow cannot exceed transmission max capacity. $$O_{\lambda,\tau} \leq T_max_cap_{\lambda,\tau}$$ |
| Transmission flow cannot be less than transmission min capacity. $$O_{\lambda,\tau} \geq T_min_cap_{\lambda,\tau}$$ |
| Transmission loss for originating (from) load zone must be greater or equal to the negative of the transmission flow times loss factor. $$T_loss_{\lambda,zf,\tau} \geq -O_{\lambda,\tau} \times T_loss_factor_{\lambda}$$ |
| Transmission loss for destination (to) load zone must be greater or equal to the transmission flow times loss factor. $$T_loss_{\lambda,zt,\tau} \geq O_{\lambda,\tau} \times T_loss_factor_{\lambda}$$ |
| Transmission loss cannot exceed maximum transmission capacity. $$T_loss_{\lambda,zf,\tau} \leq T_max_cap_{\lambda,\tau} \times T_loss_factor_{\lambda}$$ $$T_loss_{\lambda,zt,\tau} \leq T_max_cap_{\lambda,\tau} \times T_loss_factor_{\lambda}$$ |

| **System Constraints** |
|---|

| $Energy_Balance_{z,\tau}$ in zone $z$ and in timepoint $\tau$ |
|---|

$$\sum_{\gamma\in\Gamma^{\phi}\cup\Gamma^{\mu}\cup\Gamma^{\chi}} \underbrace{O_{\gamma,z,\tau}}_{\text{generator power output}} + \sum_{\gamma\in\Gamma^{\omega}} \underbrace{(B_{\gamma,\pi} \times capacity_factor_{\gamma,\tau} - CUR_{\gamma,\tau})}_{\text{candidate variable generator power output minus curtailment}}$$

$$+ \sum_{\gamma\in\Gamma^{\sigma}} \underbrace{(D_{\gamma,z,\tau} - C_{\gamma,z,\tau})}_{\text{storage net power output}} + \sum_{\lambda,zf=z} \underbrace{(O_{\lambda,zf,\tau} + T_loss_{\lambda,zf,\tau})}_{\text{transmission flow from zone}} + \sum_{\lambda,zt=z} \underbrace{(O_{\lambda,zt,\tau} - T_loss_{\lambda,zt,\tau})}_{\text{transmission flow to zone}}$$

$$+ \underbrace{UE_{z,\tau}}_{\text{unserved energy}} - \underbrace{OE_{z,\tau}}_{\text{overgeneration energy}} = \underbrace{D_{z,\tau}}_{\text{system load}}$$

Capacity credits provision from all generators for planning reserve margin in planning reserve margin zone $pz$ ($Planning_Reserve_Margin_{pz,\pi}$).

$$\sum_{\gamma\in\Gamma} capacity_credit_{pz,\gamma,\pi} \geq Planning_Reserve_Margin_{pz,\pi}$$

Spinning reserves provision from dispatchable generators, hydro, and storage in balancing area for spinning reserves in balancing area $bas$ ($Spinning_Reserves_Balance_{bas,\tau}$).

$$\sum_{\gamma\in\Gamma^{\phi}\cup\Gamma^{\sigma}\cup\Gamma^{\chi}} Spin_{\gamma,bas,\tau} \geq spinning_reserves_requirement_{bas,\tau}$$

Frequency regulation provision from dispatchable generators, hydro, and storage in balancing area $baf$ ($Frequecy_Regularion_{baf,\tau}$).

$$\sum_{\gamma\in\Gamma^{\phi}\cup\Gamma^{\sigma}\cup\Gamma^{\chi}} Freq_{\gamma,baf,\tau} \geq frequency_regularion_requirement_{baf,\tau}$$

| **Policy Constraints** |
|---|

Clean generator energy output has to meet clean energy target for that investment period ($clean_energy_target_{rz,\pi}$ in clean zone $rz$ and period $\pi$).

$$\sum_{\gamma\in\Gamma^{rz},\tau} (tp_weight_{\tau} \times O_{\gamma,\tau}) \geq clean_energy_target_{rz,\pi}$$

$CO_2$ emissions from the generators have to be below a limit for that investment period ($carbon_cap_{cz,\pi}$ in carbon zone $cz$ and period $\pi$).

$$\sum_{\gamma\in\Gamma^{cz},\tau} (tp_weight_{\tau} \times heat_rate_{\gamma} \times emission_rate_{\gamma} \times O_{\gamma,\tau}) \geq carbon_cap_{cz,\pi}$$

The total capacity of the projects in a group must add up to a certain capacity ($capacity_group_{cg,\pi}$ in group $cg$ and period $\pi$.

$$\sum_{\pi=\pi^f}^{\pi} CB_{\gamma,\pi} = capacity_group_{cg,\pi} \qquad \forall\gamma \in \Gamma^{cg}$$

| **Objective Function** |
|---|
| Generation, storage, and transmission investment costs. |

$$\sum_{\pi}\Bigg[\underbrace{\sum_{\gamma^{\epsilon}}\Big(SB_{\gamma^{\epsilon},\pi} \times annualized_investment_cost_p_mw_{\gamma^{\epsilon}}\Big)}_{\text{cost of specified generation capacity and storage power capacity}}$$

$$+ \underbrace{\sum_{\gamma^{\nu}}\Big(CB_{\gamma^{\nu},\pi} \times annualized_investment_cost_p_mw_{\gamma^{\nu},\pi}\Big)}_{\text{cost of new investments in generation capacity and storage power capacity}}$$

$$+ \underbrace{\sum_{\gamma\in\Gamma^{\epsilon}\cup\Gamma^{\sigma}}\Big(SE_{\gamma,\pi} \times annualized_investment_cost_p_mwh_{\gamma,\pi}\Big)}_{\text{cost of specified storage energy capacity}}$$

$$+ \underbrace{\sum_{\gamma\in\Gamma^{\nu}\cup\Gamma^{\sigma}}\Big(CE_{\gamma,\pi} \times annualized_investment_cost_p_mwh_{\gamma,\pi}\Big)}_{\text{cost of new investments in storage energy capacity}}$$

$$+ \underbrace{\sum_{\lambda^{\epsilon}}\Big(SL_{\lambda^{\epsilon},\pi} \times annualized_investment_cost_p_mw_{\lambda^{\epsilon},\pi}\Big)}_{\text{cost of specified transmission capacity}}$$

$$+ \underbrace{\sum_{\lambda^{\nu}}\Big(CL_{\lambda^{\nu},\pi} \times annualized_investment_cost_p_mw_{\lambda^{\nu},\pi}\Big)}_{\text{cost of new investments in transmission capacity}}\Bigg]$$

$$\times n_years_rep_{\pi} \times disc_factor_{\pi}$$

Operational costs.

$$+\sum_{\tau,\pi}\Bigg[ \underbrace{\sum_{\gamma\in\Gamma^{\phi}\cup\Gamma^{\mu}}\Big(O_{\gamma,\tau,\pi}\times var_om_{\gamma}\Big)}_{\text{dispatchable and must-run generator variable O\&M cost}}$$

$$+ \underbrace{\sum_{\gamma\in\Gamma^{\chi}\cup\Gamma^{\omega}}\Big((O_{\gamma,\tau,\pi}+CUR_{\gamma,\tau})\times var_om_{\gamma}\Big)}_{\text{curtailable (variable and hydro) generator variable O\&M cost}}$$

$$+ \underbrace{\sum_{\gamma^{\sigma}}\Big(D_{\gamma^{\sigma},\tau,\pi}\times var_om_{\gamma^{\sigma}}\Big)}_{\text{storage variable O\&M cost (for batteries)}}$$

$$+\underbrace{\sum_{\gamma\in\Gamma^{\phi}\cup\Gamma^{\mu}}\Big(O_{\gamma,\tau,\pi}\times heat_rate_{\gamma}\times fuel_price_{\gamma,\pi}\Big)}_{\text{dispatchable and must-run generator fuel cost}}$$

$$+\underbrace{\sum_{\gamma^{\phi}}\Big(unit_capacity_{\gamma^{\phi},\tau,\pi}\times startup_cost_per_capacity_{\gamma^{\phi}}\times V_{\gamma^{\phi},\tau,\pi}\Big)}_{\text{startup costs for dispatchable generators}}$$

$$+\underbrace{\sum_{z}\Big(UE_{z,\tau,\pi}\times unserved_energy_penalty_{z}\Big)}_{\text{unserved energy penalties}}$$

$$+\underbrace{\sum_{z}\Big(OE_{z,\tau,\pi}\times overgeneration_penalty_{z}\Big)}_{\text{overgeneration penalties}}\Bigg]$$

$$\times tp_weight_{\tau}\times hr_{\tau}\times n_years_rep_{\pi}\times disc_factor_{\pi}$$

## 3 Capacity Factors

**Solar.** The solar radiation measurements used to estimate the capacity factors for each solar Photovoltaic (PV) candidate site are from the National Solar Radiation Database (NSRDB) [13], maintained by the National Renewable Energy Laboratory (NREL). The PSM V3 model covers Europe, Africa, and Asia from 2017 to 2019. Both models are publicly available. The PSM V3 has a spatial resolution of $4 \times 4$ km and is available in 15, 30, or 60 min temporal resolution. We use the hourly temporal resolution (60 min). A METEOSAT IODC satellite acquired the data for the model.

The capacity factors are simulated using the System Advisor Model (*SAM*) developed by NREL as well [11]. The capacity factors for solar PV in existing sites are assumed to be fixed-tilt, but new sites can be single-axis tracking or fixed-tilt. The capacity expansion model uses data from Fiscal Year 2019 (FY19), Apr 2018 to Feb 2019. The resolution available for solar radiation measurements is 4 km from 2017-2019. If a site is partially cloudy, 4 km is still a low resolution to incorporate that information. In addition, the outage rates may be influential, and we did not make any bias correction for solar PV sites at this spatial and temporal resolution. However, we verify that the average generation simulated by *SAM* matches the generation from existing sites at the state level in FY19 (monthly/yearly).

We used *PVWatts* [52], available in the SAM Simulation Core (SSC) package for Python, *PySAM* [54]. The modules on the fixed-tilt mount have an inclination angle equal to the candidate site's latitude and have orientation to the source. We have a solar PV roof mount profile (i.e., site) per state to simulate residential PV. Their tilt is the latitude of the largest city in the state, and their orientation is south. The modules are assumed to be standard monocrystalline silicon (19% efficiency). The modules mounted on a single-axis solar tracking system rotate horizontally. We assume that if the user invests in a solar tracking system, they will also include the additional investment in premium monocrystalline silicon modules (21% efficiency). The losses are higher for utility-scale projects (14%) than for residential projects (19%) due to higher soling losses [6]. The inverter efficiency is the same (96%) for all PV systems.

**Wind.** The wind velocity measurements are from FY19 since we have high spatial resolution solar PV capacity factors from 2017 to 2019 in the PSM V3 model. However, we only have hourly wind velocity measurements with a low spatial resolution ($31 \times 31$ km) available in India from ERA5 [16]. The ERA5 weather reanalysis model has wind velocity measurements available from 1940, but we need weather-synchronized capacity factors to calculate the capacity credits. We need capacity factors from 2000 to 2014 to match the high spatial resolution solar radiation measurements available in the SUNY model.

Wind velocity measurements in ERA5 have a bias [25]. We corrected the bias with the Global Wind Atlas (GWA) to account for deviations in the capacity factors due to low resolution [32]. The GWA has a low temporal resolution, average wind velocity at 100 m from 2008 to 2017, and a very high spatial resolution of 250 m. For that, we calculate the bias in the average wind velocity at 100 m in the coordinates of each candidate site between the ERA5 and the GWA. GWA uses data from 2008 to 2017, so we calculate the bias with wind speed data from ERA5 recorded in the same years.

We used *PySAM* to simulate the capacity factors in the wind after bias-correcting the wind velocity measurements [54]. The existing and new turbines are site-specific optimal commercial 2 MW Vestas models assigned by IEC wind turbine class: Class I (V90) at high wind speed sites, Class II (V100) at mid wind speed sites, and Class III (V110) at low wind speed sites. The existing wind site capacity factors are calculated assuming high outage rates (10%) and an 80 m height hub. The capacity factors for new wind sites have low outage rates (5%), and 120 m hub height. The capacity factors simulated for offshore wind sites have a low outage rate (5%) and 135 m high hub height. The turbine is a commercial 7 MW Vestas V164 model with a 164 m diameter rotor, rated for high speeds. The system losses are (14%) the same for all wind sites onshore and offshore.

We have seen actual generation data from sites in India, and the outage rates are high (sometimes weeks long). To account for that, we increased the outage rate to 10% in existing wind sites. In addition, we want to ensure that the aggregated generation from existing wind sites matches the actual generation and correct for the potential bias. This second bias correction is made with FY19, which is the fiscal year used in the capacity expansion model. We calculate the average capacity factor at the state level from the simulated and actual capacity factor time series of existing wind sites in FY19. Then, we calculate the bias and correct the simulated capacity factor of each existing site to match the state-level average capacity factors. We apply this second bias correction only to existing sites since the average wind capacity factor at the national level is consistently increasing by approximately 0.5% every year from FY18 to FY23.

**Hydroelectric.** Monthly capacity factors for hydro Run-Of-River (ROR) and storage are the median values from annual generation reports published by the CEA in fiscal year 2019, 2020 and 2021 (April 2018 - March 2021) [57]. These reports provide monthly generation for 94 hydro ROR and 84 hydro storage generators with their nameplate capacity. We aggregate the energy generation across load zones to calculate monthly capacity factors for hydro ROR and storage at the state level. Hydro storage projects are dispatchable and have a monthly energy budget, constrained

by a monthly average capacity factor. Hydro ROR generators are non-dispatchable and must-run with hourly capacity factors set equal to monthly average capacity factors. Capacity factors for hydro ROR in Bhutan are linearly adjusted to match exports (6.35 TWh) in FY20 [18].

**Biomass.** Biomass is mainly from agricultural residues in India. The agricultural residues available amount depends on the harvest. The biomass generation depends on the crops grown in the area near the power station. In particular, bagasse is a biomass byproduct of sugar production extracted from sugarcane juice. The Central Energy Agency (CEA) reports biomass and bagasse generation separately. We model bagasse capacity within the biomass capacity. The reports contain monthly biomass and bagasse generation at the state level. The CEA also updates the existing capacity of the biomass and bagasse in the reports.

We calculate the capacity factors from the reports published for FY20. As we do not propose new biomass projects, we consider biomass generation at the state level to have the necessary granularity. However, the capacity factors are from monthly generation since we do not have hourly generation data, which is the required temporal resolution. We assign the average monthly capacity factor to each hour in the days of the month.

# 4 India-Specific Technology Cost Projections

**New Generator Cost Projections.** The Photovoltaic (PV) with single-axis tracking system cost projections are the base utility-scale estimates in the Annual Technology Baseline (ATB) from 2023 [38]. We linearly adjust them to match the auction prices from the International Renewable Energy Agency (IRENA) in 2020 [50]. The 2022 Indian Technology Catalogue (ITC) estimates are lower than the auction prices in 2020 [22], we assume them to be biased toward lower costs, so we use the mid-cost as the lower bound (2035 and 2050) and the high cost as the upper bound (2050), which matches the adjusted 2023 ATB estimates (2050). The upper bound estimate for 2035 is the average between the 2022 ITC (2035) and the adjusted 2023 ATB (2035). The mid estimate is the average between the upper and lower bounds (Figure 26A). We apply a standard 1-dimensional signal processing technique [1], Piecewise Cubic Hermit Interpolation Polynomial (PCHUP), to interpolate the trends (Table 29).

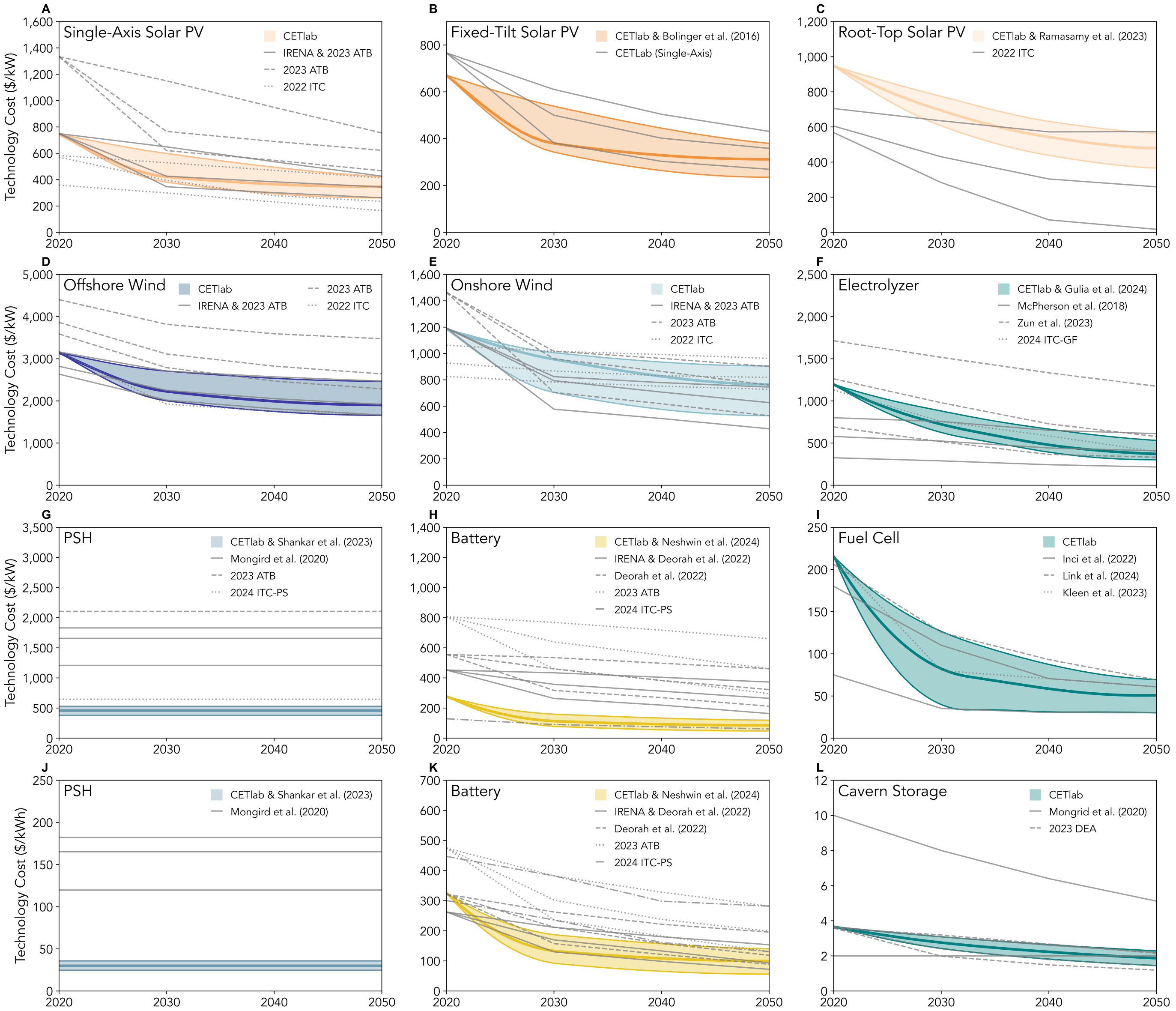


**Supplementary Figure 26. New generator and energy storage cost projections**. The gray lines represent the technology cost projections from the literature. The solid lines are the proposed technology cost projections in our study. The different colors represent a technology. The spread between the low and high-cost projections is shaded in the technology color. New renewable generator cost projections for solar PV on a single-axis tracking system (A), fixed-tilt structure (B), and mounted on a roof (C), and for wind offshore (D) and onshore (E). New short-term storage cost projections for PSH power capacity (G) and energy capacity (J), and battery power capacity (H) and energy capacity (K). New long-term storage cost projections (hydrogen) are for PEM electrolyzer (F), $H_2$ fuel cell (I), and salt cavern storage (L).

The 2023 ATB and the 2022 ITC do not provide cost estimates for solar PV with fixed tilt for utility-scale projects [22, 38]. A recent investigation provides an estimation of the ratio between the cost of both PV technologies in utility-scale plants in the US in 2022 [5]. We assume this relation is the same in India, and linearly adjust our solar PV single-axis cost projections to approximate the cost of solar PV fixed-tilt technology (Figure 26B and Table 29).

The solar PV roof-mount cost projections are based on the individual components in the 2022 ITC [22]. The hypotensions in low cost scenario in the 2022 ITC is insistent with the literature and consequentially rejected. The mid scenario is assumed the lower cost scenario. We independently adjusted the cost projections of each component using the methodology in the 2023 ATB [40]. We linearly adjusted the cost projection of the modules in India applying a scale factor between utility-scale and residential modules in the US. Similarly, the cost projection of the inverter by linearly adjusting the inverter and electrical hardware costs. The initial costs in 2020 are from the upper scenario. The lower and mid scenario are set to start with this same initial cost (Figure 26C and Table 29). The mid scenario is the average between the new high and low cost scenarios. We apply PCHUP to interpolate the trend from 2020 to 2030. The mid estimate is the average between the upper and lower bounds.

**Supplementary Table 29: Solar PV technologies**. Solar PV single-axis tracking system (SA), fixed-tilt structure (FT), and mounted on a residential top-roof (TR) cost projections. The operation and maintenance (O&M) cost projections for single-axis and fixed-tilt are the same, but top-roof has different O&M cost projections. Capital cost projections have different declining rate scenarios (low, mid and high) while O&M cost projections only have a mid-scenario.

| | **Single-Axis (USD/kW)** | | | **Fixed-Tilt (USD/kW)** | | | **Roof-Top (USD/kW)** | | | **O&M (USD/kW-yr)** | |
|---|---|---|---|---|---|---|---|---|---|---|---|
| **Period** | **low** | **mid** | **high** | **low** | **mid** | **high** | **low** | **mid** | **high** | **SA/FT** | **RT** |
| **2020** | 746 | 746 | 746 | 671 | 671 | 671 | 948 | 948 | 948 | 5 | 6 |
| **2021** | 699 | 703 | 729 | 629 | 633 | 656 | 905 | 917 | 929 | 5 | 6 |
| **2022** | 651 | 661 | 713 | 586 | 595 | 641 | 864 | 887 | 910 | 4 | 6 |
| **2023** | 605 | 619 | 697 | 545 | 557 | 627 | 825 | 858 | 892 | 4 | 6 |
| **2024** | 561 | 578 | 681 | 505 | 521 | 613 | 787 | 831 | 874 | 4 | 6 |
| **2025** | 519 | 541 | 665 | 467 | 487 | 599 | 752 | 805 | 857 | 4 | 5 |
| **2026** | 480 | 506 | 651 | 432 | 456 | 586 | 718 | 780 | 840 | 4 | 5 |
| **2027** | 446 | 476 | 637 | 402 | 429 | 573 | 686 | 756 | 823 | 3 | 5 |
| **2028** | 418 | 452 | 623 | 376 | 406 | 561 | 657 | 734 | 806 | 3 | 5 |
| **2029** | 396 | 433 | 610 | 356 | 390 | 549 | 630 | 713 | 790 | 3 | 5 |
| **2030** | 381 | 422 | 599 | 343 | 380 | 539 | 606 | 694 | 774 | 3 | 5 |
| **2031** | 370 | 415 | 587 | 333 | 373 | 529 | 584 | 675 | 758 | 3 | 4 |
| **2032** | 359 | 408 | 576 | 323 | 367 | 519 | 563 | 656 | 742 | 3 | 4 |
| **2033** | 349 | 401 | 565 | 314 | 361 | 509 | 543 | 638 | 726 | 3 | 4 |
| **2034** | 340 | 395 | 554 | 306 | 356 | 499 | 525 | 621 | 711 | 3 | 4 |
| **2035** | 331 | 389 | 544 | 298 | 350 | 489 | 507 | 605 | 695 | 3 | 4 |
| **2036** | 322 | 384 | 533 | 290 | 346 | 480 | 490 | 590 | 680 | 3 | 4 |
| **2037** | 314 | 379 | 523 | 283 | 341 | 471 | 475 | 576 | 667 | 3 | 4 |
| **2038** | 307 | 374 | 513 | 276 | 337 | 462 | 462 | 564 | 654 | 3 | 3 |
| **2039** | 300 | 370 | 504 | 270 | 333 | 453 | 449 | 553 | 642 | 2 | 3 |
| **2040** | 293 | 366 | 494 | 264 | 329 | 445 | 438 | 543 | 632 | 2 | 3 |
| **2041** | 287 | 362 | 485 | 259 | 326 | 437 | 428 | 534 | 623 | 2 | 3 |
| **2042** | 282 | 359 | 477 | 254 | 323 | 429 | 419 | 524 | 614 | 2 | 3 |
| **2043** | 277 | 356 | 468 | 250 | 321 | 422 | 409 | 516 | 606 | 2 | 3 |
| **2044** | 273 | 354 | 460 | 246 | 318 | 414 | 401 | 508 | 598 | 2 | 3 |
| **2045** | 270 | 352 | 453 | 243 | 316 | 408 | 392 | 500 | 590 | 2 | 3 |
| **2046** | 267 | 350 | 446 | 240 | 315 | 401 | 385 | 494 | 584 | 2 | 3 |
| **2047** | 264 | 348 | 439 | 238 | 314 | 395 | 379 | 488 | 578 | 2 | 3 |
| **2048** | 263 | 347 | 433 | 237 | 313 | 390 | 373 | 484 | 572 | 2 | 3 |
| **2049** | 262 | 347 | 427 | 236 | 312 | 385 | 369 | 480 | 568 | 2 | 3 |
| **2050** | 261 | 347 | 422 | 235 | 312 | 380 | 366 | 480 | 564 | 2 | 3 |

Onshore wind cost projections combine information on declining trends from the 2023 ATB with auction prices for onshore wind in India from the IRENA database [38, 50]. The 2023 ATB projections are linearly adjusted to match the auction prices in 2020. The adjusted 2023 ATB projections are higher in 2020 and lower in 2030 than the 2022 ITC cost estimates [22]. Due to this contradiction, we do not trust the adjusted trend from the 2023 ATB. Similarly, 2022 ITC projections for 2050 seem conservative compared to international declining trends. Therefore, we only use the initial cost in the trend (2020) and the estimations from the original 2023 ATB as intermediate and terminal points in the series (2030 and 2050, respectively); see Figure 26D and Table 30. We also use PCHUP to interpolate the trends.

Offshore wind cost projections combine information on declining trends from the 2023 ATB and the 2022 ITC [38, 22]. The IRENA database does not have offshore wind auction prices in India. Therefore, the 2023 ATB is linearly

adjusted using local auction prices for onshore wind. ITC offshore wind cost projections align with the upper bound of adjusted 2023 ATB projections in 2020. We use them to define the initial cost. We then use the 2022 ITC projection as a lower bound since it aligns with the adjusted 2023 ATB projections from 2035 to 2050. The adjusted 2023 ATB projections are the upper bound (2020, 2035, and 2050); see Figure 26E and Table 30. The mid estimate is the average between the upper and lower estimates. Similarly, we apply PCHUP to interpolate the trends.

**Supplementary Table 30: Wind technologies cost projections.** Onshore and offshore wind have different cost projections for capital costs, and operation and maintenance (O&M). Capital cost projections have different declining rate scenarios (low, mid and high) while O&M cost projections only have a mid-scenario.

| | Onshore (USD/kW) | | | Offshore (USD/kW) | | | O&M (USD/kW-yr) | |
|---|---|---|---|---|---|---|---|---|
| **Period** | **low** | **mid** | **high** | **low** | **mid** | **high** | **Onshore** | **Offshore** |
| **2020** | 1186 | 1186 | 1186 | 3147 | 3147 | 3147 | 10 | 94 |
| **2021** | 1124 | 1158 | 1164 | 2999 | 3028 | 3092 | 9 | 88 |
| **2022** | 1063 | 1131 | 1141 | 2850 | 2910 | 3038 | 9 | 82 |
| **2023** | 1002 | 1105 | 1120 | 2704 | 2794 | 2984 | 9 | 76 |
| **2024** | 944 | 1079 | 1099 | 2564 | 2682 | 2933 | 9 | 71 |
| **2025** | 889 | 1054 | 1079 | 2432 | 2577 | 2884 | 9 | 66 |
| **2026** | 838 | 1031 | 1060 | 2311 | 2480 | 2839 | 9 | 61 |
| **2027** | 793 | 1009 | 1043 | 2205 | 2394 | 2798 | 9 | 57 |
| **2028** | 755 | 989 | 1029 | 2116 | 2322 | 2762 | 9 | 54 |
| **2029** | 725 | 971 | 1016 | 2048 | 2265 | 2732 | 9 | 52 |
| **2030** | 704 | 956 | 1006 | 2003 | 2226 | 2708 | 9 | 50 |
| **2031** | 689 | 942 | 998 | 1971 | 2197 | 2688 | 9 | 49 |
| **2032** | 673 | 928 | 990 | 1940 | 2169 | 2669 | 9 | 49 |
| **2033** | 659 | 914 | 982 | 1911 | 2142 | 2650 | 9 | 48 |
| **2034** | 645 | 900 | 974 | 1884 | 2117 | 2631 | 9 | 47 |
| **2035** | 632 | 887 | 967 | 1857 | 2092 | 2614 | 9 | 47 |
| **2036** | 619 | 874 | 960 | 1833 | 2070 | 2597 | 9 | 46 |
| **2037** | 607 | 861 | 953 | 1810 | 2048 | 2581 | 9 | 46 |
| **2038** | 596 | 849 | 947 | 1788 | 2028 | 2565 | 9 | 46 |
| **2039** | 586 | 837 | 941 | 1768 | 2009 | 2551 | 9 | 45 |
| **2040** | 576 | 826 | 935 | 1749 | 1991 | 2537 | 9 | 45 |
| **2041** | 567 | 816 | 930 | 1733 | 1976 | 2525 | 9 | 44 |
| **2042** | 559 | 807 | 925 | 1717 | 1961 | 2513 | 9 | 44 |
| **2043** | 552 | 798 | 921 | 1704 | 1948 | 2503 | 9 | 44 |
| **2044** | 546 | 790 | 917 | 1692 | 1937 | 2494 | 9 | 43 |
| **2045** | 541 | 783 | 913 | 1682 | 1927 | 2486 | 9 | 43 |
| **2046** | 536 | 777 | 911 | 1674 | 1919 | 2480 | 9 | 43 |
| **2047** | 533 | 772 | 909 | 1667 | 1913 | 2474 | 9 | 43 |
| **2048** | 530 | 769 | 907 | 1663 | 1909 | 2471 | 9 | 42 |
| **2049** | 529 | 766 | 906 | 1660 | 1906 | 2468 | 9 | 42 |
| **2050** | 528 | 765 | 906 | 1659 | 1905 | 2468 | 9 | 42 |

The new coal (supercritical) and gas Combustion Turbine (CT) or Combined Cycle Gas Turbine (CCGT) projects are considered mature technologies. The cost projections are constant from 2020 to 2050. The coal costs are local economic proposals for new projects, and information is only available upon request. The mid and high costs in coal projects represent uncertainty in construction and equipment procurement. The CT from the 2022 ITC [22]. The 2022 ITC costs for CCGT are below the market price and lower than CT estimates and thus discarded. However, an upper bound for the CT estimations is available in the 2022 ITC; we use this estimation as the mid-scenario for CCGT. The cost projections for a new nuclear project is from the 2022 ITC [22]. These costs are similar with the assumptions in recent studies [41, 56]. Nuclear is a mature technology, we assume the same costs from 2020 to 2050 (Table 31).

**New Energy Storage Cost Projections.** The storage technologies investigated in this study are Pumped Storage Hydropower (PSH), Lithium-Ion batteries, and hydrogen. PSH and batteries are considered short-term storage, and hydrogen is long-term storage. The charge and discharge of short-term storage are optimized with a daily horizon, and long-term storage has a yearly optimization horizon.

The cost of PSH in a project is uncertain and highly depends on local economic factors, geographic conditions, and weather patterns [31, 22]. We have information about the costs in the 2023 ATB [38] and local estimations for PSH on-river and off-river in 2022 ITC and 2024 ITC-PS [22, 46]. 2023 ATB and the 2024 ITC Power Storage (PS) consider PSH a mature technology, and the costs will not decrease from 2020 to 2050. These sources provide the cost in terms of the power component in a PSH plant, and similarly to batteries, *GridPath-India* requires the cost of the power and energy components in a PSH project.

**Supplementary Table 31: Thermal technology cost projections**. Thermal generators are mature and thus technology cost are not projected to decline. The cost of a supercritical technology project may vary depending on different factors (e.g., environmental requirements, components lifetime), so two cost scenarios are considered (low and high).

| Technology | Scenario | Power Capacity (USD/kW) | O&M (USD/kW-yr) |
|---|---|---|---|
| **Nuclear** | **mid** | 1638 | 60 |
| **Supercritical Coal** | **low** | 1273 | 37 |
| **Supercritical Coal** | **high** | 1850 | 37 |
| **CCGT** | **mid** | 775 | 11 |
| **CT** | **mid** | 678 | 7 |

Another study provides the cost for the different components in a PSH project, but in the US [17]. Additionally, we have information about the auctioned PSH projects in India in 2023 [42], and we have costs for distinctly low ("Kurukutti PSH MP30"), mid ("Gandhi Sagar PSH"), and high ("Ghatghar PSH") cost projects. We linearly adjust the power and energy capacity costs (from the US) to derive low, mid, and high cost estimates with the auction information from India (Figure 26G and J, and Table 32). We consider PSH a mature technology similar to the 2023 ATB and 2024 ITC-PS. Therefore, we assume the cost projections from 2020 to 2030 are constant.

**Supplementary Table 32: Pumped Storage Hydro (PSH) technology cost projections**. PSH is a mature technology and projections costs are not expected to decline. However, the cost of a PSH project depends on the location. We propose three different project cost scenarios (low, mid and high).

| Scenario | Power Capacity (USD/kW) | Energy Capacity (USD/kWh) | O&M (USD/kW-yr) |
|---|---|---|---|
| **low** | 379 | 25 | 12 |
| **mid** | 458 | 30 | 13 |
| **high** | 528 | 36 | 15 |

The battery cost combines information from the 2024 ITC-PS and the 2022 ATB [46, 28]. We do not use the 2023 ATB for batteries since India did not observe the projected cost increase in 2024 [38]. The 2022 ATB cost projections are in terms of the power capacity (kW) of a 4 h duration storage, and *GridPath-India* optimizes power (kW) and energy (kWh) capacity. We calculate the cost projections for multiple storage times (2 h, 4 h, 6 h, and 8 h) and 1 MW and estimate the relation between energy (battery) and power capacity (electronic) in the 2022 ATB projections; see Figure 26H and K. Then, we implement a method available in the literature to adjust the battery cost projections in the ATB to cost projections in India, based on international auctions [23]; see Figure 26H and K. However, recent battery auctions in the IRENA database show that the battery cost in 2020 [50] was lower than the lower bound in 2050 estimated following this method. We linearly adjust the projections to incorporate this information (Figure 26H and K).

The power capacity component of these projections is still much lower than the 2024 ITC-PS [46], and the energy capacity component is much higher (Figure 26H and K). This shows a great disparity between the projections from different sources and the actual auction costs. Due to this, we are inclined not to trust these projections and relay auction costs, and ongoing tenders (FDRE, RTC, and GUVNL BESS) show that these cost estimates will also be high in 2030 [53]. We linearly adjust the upper and lower bounds from 2030 to 2050 and interpolate the trends using PCHUP from 2020. The mid estimate is the average between the upper and lower bounds (Table 33).

We assume hydrogen will be generated from water via electrolysis with a Proton Exchange Membrane (PEM) electrolyzer [14]. The hydrogen is liquified and stored in tanks or nonporous caverns, and the hydrogen is oxidized in fuel cells to generate electricity and water again. As of today, hydrogen open-cycle combustion turbines are cheaper than fuel cells [9]. However, hydrogen storage will not be a feasible alternative until 2030 or 2040, when the cost projections for fuel cells are lower than for combustion turbines.

The PEM electrolyzer base costs are from the 2024 ITC Green Fuels (GF) [45], but do not provide uncertainty. We also have prospect studies projections with upper and lower bounds [43, 12]. The initial point is NREL's projection (2020), similar to the 2024 ITC-GF (2020). The prospect provides extremely low [12] and high estimates in 2020 [43]. We discard these estimates and focus on the uncertainty bounded by the region between the mid and low estimates in [43] and the upper and mid estimates in [12] from 2030 to 2050. We define our lower bound (2035) as the lower bound in [12] and our upper bound as their mid estimates in 2035 and 2050. Our lower bound is linearly adjusted to match local auction prices for electrolyzers expected to come online before 2030 without considering incentives [48]; see Figure 26F and Table 34.

We used international projections of hydrogen storage in salt caverns from the US [17] and Denmark per lack of local information [33]; see Figure 26L. Previous studies have independently trusted these projections [39]. However, we

**Supplementary Table 33: Battery energy storage cost projections**. The cost of a battery project depends on the power capacity (kW) and the energy storage duration (kWh). Li-ion battery technology is in early stage of adoption and thus three different declining rates are considered (low, mid and high). Similarly, operation and maintenance (O&M) cost projections are uncertain, hence three scenarios are proposed (low, mid and high).

| | Power Cap. (USD/kW) | | | Energy Cap. (USD/kWh) | | | O&M (USD/kW-yr) | | | O&M (USD/kWh-yr) | | |
|---|---|---|---|---|---|---|---|---|---|---|---|---|
| **Period** | **low** | **mid** | **high** | **low** | **mid** | **high** | **low** | **mid** | **high** | **low** | **mid** | **high** |
| **2020** | 277 | 277 | 277 | 326 | 326 | 326 | 4 | 4 | 4 | 5 | 5 | 5 |
| **2021** | 243 | 249 | 258 | 286 | 293 | 303 | 3 | 3 | 4 | 4 | 4 | 4 |
| **2022** | 213 | 224 | 240 | 250 | 264 | 283 | 3 | 3 | 3 | 3 | 4 | 4 |
| **2023** | 185 | 201 | 225 | 218 | 237 | 264 | 3 | 3 | 3 | 3 | 3 | 4 |
| **2024** | 160 | 181 | 210 | 189 | 213 | 248 | 2 | 3 | 3 | 3 | 3 | 3 |
| **2025** | 139 | 164 | 198 | 164 | 192 | 233 | 2 | 2 | 3 | 2 | 3 | 3 |
| **2026** | 121 | 148 | 187 | 142 | 175 | 220 | 2 | 2 | 3 | 2 | 2 | 3 |
| **2027** | 105 | 136 | 178 | 124 | 160 | 209 | 1 | 2 | 2 | 2 | 2 | 3 |
| **2028** | 93 | 126 | 170 | 110 | 148 | 200 | 1 | 2 | 2 | 2 | 2 | 3 |
| **2029** | 84 | 118 | 164 | 99 | 139 | 193 | 1 | 2 | 2 | 1 | 2 | 3 |
| **2030** | 79 | 113 | 159 | 92 | 133 | 187 | 1 | 2 | 2 | 1 | 2 | 3 |
| **2031** | 76 | 110 | 156 | 89 | 130 | 184 | 1 | 2 | 2 | 1 | 2 | 3 |
| **2032** | 73 | 108 | 153 | 86 | 127 | 181 | 1 | 1 | 2 | 1 | 2 | 2 |
| **2033** | 70 | 106 | 151 | 83 | 124 | 177 | 1 | 1 | 2 | 1 | 2 | 2 |
| **2034** | 68 | 103 | 148 | 80 | 122 | 174 | 1 | 1 | 2 | 1 | 2 | 2 |
| **2035** | 65 | 101 | 145 | 77 | 119 | 171 | 1 | 1 | 2 | 1 | 2 | 2 |
| **2036** | 63 | 99 | 143 | 74 | 117 | 168 | 1 | 1 | 2 | 1 | 2 | 2 |
| **2037** | 61 | 97 | 140 | 72 | 114 | 165 | 1 | 1 | 2 | 1 | 2 | 2 |
| **2038** | 59 | 96 | 138 | 70 | 112 | 162 | 1 | 1 | 2 | 1 | 2 | 2 |
| **2039** | 57 | 94 | 136 | 67 | 110 | 160 | 1 | 1 | 2 | 1 | 2 | 2 |
| **2040** | 56 | 92 | 133 | 66 | 109 | 157 | 1 | 1 | 2 | 1 | 2 | 2 |
| **2041** | 54 | 91 | 131 | 64 | 107 | 154 | 1 | 1 | 2 | 1 | 1 | 2 |
| **2042** | 53 | 90 | 129 | 62 | 106 | 152 | 1 | 1 | 2 | 1 | 1 | 2 |
| **2043** | 52 | 89 | 127 | 61 | 104 | 150 | 1 | 1 | 2 | 1 | 1 | 2 |
| **2044** | 51 | 88 | 126 | 60 | 103 | 148 | 1 | 1 | 2 | 1 | 1 | 2 |
| **2045** | 50 | 87 | 124 | 59 | 102 | 146 | 1 | 1 | 2 | 1 | 1 | 2 |
| **2046** | 49 | 86 | 123 | 58 | 101 | 145 | 1 | 1 | 2 | 1 | 1 | 2 |
| **2047** | 49 | 86 | 122 | 57 | 101 | 143 | 1 | 1 | 2 | 1 | 1 | 2 |
| **2048** | 48 | 85 | 121 | 57 | 100 | 142 | 1 | 1 | 2 | 1 | 1 | 2 |
| **2049** | 48 | 85 | 120 | 56 | 100 | 141 | 1 | 1 | 2 | 1 | 1 | 2 |
| **2050** | 48 | 85 | 119 | 56 | 100 | 140 | 1 | 1 | 2 | 1 | 1 | 2 |

propose to combine both studies. These projections have a similar initial cost in 2020, but estimations for 2050 are considerably different. The lower estimates from the DEA are very similar to their mid-cost, so we use the mid-cost as our lower estimate for simplification [33]. The low cost from the US matches the mid-cost from the DEA. However, US high-cost estimates are 5 times larger than the mid-cost and thus inconsistent with the rest of the literature [17]. We define US estimates as high projections and DEA as low projections, and use their average as our mid projections. We interpolate upper, mid, and lower trends using PCHUP to create monotonically decreasing trends (Table 34).

The fuel cell projections follow the upper bound proposed in previous studies from 2020 to 2050 [51]. The mid-cost projections (2020-2030) are from targets set by the Department of Energy (DOE) and use actual fuel cell costs [35]. Notice that the initial projections based on the actual cost in 2020 considerably differ from previous studies [12]. The lower bound is the same lower bound from a recent study from 2030 to 2050 [26]. We do not use this lower bound from 2020 to 2030, as it considerably differs from the DOE study with actual costs. The mid projections from 2030 to 2050 are the average between the lower and upper bounds (Figure 26I and Table 34). We interpolate the points by applying PCHUP to define the trends.

**New Project Operational Characteristics.** The Operation and Maintenance (O&M) cost projections are from international estimates available on the 2023 ATB, the 2022 ITC, the 2024 ITC-PS, and the 2024 ITC-GF [22, 38, 46, 45]. We only consider mid-cost projections and the costs are assumed to decrease slightly from 2020 to 2050 for emerging technologies (hydrogen, storage, solar PV, and wind), except for offshore since the 2023 ATB indicates a considerable decrease [38], and the battery is steady (due to an increase in labor decrease in technology cost). The O&M cost from mature technologies is expected to increase (coal and nuclear) or remain constant (CT, CCGT, and PSH). In particular, we consider O&M capacity cost (USD/kW) for all technologies and O&M energy cost (USD/kWh) for battery storage (Table 29-34). We assume PSH for the energy component (USD/kWh) included in the power component. We assume the O&M cost for a hydrogen tank and cavern storage is equivalent they are in the PEM electrolyzer and fuel cell O&M

**Supplementary Table 34: Long-term storage (hydrogen) technology cost projections**. $H_2$ energy storage is in the initial phase of development and thus cost are uncertain. Three different declining cost trends are proposed (low, mid, and high). The $P2H_2$ project in an hydrogen ($H_2$) energy storage system has bidirectional power capacity (i.e., PEM for charging and $H_2$ fuel cell for discharging). The $H_2$ storage project utilized a tank or a salt cavern storage technology. Similarly, the operational and maintenance (O&M) are uncertain and propose three non-declining scenarios (low, mid and high).

| | Power Cap. (USD/kW) | | | Tank (USD/kWh) | | | Cavern (USD/kWh) | | | O&M (USD/kW-yr) | | |
|---|---|---|---|---|---|---|---|---|---|---|---|---|
| **Period** | **low** | **mid** | **high** | **low** | **mid** | **high** | **low** | **mid** | **high** | **low** | **mid** | **high** |
| **2020** | 1415 | 1415 | 1415 | 74 | 74 | 74 | 4 | 4 | 4 | 26 | 29 | 32 |
| **2021** | 1321 | 1340 | 1368 | 72 | 72 | 72 | 4 | 4 | 4 | 26 | 29 | 32 |
| **2022** | 1226 | 1266 | 1321 | 70 | 70 | 70 | 3 | 3 | 4 | 26 | 29 | 32 |
| **2023** | 1133 | 1194 | 1275 | 67 | 68 | 68 | 3 | 3 | 3 | 26 | 29 | 32 |
| **2024** | 1044 | 1123 | 1230 | 65 | 66 | 66 | 3 | 3 | 3 | 26 | 29 | 32 |
| **2025** | 959 | 1057 | 1186 | 63 | 64 | 64 | 3 | 3 | 3 | 26 | 29 | 32 |
| **2026** | 881 | 994 | 1145 | 60 | 62 | 63 | 3 | 3 | 3 | 26 | 29 | 32 |
| **2027** | 812 | 937 | 1106 | 58 | 60 | 61 | 3 | 3 | 3 | 26 | 29 | 32 |
| **2028** | 752 | 886 | 1070 | 56 | 58 | 60 | 3 | 3 | 3 | 26 | 29 | 32 |
| **2029** | 704 | 843 | 1036 | 54 | 56 | 58 | 3 | 3 | 3 | 26 | 29 | 32 |
| **2030** | 668 | 808 | 1006 | 52 | 54 | 57 | 2 | 3 | 3 | 26 | 29 | 32 |
| **2031** | 641 | 777 | 978 | 49 | 52 | 55 | 2 | 3 | 3 | 26 | 29 | 32 |
| **2032** | 614 | 747 | 950 | 47 | 51 | 54 | 2 | 3 | 3 | 26 | 29 | 32 |
| **2033** | 587 | 718 | 923 | 45 | 49 | 53 | 2 | 3 | 3 | 26 | 29 | 32 |
| **2034** | 562 | 689 | 896 | 44 | 48 | 51 | 2 | 3 | 3 | 26 | 29 | 32 |
| **2035** | 537 | 661 | 870 | 42 | 46 | 50 | 2 | 2 | 3 | 26 | 29 | 32 |
| **2036** | 514 | 633 | 844 | 40 | 45 | 49 | 2 | 2 | 3 | 26 | 29 | 32 |
| **2037** | 491 | 607 | 819 | 39 | 43 | 48 | 2 | 2 | 3 | 26 | 29 | 32 |
| **2038** | 470 | 582 | 795 | 37 | 42 | 47 | 2 | 2 | 3 | 26 | 29 | 32 |
| **2039** | 449 | 558 | 772 | 36 | 41 | 45 | 2 | 2 | 3 | 26 | 29 | 32 |
| **2040** | 430 | 536 | 749 | 35 | 39 | 44 | 2 | 2 | 3 | 26 | 29 | 32 |
| **2041** | 413 | 515 | 728 | 34 | 38 | 43 | 2 | 2 | 3 | 26 | 29 | 32 |
| **2042** | 397 | 496 | 708 | 33 | 37 | 42 | 2 | 2 | 3 | 26 | 29 | 32 |
| **2043** | 382 | 478 | 688 | 32 | 36 | 41 | 2 | 2 | 3 | 26 | 29 | 32 |
| **2044** | 370 | 463 | 670 | 31 | 35 | 40 | 2 | 2 | 2 | 26 | 29 | 32 |
| **2045** | 358 | 449 | 654 | 30 | 34 | 39 | 2 | 2 | 2 | 26 | 29 | 32 |
| **2046** | 349 | 438 | 638 | 29 | 33 | 38 | 2 | 2 | 2 | 26 | 29 | 32 |
| **2047** | 342 | 429 | 624 | 28 | 33 | 37 | 2 | 2 | 2 | 26 | 29 | 32 |
| **2048** | 337 | 422 | 612 | 28 | 32 | 36 | 2 | 2 | 2 | 26 | 29 | 32 |
| **2049** | 333 | 418 | 601 | 27 | 31 | 35 | 1 | 2 | 2 | 26 | 29 | 32 |
| **2050** | 332 | 417 | 592 | 27 | 30 | 34 | 1 | 2 | 2 | 26 | 29 | 32 |

costs.

**Supplementary Table 35: New projects operational characteristics**. The summarized characteristics include: project lifetime, fuel and variable O&M costs, financial discount rate, charging and discharging efficiency of a storage system, thermal generator minimum stable capacity, minimum up and down time, and start-up cost. We indicate when public data from Tariff Orders (TO) or the MERIT program is available.

| Technology | Lifetime (yr) | Fuel (USD/MWh) | Discount Rate (%) | Min. Capacity (%) | Efficiency In/Out (%) | Min. Up/Down Time (hr) | Start-up Cost (USD/MWh) |
|---|---|---|---|---|---|---|---|
| **Single-axis solar PV** | 25 | 0 | 9 | - | - | - | - |
| **Fixed-tilt solar PV** | 25 | 0 | 9 | - | - | - | - |
| **Roof-top solar PV** | 25 | 0 | 9 | - | - | - | - |
| **Onshore wind** | 25 | 0 | 9 | - | - | - | - |
| **Offshore wind** | 30 | 0 | 9 | - | - | - | - |
| **Supercritical coal** | 45 | state-specific | 9 | 60 | - | 12/12 | 416.67 |
| **CCGT** | 35 | 94.93 | 9 | 50 | - | 6/6 | 108.08 |
| **CT** | 35 | 148.73 | 9 | 40 | - | 1/1 | 108.08 |
| **Nuclear** | 30 | state-specific | 9 | 90 | - | - | - |
| **Battery** | 15 | 0 | 9 | - | 92/92 | - | - |
| **Hydrogen** | 20 | 0 | 9 | - | 75/50 | - | - |
| **PSH** | 40 | 0 | 9 | - | 87/87 | - | - |

The technical and financial characteristics of the new projects are in Table 35. The fuel assumptions for new thermal generators are publicly available in state-level tariff orders or published by the Ministry of Power through the "Merit Order of Despatch for the Rejuvenation of Income and Transparency (MERIT) program". The fuel and variable O&M costs (USD/MWh) for a new thermal project are the average of the generators with the same fuel type in the state of the project. However, we consider that a new gas project will not consume local gas. The cost of gas is uncertain, but we assume international liquefied gas costs for simplification. We annualized the capital investment costs with a discount rate of 9% and aggregated the fixed O&M cost projections (kW/yr) for each technology. The minimum capacity factor and minimum up/down time are generic values for thermal generators (coal, gas, nuclear). CT technology is more flexible than CCGT. Storage efficiencies and project lifetime are current estimates and do not consider improvements over time. We assume different start-up cost (USD/MWh) from the 2022 ITC for coal (subcritical small and large, and supercritical) and gas (CT and CGT) projects but the same for existing and new projects [22]; see Table 35 and 36.

**Specified Projects.** The annualized capital investment and the fuel plus variable costs O&M (USD/MWh) for each specified thermal project (coal, gas, and oil) are publicly available through the MERIT program. This research considers generations scheduled to come online between 2020 and 2025 (nuclear, coal, and hydroelectric). The hydroelectric projects in the pipeline are ROR. For these projects, we consider 2022 ITC costs [22]; see Table 36. Similarly, specified biomass/bagasse is from 2022 ITC costs [22]. This investigation does not differentiate between bagasse and biomass projects. We did not find concise information for Waste Heat Recovery (WHR) projects in India [29]. Thus, we use the information from a recently published viability study for WHR projects in Pakistan. We do not propose any new project on top of the capacity in the pipeline for ROR technology.

**Supplementary Table 36: Specified projects operational characteristics**. The summarized characteristics include: project lifetime, investment capital cost, fuel and variable O&M costs, financial discount rate, charging and discharging efficiency of a storage system, thermal generator minimum stable capacity, minimum up and down time, and start-up cost. We indicate when public data from Tariff Orders (TO) or the MERIT program is available.

| Technology | Lifetime (yr) | Cost (USD/kW) | Fuel (USD/MWh) | Discount Rate (%) | Min. Capacity (%) | Efficiency In/Out (%) | Min. Up/Down Time (hr) | Start-Up Cost (USD/MWh) |
|---|---|---|---|---|---|---|---|---|
| **PSH** | 40 | 1079.97 | - | 9 | - | 85/85 | - | - |
| **ROR hydro** | 40 | 2236.40 | - | 9 | - | - | - | - |
| **Storage hydro** | 40 | 2755.39 | - | 9 | - | - | - | - |
| **Fixed-tilt solar PV** | 25 | 789.13 | - | 9 | - | - | - | - |
| **Onshore wind** | 25 | 1072.60 | - | 9 | - | - | - | - |
| **Subcritical coal (large)** | 45 | TO/MERIT | TO/MERIT | 9 | 55 | - | 12/12 | 416.67 |
| **Subcritical coal (small)** | 45 | TO/MERIT | TO/MERIT | 9 | 55 | - | 12/12 | 416.67 |
| **Supercritical coal** | 45 | TO/MERIT | 9 | 60 | - | 12/12 | 416.67 | |
| **Nuclear** | 30 | 2301.85 | TO/MERIT | 9 | 90 | - | - | - |
| **CCGT** | 35 | 958.18 | TO/MERIT | 9 | 50 | - | 6/6 | 108.08 |
| **Diesel** | 25 | 668.35 | TO/MERIT | 9 | 40 | - | 1/1 | - |
| **WHR** | 30 | 958.17 | 53.86 | 9 | 60 | - | 12/12 | - |
| **Biomass** | 20 | 818.02 | 69.54 | 9 | - | - | 12/12 | - |

The specified solar PV projects are utility-scale fixed-tilt technology, and the wind projects are onshore wind with the typical model turbine model 5 to 10 years ago (Vesta V80 with a hub at 80 m high). We aggregate fuel and variable O&M cost (USD/MWh), annualized capital investment with a 9% discount rate, and aggregate fixed O&M cost (USD/kW).

We use average exchange rates from 2020 to preserve the infrastructure expansion investment's NPV in 2020 USD. The EUR to USD exchange rate is 1.142, and the USD to IND is 72.

## 5 Demand Projection Scenarios

The demand projections are based on the model Perspectives on Indian Energy based on Rumi version 2.0 (*PIER v2.0*), which estimates the electricity demand from cost-optimal supply options at the wide energy system level [55]. *PIER v2.0* demand projections are alternative to the demand projections published on the Electric Power Survey (EPS) [21]. The main difference is in the assumption of the solarization of the agricultural sector in *PIER v2.0* projections and climate change impacts on the residential sector. *PIER v2.0* demand projections include state-level captive generation demand projections.

The temperature projections used in *PIER v2.0* for the bottom-up demand projections for the residential sector are based on the *SSP2-4.5* scenario [7, 8], from the *IPSL-CM6A-LR* climate model [15], available on the *CMIP6* dataset [19]. The *SSP2-4.5* scenario combines the Shared Socioeconomic Pathway 2 (SSP2) development scenario ("middle of the road"), with the assumption that Earth's surface will receive an additional increase of $4.5W/m^2$ radiation forcing by 2100. This scenario assumes development and income growth persist in being unequal, leading to a moderate increase in GHG emissions by 2050 and eventually achieving net-zero by 2100.

The temperature projections from the *SSP2-4.5* scenario are used to determine population-weighted cooling degree hours and cooling hours for households modeled at the state urban/rural granularity. The cooling degree hours and cooling hours are used to determine electricity demand due to air conditioners and fans/coolers, respectively, based on the appliance characteristics. Climate impacts are not considered in any other sectors.

*PIER v2.0* demand projections are at the consumer (i.e., ex-bus), and do not consider transmission and distribution (T&D) losses. Projections of transmission losses are available at the state level. We project them to FY50 and linearly adjust the projections to account for different transmission losses for each FY and state. The losses are projected to FY50 based on past trends of loss reduction, assuming a maximum reduction rate of 3% per year to reach a minimum T&D loss of 7% at the state level. The historical T&D loss data is available in the "All India Electricity Statistics - General Review" reports [30].

*GridPath-India* includes transmission corridors and losses due to energy exchanges. It does not account for all losses at the transmission and distribution level. The transmission losses modeled by *GridPath-India* are approximately 2.5% at the national level. We subtract this number from the transmission losses projections (before linearly adjusting the profiles) to properly characterize the additional generation necessary to balance the transmission losses.

We have three scenarios in *PIER v2.0*. The different narratives are explained below.

- "Reference" (mid growth): This is a combination of the sector-specific reference scenarios, and represents our best guess of what is likely to happen in the future based on past and likely future trends.
- "Vikasit Bharat" (high growth): This scenario can represent the equivalent of a *sustainable development* scenario. It reflects the aspirations of a *developed India* by 2047 (FY48) announced by the Government of India. We assume that this also means a more equitable development, with greater focus on sustainability.
- "Vichalit Bharat" (low growth): This scenario is a counterpoint to "Vikasit Bharat" and represents a scenario in which development is a bit more haphazard. Economic growth is lower than the "Reference", it is less equitable, and it is also less environmentally sustainable due to lower investments in efficiency and new technologies.

Further details about the bottom-up assumptions in each economic sector for each scenario ("Vikasit Bharat", "Reference", and "Vichalit Bharat") in *PIER v2.0* are in Table 37.

**Supplementary Table 37: Demand projection scenarios**. Description of the assumptions behind the different scenarios proposed in *PIER v2.0*.

| Component | Reference | Vikasit Bharat | Vichalit Bharat |
|---|---|---|---|
| **GDP growth** | • Per-capita gross domestic product (GDP) growth rate (GR) based on IMF projections until 2027 and continued at the 2027 rate throughout the model period. | • Per-capita GDP growth is increased so that it reaches USD 5,000 by FY48 (still falling in Upper Middle Income country category).<br>• This is done so that "poorer states" (those with lower per-capita GSDP) increase faster than average, and richer states increase slower than average. | • Per-capita GDP growth in this scenario is kept at the same rate as it was in the FY14-FY24 period (which included demonetization, pandemic, etc.). This is lower than in the "Reference" scenario.<br>• State per-capita GDP and GRs are kept the same as in the past decade, so the same inequities will carry forward. |
| **Residential** | • Penetration of appliances based on assumed gross state domestic product (GSDP) GRs.<br>• Number of appliance instances per household grows as per past trends.<br>• Specific energy consumption same as likely efficiency trend (LET):<br>– Efficiency standards are revised every four years for all appliances except fans (revised once in six years).<br>– "Actual" appliance efficiency is 60% of the notified efficiency.<br>– Gradual shift to higher star-rated appliances. | • Higher penetration level of appliances due to higher GDP GR.<br>• Share of ACs in AC+cooler is the same for rural and urban households.<br>• Increased number of appliance instances per household.<br>• Improve SEC similar to desired efficiency trend (DET):<br>– Efficiency standards are revised once in four years for all appliances.<br>– "Actual" appliance efficiency is 80% of notified efficiency.<br>– Slightly faster shift to higher star-rated appliances.<br>– Electricity demand for non-modeled residential energy services increases slightly faster than the "Reference". | • Penetration of appliances follows from the revised GSDP GRs.<br>• Number of appliance instances per household grows a little slower than the "Reference".<br>• Appliance lifetime is a little greater than the "Reference" for new purchases.<br>• Efficiency standard revision is a little slower than LET.<br>• "On-field" efficiencies of appliances are a little worse than LET.<br>• Electricity demand for non-modeled residential energy services is the same as the "Reference", assuming fewer appliances but with lower efficiency. |
| **Transport** | • Same as Reference. | • Lower activity for private vehicles (4W and 2W).<br>• Higher public transport stock and flow than Ref, and with higher load factors and activity.<br>• Electrification and SEC similar to the "Improve" scenario.<br>• Net result should be a higher modal share for passenger transport (PT) and non-motorised transport (NMT), with perhaps a little higher passenger-km for short distance transport (SDT).<br>• For long-distance transport (LDT), there would be higher passenger-km corresponding to more holidays and business travel, but with modal shifts to rail from bus, and an increase in GRs of air and rail.<br>• Freight: an increase in tonne-km compared to Base to account for higher GDP, but with a greater share for rail. On roads, the number of vehicles increases, GR for passenger-km in rail increases, the same for air, whereas water remains the same.<br>• Freight SECs similar to improve. | • Lower vehicle flow rate due to lower purchases— to roughly mirror "Vikasit Bharat". The GRs of the private vehicle reduce stock.<br>• For both Passenger and Freight, higher vehicle lifetimes in Internal combustion engine (ICE) - stock is the same as Reference.<br>• Per-vehicle activity same.<br>• Electrification, SEC improvement slightly worse than Reference.<br>• Modal share of PT, NMT increase (and overall passenger-km is lower).<br>• Modal share of intermediate public transport (IPT) - also stands for taxis, cabs, and other hired transport, reduced.<br>• LDT: All stock reduces while the modal share remains roughly the same.<br>• Freight: Lower flow and greater lifetime for trucks, greater activity – to make sure that freight tonne-km is appropriate for the GDP. Modal share remains similar.<br>• SEC improvement, electrification is a little slower than the "Reference". |
| **Bottom-up industry** | • Levers tweaked: production, specific energy consumption (SEC), and technology adoption.<br>• Continue past trends and announced targets of production, SEC, and technology adoption. | • Production & per capita consumption is higher than the "Reference" by 2041.<br>• SEC improvement rate is higher than the "Reference".<br>• Higher share for electric arc route (EAR) in steel compared to blast furnace-basic oxygen furnace (BF-BOF) - iron and steel making technology largely using coking and thermal coal.<br>• Higher share for blended cements.<br>• Higher uptake of decarbonization technologies such as green hydrogen (G-$H_2$) and Carbon Capture Utilization and Storage (CCUS) as compared to the "Reference". | • Production & per capita consumption is lower than the "Reference" by 2041.<br>• SEC improvement rate is the same as the "Reference" for steel, and lower for cement and aluminium.<br>• Lower uptake than the "Reference" of G-$H_2$ and CCUS. |
| **"Other" industry and "Other" sectors** | • Same elasticity of electricity demand GR for GDP GR as in the past (with adjustments for efficiency, electrification, etc.) | • Elasticity of electricity demand GR for GDP GR reflects a combination of increased electrification and better efficiency than the "Reference". | • Elasticity of electricity demand GR for GDP GR based on a combination of lower electrification and lower efficiency improvement (compared to the "Reference"). |

## 6 Demand Profiles

The demand profiles from *PIER v2.0* are an the average day ($h = \{1, \ldots, 24\}$) for each month ($m = \{1, \ldots, 12\}$) in an investment period ($\pi = \{2024, \ldots, 2040\}$). The average demand monthly profiles ($\bar{D}_{h,m,\pi,z}$) are at the state level ($z = \{1, \ldots, 34\}$), i.e., a profile per load zone. The VRE capacity factors have hourly resolution ($\tau = 1, \cdots, 8,760$) and are from FY19. We propose to weather-synchronize the demand profiles ($\hat{D}_{\tau,\pi,z}$) to obtain hourly profiles for each investment period.

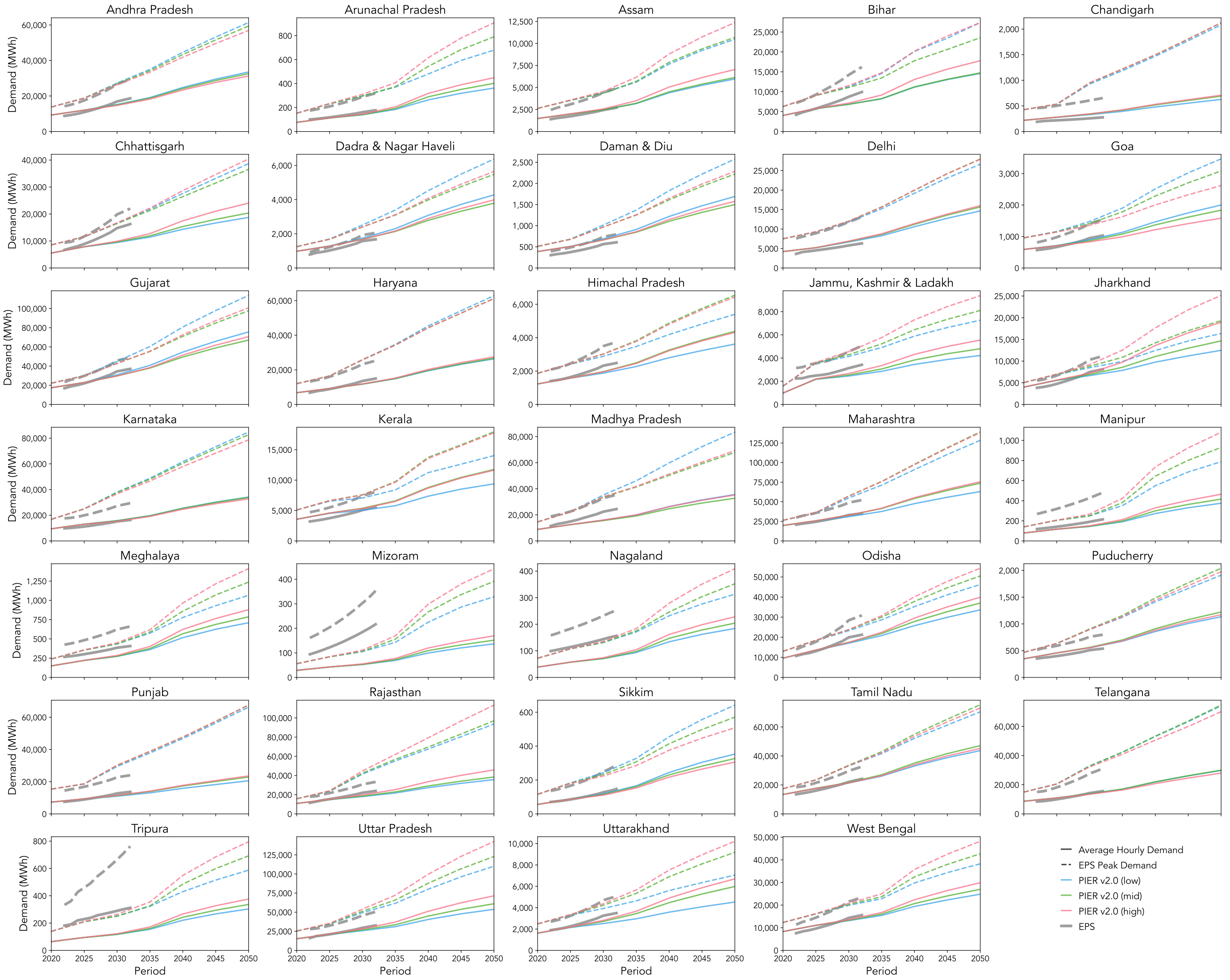


**Supplementary Figure 27. 2020-2050 average and peak demand in *PIER v2.0* "bottom-up" and *FY19 ICED* "linearly-scaled" (mid) scenario in each state compared to Electric Power Survey**. The colors represent a different month in the Indian fiscal year. The states are the load zones considered in the *GridPath-India* model.

Originally, the profiles from *PIER v2.0* are from FY24 to FY41. However, we need demand profiles until FY50. For that, we linearly project the demand for each hour in a month at the state level ($\bar{D}_{h,m,\pi,z}$). The proposed linear model is,

$$\bar{D}_{h,m,\pi,z} = \beta_{h,m,z} \cdot \pi + \alpha_{h,m,z} + \epsilon_{h,m,z}, \tag{1}$$

where $\beta_{h,m,z}$ is the coefficient, $\alpha_{h,m,z}$ is the intercept, and $\epsilon_{h,m,z}$ is the error. We predict the profiles ($\pi^\star = \{2040, \ldots, 2050\}$), after estimating $\beta_{h,m,z}$ and $\alpha_{h,m,z}$ as $\bar{D}_{h,m,\pi^\star,z} = \beta_{h,m,z} \cdot \pi^\star + \alpha_{h,m,z}$. We call the whole set $\bar{D}_{h,m,\pi,z}$ for simplification, but it includes $\bar{D}_{h,m,\pi,z}$ and $\bar{D}_{h,m,\pi^\star,z}$. These are the "bottom-up" demand profiles.

We also propose normalizing the *FY19 ICED* daily demand profiles using the *PIER v2.0* profiles to weather-synchronize the demand projections. There are the "linearly-scaled" demand profiles. The *FY19 ICED* profiles have hourly timepoints ($D_{\tau,\pi=2019,z}$), and we have a demand profile for each load zone. We apply the following formula to

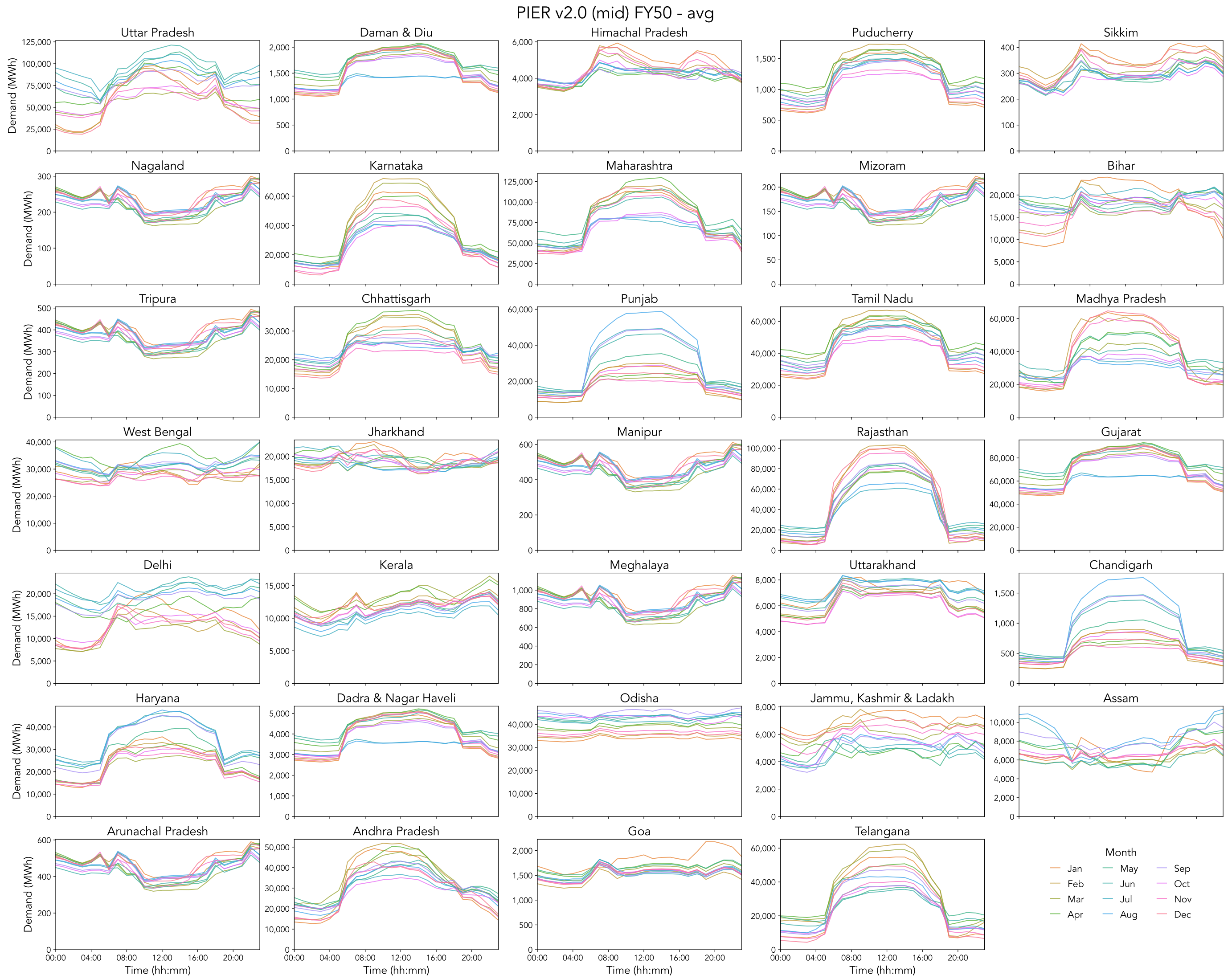


**Supplementary Figure 28. 2050 average monthly demand in *PIER v2.0* "bottom-up" (mid) scenario in each state**. The colors represent a different month in the Indian fiscal year. The states are the load zones considered in the *GridPath-India* model.

the *PIER v2.0* profiles ($\bar{D}_{h,m,\pi,z}$), and obtain weather-synchronized ($\tilde{D}_{\tau,\pi,z}$) profile projections,

$$\tilde{D}_{\tau,\pi,z} = D_{h,m,\pi,z} \frac{\bar{D}_{\tau_{h,m},\pi=2019,z}}{\frac{1}{\mathcal{C}(\tau_{h,m})} \sum_{\tau' \in \tau_{h,m}} \bar{D}_{\tau',\pi=2019,z}}. \tag{2}$$

where $\mathcal{C}(\cdot)$ represents the cardinality in a set, $\tau_{h,m}$ represents the set containing the $\tau'$ timepoints for each hour $h$ in month $m$.

As *PIER v2.0* profiles are from FY24 to FY41, so we adjust the peak and total energy of the profiles to match the peak ($P_{\pi^\star,z}$), and total demand ($E_{\pi^\star,z}$) reported from FY20 to FY23 ($\pi^\star = \{2020,\ldots,2023\}$). For that, we use the *FY19 ICED* profiles at the state level (i.e., load zone) and propose a two-step approach to project the profiles.

First, linearly adjust the total energy in the missing periods $\pi^\star$,

$$\tilde{D}_{\tau,\pi^\star,z} = E_{\pi^\star,z} \cdot \frac{D_{\tau,pi=2019,z}}{\sum_{\tau} D_{\tau,\pi=2019,z}}. \tag{3}$$

Second, we adjust the peak demand without changing the total energy for all periods $\pi$,

$$\tilde{D}_{\tau,\pi,z} = \text{sort}^{-1}\left[\delta_\tau \cdot \left(P_{\pi,z} - \max\left[\tilde{D}_{\tau,\pi,z}\right]\right) + \text{sort}\left[\tilde{D}_{\tau,\pi^\star,z}\right]_{\tau',\pi,z}\right], \tag{4}$$

where sort[·] is a function that sorts the elements in a vector in ascending order, and $\text{sort}^{-1}[\cdot]$ is a function that restores the vector to the original order. $\delta$ is a set with $\tau$ 8,760 linearly increasing elements ($\delta = -1, \cdots, 1$). $\tilde{D}_{\tau,\pi,z}$ is the set of weather-synchronized demand profiles from all periods ($\pi = 2020, \ldots, 2050$)

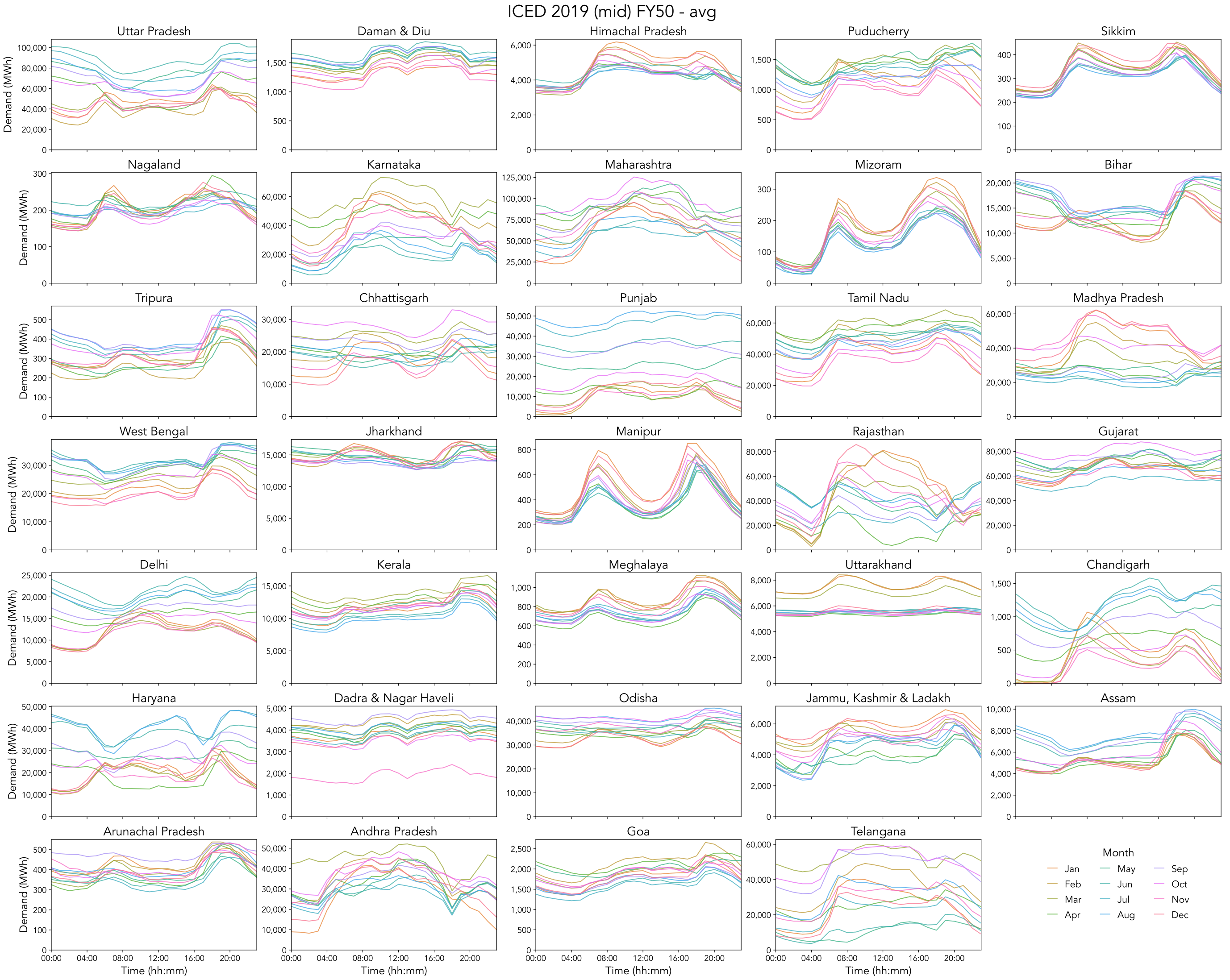


**Supplementary Figure 29. 2050 average monthly demand in *FY19 ICED* "linearly-scaled" (mid) scenario in each state**. The colors represent a different month in the Indian fiscal year. The states are the load zones considered in the *GridPath-India* model.

We compared the demand projections in *PIER v2.0* with the demand projections in the Electric Power Survey (EPS) [21]; see Figure 28. The EPS projections also include captive electricity demand.

## 7 Capacity Credits

The top net-load hours (TNLH) method is used to approximate the capacity credits (CC) for VRE assets (i.e., projects). For this method, it is necessary to implement a 2-year resource adequacy study based on the cumulative distribution of the net-load to approximate capacity credits of a VRE asset. We propose an 18-year resource adequacy to improve the accuracy.

To implement this approach, we first sort the net load timepoints ($\tau = \{1, \dots, 8760\}$) in each resource adequacy year ($y = \{1, \dots, 18\}$) in descending order. Then, we select the first $T = 250$ samples, and obtain the capacity $B_{\gamma,y,\pi}$ and generation $O_{\gamma,y,\pi}$ of each VRE generator $\gamma \in \Gamma^{\omega}$ in a given investment period $\pi$. Then, we calculate the NLCC $capacity_credit_{\gamma,y,\pi}$ of project $\gamma$, and average for each year $y$,

$$capacity_credit_{\gamma,\pi} = \frac{1}{Y} \sum_{y} \left( \frac{\sum_{\tau'} O_{\gamma,\tau',y,\pi}}{\sum_{\tau'} B_{\gamma,y,\pi}} \right) = \frac{1}{Y} \sum_{y} \left( \frac{1}{T} \sum_{\tau'} capacity_factor_{\gamma,\tau',y} \right). \tag{5}$$

If $B_{\gamma,\tau',y,\pi}$ is constant in all the samples in the set $\tau' \in T$ (i.e., there are no forced or unforced outages), the TNLH $capacity_credit_{\gamma}$ for generator $\gamma$ is equivalent to the average capacity factors ($capacity_factor_{\gamma,\tau',y,\pi}$) on timepoints $\tau'$ in year $y$.

**Weather-Synchronized Capacity factors Time Series.** We utilize weather data from FY19 to weather-synchronize the capacity factors with the demand time series. However, estimating the capacity credits requires weather data from multiple years [20].

As the solar radiation data from the PSM V3 model is only available from 2017 to 2019, we combined these data with data from the SUNY International model available from 2010 to 2014 and maintained by NREL [13]. SUNY has a spatial resolution of $10 \times 10$ km and hourly temporal resolution (60 min). The spatial resolution is lower than PSM V3 but higher than ERA5 [16], so we assume we do not need bias correction. However, the onshore and offshore time series are from ERA5 (2010 to 2014 and 2017 to 2019), and we applied the same bias correction explained in Section 3.

In addition to wind speed, we use other weather features from ERA5 (precipitation, temperature, humidity, and atmospheric pressure) combined with elevation data from GMTED2010 maintained by U.S. Geological Survey [4] to predict hydro ROR capacity factor time series and hydro storage month availability for 2010 to 2014 and 2017 to 2019. The proposed model combines topological details and weather information to apply a distance-based weight and learn local non-linear water runoff and evaporation rate patterns.

**Weather-Synchronized Demand Time Series.** The demand profiles also require weather synchronization to calculate the capacity credits effectively. We utilized the method proposed in Section 6 to generate profiles for 2030, 2040, and 2050 from *PIER* projections, and we weather synchronized with the demand profiles from FY18 and FY19 provided by ICED [36]. The demand profile synchronization predicts state-level demand profiles with the weather features from ERA5 utilized to predict the hydro ROR and storage capacity factors. In this case, we proposed a model that combines the weather features with the "Gridded Population of the World" dataset (GPWv4) [10] to weight the observations in the learning function by space coordinate with a higher population.

## 8 Electricity System Representation

**Geographical Representation.** The electricity transmission network in India (i.e., the national grid) is formed by five interconnected regional power grids: Northern, Eastern, Western, Northeastern, and Southern. The national grid has a synchronous interconnection with the neighboring country of Bhutan. The "National Load Despatch Centre" and the "Regional Load Despatch Centres" coordinate the operations at the national and regional levels, respectively. The "State Load Despatch Centres" balance the grids at the state and union territory level (Supplementary Table 21).

We model 35 load zones where each state and union territory that is part of contiguous India is considered a separate load zone with the following exceptions (Supplementary Table 21). Union territories of Jammu and Kashmir and Ladakh are combined into a single zone. Dadra Nagar Haveli and Diu and Daman are administratively a single union territory but because they are geographically discontinuous, they are treated as separate nodes in this model. In addition, we include the country of Bhutan as an energy exporting load zone but assume its own electricity demand to be zero in the model to separate energy exporting generators from generators meeting domestic demand.

**Temporal Representation.** We select the median day per month to capture seasonal trends and the peak demand day per month to capture extreme regional characteristics. We assign them a different weight ($\tau$) in the optimization function. The weight for peak day is $\tau = 1$ and weight for median day is $\tau = N_m - 1$, where $N_m$ is the number of days ($N$) in the month ($m = \{1, \ldots, 12\}$). We have 4 periods with $4 \times 2 \times 12 = 96$ representative days and $96 \times 24 = 2304$ total time points. Days are selected independently for each investment period and demand scenario (Supplementary Note 6).

**Transmission Representation.** The transmission losses in existing and new transmission corridors depend on the distance between the two interconnected load zones, and it is proportional to the energy transfer. We define our transmission loss factor as 0.016 %/km. The new transmission capacity costs are divided into line and substation costs. The average transmission line cost is 0.96 USD/kW-km, and the average substation cost is 22.5 USD/kW in India. We assume 50 years as the average transmission line lifetime. In addition, we exogenously add existing transmission capacity cost of USD 5.79 Billion, reported by the Central Electricity Regulatory Commission for FY22 in their annual report on "Short-Term Power Market in India" [47].

The "General Review" annually published by CEA contains information about electricity theft in transmission losses [44]. We estimate total electricity theft using the simulated transmission losses and the transmission losses reported by CEA (Supplementary Table 25). The transmission losses report shows a steady decrease in losses in the last 10 fiscal years. We projected electricity theft and included demand projections (Supplementary Table 26).

## 9 Availability Factors

The objective is to seasonally derate the dispatchable generators ($\gamma \subset \Gamma^{\mu} \cap \Gamma^{\phi}$) by their respective fuel availability and forced and unforced maintenance (i.e., outage rates). CEA provides information about the outage rates ($outage_rate_{\gamma,\tau}$) in their daily outage reports. We have historical outage reports from multiple periods ($\pi^{\star} = \{2018, \ldots, 2023\}$). The availability factors ($availability_factor_{\gamma,\tau,\pi^{\star}}$) for a generator $\gamma$ with capacity ($B_{\gamma}$) can be derived from the outage rates,

$$availability_factor_{\gamma,\tau,\pi^{\star}} = 1 - \frac{outage_rate_{\gamma,\tau,\pi^{\star}}}{B_{\gamma,\pi^{\star}}}. \tag{6}$$

Similarly, the plant load factor ($plant_load_factor_{\gamma,\tau,\pi^{\star}}$) for a generator $\gamma$ can be derived from historical generation reports available for multiple periods ($\pi^{\star}$). Now, we can linearly model the relation between the availability factors and plant load factors, using average availability factor ($availability_factor_{\gamma}$) and plant load factor ($plant_load_factor_{\gamma}$),

$$\alpha_{\gamma} = \frac{\frac{1}{\mathcal{C}(\pi^{\star})\cdot\mathcal{C}(\tau)} \sum_{\pi^{\star}} \sum_{\tau} capacity_factor_{\gamma,\tau,\pi^{\star}}}{\frac{1}{\mathcal{C}(\pi^{\star})\cdot\mathcal{C}(\tau)} \sum_{\pi^{\star}} \sum_{\tau} plant_load_factor_{\gamma,\tau,\pi^{\star}}}. \tag{7}$$

However, we want to derate a generator ($\gamma$), assuming it operates at maximum plant load factor ($plant_load_factor_{\gamma}^{\max}$), which is equivalent to the maximum feasible availability factor for the generator. To do that, we introduce an incremental term in the plant load factor ($\Delta plant_load_factor_{\gamma}$) to model the increase in the average plant load factor necessary to reach the maximum plant load factor $plant_load_factor_{\gamma}^{\max} = plant_load_factor_{\gamma} + \Delta plant_load_factor_{\gamma}$. Equivalently, we want to calculate the maximum feasible availability factor ($availability_factor_{\gamma}^{\max}$) assuming a linear decrease in the availability factor ($availability_factor_{\gamma}$) when a generator is operated at maximum plant load factor ($availability_factor_{\gamma}^{\max} = availability_factor_{\gamma} - \alpha_{\gamma} \cdot \Delta plant_load_factor_{\gamma}$).

Applying the equal equilibrium equation that assumes $availability_factor_{\gamma}^{\max} = plant_load_factor_{\gamma}^{\max}$, we obtain

$$\begin{aligned} availability_factor_{\gamma} - \alpha_{\gamma} \cdot \Delta plant_load_factor_{\gamma} &= plant_load_factor_{\gamma} + \Delta plant_load_factor_{\gamma} \\ \Delta plant_load_factor_{\gamma} &= \frac{availability_factor_{\gamma} - plant_load_factor_{\gamma}}{1 + \frac{availability_factor\gamma}{plant_load_factor_{\gamma}}}. \end{aligned} \tag{8}$$

The availability factor for a generator ($\gamma$) is equal to the historical plant load factor plus the increase expected if it were operated at the maximum plant load factor,

$$availability_factor_{\gamma}^{\max} = plant_load_factor_{\gamma} + \Delta plant_load_factor_{\gamma}. \tag{9}$$

In *GridPath-India*, we estimate seasonal availability factors calculating the availability factor ($availability_factor_{\gamma,\tau'}$) and plant load factor ($plant_load_factor_{\gamma,\tau'}$) with the time points ($\tau'$) corresponding to a given season (winter, summer, monsoon and post-monsoon). The availability factors are calculated separately for each technology (e.g., nuclear, subcritical coal, supercritical coal, CCGT, CT, or diesel) existing in a load zone ($z$). For that, the outage rates ($\sum_{\gamma \subset \Gamma_z^{\mu} \cap \Gamma_z^{\phi}} outage_rate_{\gamma,\tau,\pi^{\star}}$) and capacity ($\sum_{\gamma \subset \Gamma_z^{\mu} \cap \Gamma_z^{\phi}} B_{\gamma,\pi^{\star}}$) in Eq. (6) are aggregated for each technology in a load zone.

These availability factors apply to nuclear and coal generators, as fuel availability is not a constraint in India. However, diesel and gas generators generally have fuel availability constraints. We define the maximum availability factor for all dispatchable generators ($\gamma \subset \Gamma^{\mu} \cap \Gamma^{\phi}$) as,

$$availability_factor_{\gamma}^{\max} = \begin{cases} 0.5 & plant_load_factor_{\gamma} \leq 50 \\ 0.25 & plant_load_factor_{\gamma} \leq 25 \\ 0 & plant_load_factor_{\gamma} = 0 \end{cases} \tag{10}$$

The availability factors for storage technologies ($\gamma \subset \Gamma^{\sigma}$) are provided in the 2024 ITC Power Storage [46]. The availability factor for battery and PSH is 0.99, whereas 0.98 is for hydrogen storage.

## 10 Operational Reserves Projections

There are three levels of operational reserves [2]. Primary reserves are provided by online generators and are automatically dispatched (less than 1 sec) for grid stability. Secondary reserves are spinning reserves provided by a fraction of the committed generation and must respond quickly to imbalances between demand and supply (1 to 5 min). Tertiary reserves are non-spinning reserves for frequency regulation, activated to restore the secondary reserves and must respond in 5 to 15 min. We do not model primary reserves because they are provided by running generators, and their response is faster than the temporal resolution of our model.

The reserve requirements are determined based on the procedure available in "Detailed Procedure on Interim Methodology for Estimation of Reserves Under CERC (Ancillary Services) Regulations, 2022" [24]. In this manual, the description of the secondary and tertiary reserves is equivalent to the spinning reserves ($Freq$) and frequency response ($Spin$) functionalities available in *GridPath*. The secondary and tertiary reserve requirements are specified quarterly by CEA. The requirements are defined at the state-level (i.e., load zone $z$) and regional-level ($r = \{1, \cdots, 5\}$).

We calculate the reserve requirements based on secondary and tertiary reserves estimation implemented in the "Quarter-Ahead SRAS and TRAS Reserve Requirement for Quarter 2 (July - September) of the Year 2023-24 (Scaled for Reference Contingency of 4500 MW)" [37]. In this report, the secondary reserves are determined based on the positive and negative regulations in the last quarter, the peak demand, and the state generation at the peak demand. The tertiary reverses are determined based on the secondary reserves and the largest unit capacity in that state. This information is not available for future fiscal years, so we assume unit capacity remains constant—thus no larger units will be commissioned in future investment periods— and that the magnitude of the regulations (i.e., net demand forecasting error) will grow as a function of the peak demand for simplification. Additionally, we project the reserve requirements for each investment period, instead of quarterly.

Therefore, we propose to model the state-level secondary reserves in *GridPath-India* as frequency regulation ($Freq$). We calculate $Freq_{\tau,\pi,z}$ with the ratio between the state (i.e., load zone) secondary reserves requirement in the FY24 Jul. to Sep. report and the peak load $\rho_z = Freq_{\pi=FY24,z} / \max\left[D_{\tau,\pi=FY24,z}\right]$ for each zone ($z$) in the investment periods ($\pi$) from FY20 to FY50. In this way, we project the secondary reserves for each time point ($\tau$) as

$$Freq_{\tau,\pi,z} = \rho_z \cdot \tilde{D}_{\tau,\pi,z}, \tag{11}$$

since we know the peak demand $\max\left[\tilde{D}_{\tau,\pi,z}\right]$ in each zone in a investment period $\pi$. Additionally, we propose to model the state-level tertiary reserves in *GridPath-India* as spinning reserves ($Spin$), and project them as

$$Spin_{\tau,\pi,z} = Freq_{\tau,\pi,z} + 0.5 \cdot B_z^{\max}, \tag{12}$$

the secondary reserves plus half the largest unit capacity ($B_z^{\max}$) in each zone $z$ and for each time point ($\tau$).

The projections for regional secondary and tertiary reserves are equal to $Spin_{\tau,\pi,r} = Freq_{\tau,\pi,r} = \rho_r \cdot \max\left[\tilde{D}_{\tau,\pi,r}\right]$ in each period, and we use the ratio between the regional secondary reserves and the peak demand in FY23, $\rho_r = Freq_{\pi=2023,r} / \max\left[D_{\tau,\pi=2023,r}\right]$ for each zone $r$ to determine them.



\end{document}